\documentclass[preprint,authoryear,11pt]{elsarticle}

\usepackage{graphics}
\usepackage{graphicx}
\graphicspath{{Figures/}}
\usepackage{epsfig}
\usepackage{subfigure}
\usepackage{amssymb}
\usepackage{amsthm}
\usepackage{bm}
\usepackage{amsmath}
\usepackage{color}
\usepackage{geometry}
\usepackage{setspace}

\allowdisplaybreaks
\usepackage{enumerate}
\usepackage{booktabs}

\usepackage{hyperref}
\hypersetup{
	colorlinks=true,       
}

\biboptions{}

\journal{}

\begin{document}

\begin{frontmatter}



\title{Effects of strain stiffening and electrostriction on tunable elastic waves  \\ in compressible dielectric elastomer laminates}


\author[1]{Yingjie Chen}
\author[2]{Bin Wu\corref{cor1}}
\ead{bin.wu@nuigalway.ie}
\author[2]{Yipin Su}
\author[1,3]{Weiqiu Chen}

\cortext[cor1]{Corresponding author. }
\address[1]{Key Laboratory of Soft Machines and Smart Devices of Zhejiang Province \\ and Department of Engineering Mechanics,\\ Zhejiang University, Hangzhou 310027, P.R. China;\\[6pt]}
\address[2]{School of Mathematics, Statistics and Applied Mathematics,\\ NUI Galway, University Road, Galway, Ireland; \\[6pt]}
\address[3]{Soft Matter Research Center, Zhejiang University, Hangzhou 310027, China. }

\begin{abstract}
	
This paper presents an electromechanical analysis of the nonlinear static response and the superimposed small-amplitude wave characteristics in an infinite periodic compressible dielectric elastomer (DE) laminate subjected to electrostatic excitations and prestress in the thickness direction. The enriched Gent material model is employed to {\color{red}account for} the effects of strain stiffening and electrostriction of the DE laminate. The theory of nonlinear electroelasticity and related linearized incremental theory are exploited to derive the governing equations of nonlinear response and the dispersion relations of incremental shear and longitudinal waves.  Numerical results reveal that the snap-through instability of a Gent DE laminate resulting from geometrical and material nonlinearities can be used to achieve a sharp transition in the position and width of wave band gaps. Furthermore, the influence of material properties (including Gent constants,  {\color{red}the second strain invariant} and electrostrictive parameters) and that of prestress on the snap-through instability and the electrostatic tunability of band gaps for both shear and longitudinal waves are discussed in detail. The electrostrictive effect and prestress {\color{red}are} beneficial to stabilizing the periodic DE laminate. Depending on whether the snap-through instability occurs or not, a continuous variation or a sharp transition in wave band gaps can be realized by varying the electric stimuli.  Our numerical findings are expected to provide a solid guidance for the design and manufacture of soft DE wave devices with tunable band structures.

\end{abstract}


\end{frontmatter}




\section{Introduction}

Dielectric elastomers (DEs) are one type of smart soft materials that can deform and change their physical properties dramatically in response to an external electric stimulus. Owing to their fascinating characters, such as large controllable deformation, light weight, high fracture toughness, and high sensitivity, DEs have been widely used in various engineering applications ranging  {\color{red} from energy harvesters, soft robots and sensing devices to tunable resonators \citep{koh2010dielectric, huang2013maximizing, viry2014flexible, li2017fast, lu2013acoustic, lu2015electronically, sugimoto2013lightweight, yu2017tunable, chen2019dielectric}. In particular, \citet{lu2013acoustic} designed and investigated experimentally a DE absorber with a back cavity, and achieved broadband frequency absorptions of noise. The same authors \citep{lu2015electronically} developed a new type of duct silencer made of DE actuators whose acoustic characteristics can be actively tuned by voltage. A semicylindrical push-pull acoustic transducer composed of DE films was proposed by \citet{sugimoto2013lightweight} for the supression of harmonic distortion. \citet{yu2017tunable} realized the broadband sound attenuation in a novel acoustic metamaterial consisting of an array of DE resonators, which can be tuned by either electric or mechanical biasing fields. Recently, a comprehensive electromechanical characterization of different DE materials was conducted experimentally by \citet{chen2019dielectric} to promote practical applications of DE actuators and generators.}

It is well-known that large deformations induced by the electric field will significantly affect the electromechanical properties of DEs, which may provide an effective approach to manipulating the acoustic/elastic waves in these soft matters. In recent years, many efforts have been devoted to studying small-amplitude wave propagation characteristics in DEs under the application of electric and mechanical biasing fields following the theory of nonlinear electroelasticity and its associated linearized incremental theory developed by \citet{dorfmann2005nonlinear, dorfmann2010electroelastic}. For example, \citet{dorfmann2010electroelastic} investigated the propagation of Rayleigh-type surface waves superimposed on a finitely deformed electroactive (EA) half-space. The longitudinal axisymmetric waves propagating in solid and hollow DE cylinders under combined action of electric and mechanical loadings {\color{red}were} explored by {\color{red}\citet{chen2012waves} and \citet{shmuel2013axisymmetric}, respectively}. Then, \citet{shmuel2015manipulating} discussed the possibility to exploit the applied voltage to manipulate torsional elastic waves in DE tubes. Adopting proper displacement functions, the non-axisymmetric wave motions in an infinite incompressible soft EA hollow cylinder with uniform biasing fields were studied by \citet{su2016propagation}. For soft EA tubes under inhomogeneous electromechanical biasing fields, \citet{wu2017guided} employed the State Space Method to analyze the guided circumferential waves propagation and pointed out that it is feasible to design self-sensing EA actuator and conduct ultrasonic non-destructive detection of soft EA structures via the circumferential waves. Recently, an experimental apparatus was designed by \citet{ziser2017experimental} to excite and measure elastic waves in a DE film, and the experimental results demonstrated that the wave velocity of the fundamental flexural wave could be slowed down under voltage.

Most of the existing works on wave propagation adopt the ideal dielectric model with constant dielectric permittivity to characterize the soft EA materials. In fact, the electrostrictive effect describing the deformation-dependent permittivity has been experimentally observed \citep{wissler2007electromechanical,li2011effect} and plays an important role in the electromechanical analysis of DEs. \citet{zhao2008electrostriction} proposed a thermodynamic electrostrictive model for DEs and pointed out that the electrostrictive effect may become pronounced at large deformations and should not be neglected. Utilizing the enriched DE model, \citet{gei2014role} discussed the role of electrostriction on the stability of a homogeneous DE actuator and made a conclusion that electrostriction has a trend to stabilizing the DE actuator. \citet{liu2015stability} developed a theoretical material model coupling electrostriction with polarization, and proved that the stability of DE/carbon nanotube composites can be affected significantly by the electrostrictive effect. Based on the Hessian matrix method, \citet{su2018optimizing} presented a theoretical analysis of parameter optimization (including the strain-stiffening and electrostrictive parameters) to achieve giant deformations of an incompressible DE plate. {\color{red}As regards the investigation on elastic waves}, \citet{galich2016manipulating} found that whether accounting for the electrostrictive effect into the material model or not evidently influences shear and pressure bulk wave propagation behaviors in compressible DEs subjected to an electrostatic field. {\color{red}Although the existing experimental results of the electrostrictive effect are still contradictory, there is no definite conclusion yet. For example, \citet{chen2019dielectric} argued that the dielectric permittivity parameters for three DE materials are deformation-independent, but they also pointed out that there exist experimental errors and the measurement of electrostrictive parameters also depends on the electrode and DE type, DE size and shape, interfacial integrity, edge effects and other factors. Thus, the electrostrictive effect may appear in different DE materials and it should be of practical value to study its effect on the nonlinear response and wave propagation behaviors in DE structures.} 

Since the concept of phononic crystals (i.e. PCs, which are intrinsically periodic composites) was first proposed in the 1990s \citep{kushwaha1993acoustic}, their unique properties have attracted intensive research interests and provided new avenues for the manipulation of acoustic/elastic waves. Due to the Bragg scattering mechanism or the locally resonant mechanism \citep{liu2000locally}, band gap is the most important feature of PCs, which corresponds to the frequency range where waves are forbidden to propagate in media \citep{li2012elastic,yuan2019tuning,wang2020tunable}. Many acoustic devices with desired superior functions have been designed on the foundation of band gaps \citep{soukiassian2007acoustic,liang2009acoustic, gao2017low, ma2018bilayer, huang2018extension, gao2018robustly, chen2019tunable}. However, once the structures are fabricated with specified geometrical and material parameters, the operating frequencies of most traditional PCs are generally fixed. Therefore, in order to realize real-time control of waves, a large number of researches have been carried out to acquire high-performance PCs with broad tunable band gaps. For example, in a two-dimensional PC composed of air and quartz, \citet{huang2005temperature} found that the band structure of both bulk modes and surface modes can be tuned by the temperature variation. \citet{yang2014significant} showed experimentally that the band gap of a magneto-mechanical PC is highly sensitive to a small magnetic field due to the magnetic torque effect on the coupling between Bragg scattering and local resonant modes. \citet{lian2016enhanced} proposed a piezoelectric PC connected with resonant shunting circuits and indicated that both central frequency and bandwidth of the locally resonant band gaps decrease with the increase of resistance and inductance. A soft membrane-type acoustic metamaterial was designed by \citet{zhou2018actively}, and its band gap strongly depends on the applied pre-stretch when choosing Gent material model. By employing periodic electrical boundary conditions, an innovative design to generate actively tunable topologically protected interface mode in a homogeneous piezoelectric rod system was proposed by \citet{zhou2020actively}. \citet{matar2013tunable} and \citet{wang2020tunable} provided literature overview on tunable PCs by diverse tuning methods and demonstrated a brilliant prospect in optional manipulation of waves via tunable PCs. More recently, the research status, design principles, future development and challenges were reviewed by \citet{bertoldi2017flexible} for soft or flexible mechanical PCs and metamaterials (MMs).

Among the researches on tunable PCs, some studies have been conducted to use large deformation and electric stimuli to tune wave propagation behaviors in PCs composed of soft EA materials. \citet{yang2007tunable} carried out the first piece of work to explore the electro-mechanical coupling effects in soft DE PCs, and achieved the tunable band gaps in a two-dimensional PCs consisting of hollow DE cylinders immersed in an air background. A pre-stretched DE membrane-type waveguide was proposed by \citet{gei2010controlling} for the purpose of tuning bandgap position at will. \citet{shmuel2012band} investigated the incremental thickness-shear waves propagating in a DE laminate composed of two alternating neo-Hookean (N-H) phases, and analyzed the tunability of band gaps under electric biasing field. Then, \citet{getz2017band} found that the band gaps of DE composite plates can be adjusted by the applied voltage and suggested its application to actively tunable waveguides and isolators.  \citet{galich2017shear} re-checked the results in \citet{shmuel2012band} and a different conclusion was obtained that the band gaps of shear wave propagating in periodic Gent DE laminates may be widened and lifted up via the applied electric stimuli, while the band gaps keep unchanged for the N-H material model. Recently, \citet{zhu2018tunable} obtained the dispersion relation and transmission behaviors of shear horizontal wave propagating in the periodic DE laminate and demonstrated the tunable effect of external electric stimuli on shear wave band gaps. In addition, by adopting a genetic algorithm approach, \citet{bortot2018topology} utilized the topological optimization to maximize the band-gap width of anti-plane shear waves propagating in DE fiber composites.

Nevertheless, all of the above-mentioned works take no account of the instabilities that commonly occur in soft EA structures. It is worth noting that different kinds of instabilities can be used to enhance the electrostatic manipulability for acoustic/elastic waves in soft EA PCs, which is an interesting topic and has received some attention. \citet{bortot2017tuning} made use of the plane wave expansion method to explore the electrostatic tunability of an array of DE tubes surrounded by air, and showed that the snap-through instability can be used to achieve an abrupt transition in band gaps. Recently, \citet{wu2018tuning} investigated the longitudinal waves propagating in a one-dimensional PC cylinders tuned by axial force and voltage, and obtained a sudden and enormous change in the band gaps through the snap-through transition emerging only in the Gent DE materials. Using the finite-element-based numerical simulations,
\citet{jandron2018numerical} presented the possibility to harness electromechanical pull-in instability and macroscopic instability to achieve enhanced electrical tunability of waves in soft DE composites. The snap-through instability induced by both electric voltage and internal pressure was also exploited by \citet{mao2019electrostatically} to realize the resonant frequency jump of soft EA balloons.


The primary objective of the present investigation is to shed light on the effects of material properties (including strain stiffening and electrostriction) and external loadings (including electric stimuli and prestress) on the nonlinear response and the superimposed shear and longitudinal waves in a periodic two-phase compressible DE laminate. The research motivations are the following {\color{red}four} important aspects: (1) The material compressibility needs to be considered when analyzing the longitudinal wave propagation; (2) How the snap-through instability due to the strain-stiffening effect influences the static and dynamic behaviors of periodic DE laminate; {\color{red}(3) How the second strain invariant affects the nonlinear static response and wave propagation in the DE laminate;} (4) To what extent of the electric stimuli, the electrostrictive effect will become significant.

This paper is outlined as follows. A brief summary of basic formulations of the nonlinear electroelasticity theory and related linearized incremental theory is described in Section~\ref{section2}. The nonlinear deformation of a periodic compressible DE laminate is considered in Section~\ref{Sec3} for the enriched Gent material model. Combining the transfer matrix method with the Bloch-Floquet theorem, Section~\ref{Sec4} derives the dispersion relations for the incremental shear and longitudinal waves. Then numerical discussions in Section~\ref{section5} elucidate the effects of strain stiffening, {\color{red}the second strain invariant}, electrostriction and prestress on the nonlinear deformation and the band structures of periodic DE laminate subjected to the electric stimuli. Some conclusions are finally made in Section~\ref{section6}.

\section{Preliminary formulations}
\label{section2}

This section briefly reviews the general theory of nonlinear electroelasticity and the associated linearized incremental theory. Interested readers are referred to the papers of \citet{dorfmann2005nonlinear,dorfmann2006nonlinear,dorfmann2010electroelastic} and the monograph by \citet{dorfmann2014nonlinear} for more detailed discussions about the basic ideas. 


\subsection{Nonlinear electroelasticity theory}\label{section2.1}


Consider a deformable electroelastic body that occupies a region ${{\mathcal{B}}_{r}}$ in the Euclidean space with the boundary $\partial {{\mathcal{B}}_{r}}$ and the outward unit normal $\mathbf{N}$ in the undeformed `reference configuration' at time ${{t}_{0}}$. An arbitrary material point in this state is labelled by its position vector $\mathbf{X}$. The body moves to a region ${{\mathcal{B}}_{t}}$ with the boundary $\partial {{\mathcal{B}}_{t}}$ and the outward unit normal $\mathbf{n}$ at time $t$, if subjected to a motion $\mathbf{x}=\bm{\chi }\text{(}\mathbf{X},t\text{)}$, where $\bm{\chi }$ is a vector field function with a sufficiently regular property and $\mathbf{x}$ is the new location of the material point associated with $\mathbf{X}$ in the `current configuration'. The deformation gradient tensor is defined as $\mathbf{F}=\text{Grad }\bm{\chi }$, where Grad is the gradient operator with respect to ${{\mathcal{B}}_{r}}$, and ${{\mathbf{F}}_{i\alpha }}=\partial {{x}_{i}}/\partial {{X}_{\alpha }}$ in Cartesian components. Note that Greek and Roman indices are associated with ${{\mathcal{B}}_{r}}$ and ${{\mathcal{B}}_{t}}$, respectively. We will adopt the summation convention for repeated indices. The relations between the infinitesimal line element $\text{d}\mathbf{X}$, surface element $\text{d}A$ and volume element $\text{d}V$ in ${{\mathcal{B}}_{r}}$ and those in ${{\mathcal{B}}_{t}}$ are connected by $\text{d}\mathbf{x}=\mathbf{F}\text{d}\mathbf{X}$, the well-known Nanson's formula $\mathbf{n}\text{d}a=J{{\mathbf{F}}^{-\text{T}}}\mathbf{N}\text{d}A$ and $\text{d}v=J\text{d}V$, respectively. Here the superscript $^{\text{T}}$ signifies the transpose of a second-order tensor if not otherwise stated and $J=\det \mathbf{F}>0$ denotes the local measure of volume change. The left and right Cauchy-Green tensors are defined as $\mathbf{b=F}{{\mathbf{F}}^{\text{T}}}$ and $\mathbf{C=}{{\mathbf{F}}^{\text{T}}}\mathbf{F}$ that will be employed as the deformation measures.
 
Here, under the assumption of the quasi-electrostatic approximation, Gauss's law and Faraday's law in the absence of free charges and currents can be read as
\begin{equation} \label{1}
 {\rm{div}}{\kern 1pt} {\bf{D}} = 0,\quad {\rm{   curl}}{\kern 1pt} {\bf{E}} = {\bf{0}},
\end{equation}
where `curl' and `div' denote the curl and divergence operators in the current configuration ${{\mathcal{B}}_{t}}$; $\mathbf{D}$ and $\mathbf{E}$ are the electric displacement and electric field vectors in ${{\mathcal{B}}_{t}}$, respectively. In addition, in the absence of mechanical body forces, the equations of motion can be written as
\begin{equation} \label{2}
 {\rm{div}}{\kern 1pt} {\bm{\tau }} = \rho {{\bf{x}}_{,tt}},
\end{equation}
where $\bm{\mathbf{\tau }}$ is the so-called `total Cauchy stress tensor' including the contribution of electric body forces, $\rho ={{\rho }_{0}}{{J}^{-1}}$ is the mass density in the deformed configuration ${{\mathcal{B}}_{t}}$ (${{\rho }_{0}}$ is the initial material mass density in ${{\mathcal{B}}_{r}}$), and the subscript $t$ following a comma represents the material time derivative. The conservation of angular momentum ensures the symmetry of $\bm{\mathbf{\tau }}$.
 
In this paper, {\color{red}we will consider a periodic DE laminate that is coated with flexible electrodes on the leftmost and rightmost surfaces with equal and opposite surface free charges, as shown in Fig.~\ref{1}. Then, there is no electric field in the surrounding vacuum by Gauss's theorem.} As to be shown in Subsec.~\ref{Sec3}, the electric displacement field in the DE laminate is uniformly distributed in the thickness direction. Therefore, the electric boundary conditions on $\partial {{\mathcal{B}}_{t}}$ yield
\begin{equation} \label{3}
{\bf{E}} \times {\bf{n}} = 0,\quad {\rm{   }}{\bf{D}} \cdot {\bf{n}} =  - {\sigma _{\rm{f}}},
\end{equation}
where ${{\sigma }_{\text{f}}}$ is the free surface charge density on $\partial {{\mathcal{B}}_{t}}$. Furthermore, the mechanical boundary condition can be expressed in Eulerian form as
\begin{equation} \label{4}
 {\bm{\tau n}} = {{\bf{t}}^a},
\end{equation}
where ${{\mathbf{t}}^{a}}$ is the prescribed mechanical traction vector per unit area of $\partial {{\mathcal{B}}_{t}}$. For the corresponding Lagrangian counterparts of Eqs.~\eqref{1}-\eqref{4}, we refer to the works of \citet{dorfmann2005nonlinear,dorfmann2006nonlinear}.
 
According to the theory of nonlinear electroelasticity \citep{dorfmann2006nonlinear}, the nonlinear constitutive relations for a compressible electroelastic material in terms of the total energy density function $\Omega (\mathbf{F},\bm{\mathcal{D}})$ per unit reference volume are given by
\begin{equation} \label{5}
{\bf{T}} = \frac{{\partial \Omega }}{{\partial {\bf{F}}}},\quad \bm{{\cal E}} = \frac{{\partial \Omega }}{{\partial \bm{{\cal D}}}},
\end{equation}
where $\mathbf{T}=J{{\mathbf{F}}^{-1}}\mathbf{\bm{{\tau }}}$, $\mathcal{\bm{{\cal D}}}=J{{\mathbf{F}}^{-1}}\mathbf{D}$ and $\mathcal{\bm{{\cal E}}}={{\mathbf{F}}^{\text{T}}}\mathbf{E}$ are the `total nominal stress tensor', Lagrangian electric displacement and electric field vectors, respectively. Thus, the nonlinear constitutive relations expressed by $\bm{{\tau }}$ and $\mathbf{E}$ can be derived as $\bm{{\tau }} = J^{-1}\mathbf{F}{{\partial \Omega }}/{{\partial {\bf{F}}}}$ and $\mathbf{E} = {{\mathbf{F}}^{-\text{T}}}{{\partial \Omega }}/{{\partial \bm{{\cal D}}}}$. For a compressible isotropic electroelastic material, the form of $\Omega $ can be written as $\Omega=\Omega(I_1,I_2,I_3,I_4,I_5,I_6)$ depending on the following six invariants:
\begin{equation} \label{6}
\begin{split}
{I_1} &= {\rm{tr}}{\bf{C}},\quad {I_2} = [{({\rm{tr}}{\bf{C}})^2} - {\rm{tr}}({{\bf{C}}^2})]/2,\quad {I_3} = \det {\bf{C}} = {J^2},\\
{I_4} &= {\bm{{\cal D}}} \cdot {\bm{{\cal D}}},\quad {I_5} = {\bm{{\cal D}}} \cdot ({\bf{C}}{\bm{{\cal D}}}),\quad {I_6} = {\bm{{\cal D}}} \cdot ({{\bf{C}}^2}{\bm{{\cal D}}}),
\end{split}
\end{equation}
which are substituted into Eq.~\eqref{5} to yield the explicit forms of total stress tensor $\mathbf{\bm{\tau} }$ and electric field vector $\mathbf{E}$ as \citep{dorfmann2010electroelastic}
\begin{equation} \label{7}
\begin{split}
J{\bm{\tau }} &= 2{\Omega _1}{\bf{b}} + 2{\Omega _2}({I_1}{\bf{b}} - {{\bf{b}}^2}) + 2{I_3}{\Omega _3}{\bf{I}} + 2{I_3}{\Omega _5}{\bf{D}} \otimes {\bf{D}} + 2{I_3}{\Omega _6}({\bf{D}} \otimes {\bf{bD}} + {\bf{bD}} \otimes {\bf{D}}), \\
{\bf{E}} &= 2J({\Omega _4}{{\bf{b}}^{ - 1}}{\bf{D}} + {\Omega _5}{\bf{D}} + {\Omega _6}{\bf{bD}}),
\end{split}
\end{equation}
where ${{\Omega }_{m}}=\partial \Omega /\partial {{I}_{m}} (m=1-6)$. We note that the last two terms of the expression for $J\mathbf{\bm{\tau} }$ in \citet{dorfmann2010electroelastic} contain typographical errors, which have been corrected in Eq.~\eqref{7}.
 
\subsection{The linearized incremental theory}\label{Sec2-2}
 
Based on the theoretical framework developed by \citet{dorfmann2010electroelastic}, we reproduce the governing equations for time-dependent infinitesimal incremental changes in both motion $\dot{\mathbf{x}}\left( \mathbf{X},t \right)$ and electric displacement vector superimposed on a deformed soft electroelastic body with a static finite deformation $\mathbf{x}=\bm{\chi }\left( \mathbf{X} \right)$ and an electric displacement vector $\mathbf{D}\left( \mathbf{X} \right)$. A linearized incremental theory can be established by employing the perturbation method due to the infinitesimality of incremental mechanical and electrical fields. Here and thereafter a superposed dot signifies the increment in the quantities concerned. 
 
In updated Lagrangian form, the incremental governing equations can be written as
\begin{equation} \label{8}
{\rm{div}}{\kern 1pt} {\dot {\bm{{\cal D}}}_0} = 0,\quad {\rm{   curl}}{\kern 1pt}{\dot {\bm{{\cal E}}}_0} = \bf{0},\quad {\rm{   div}}{{\bf{\dot T}}_0} = \rho {{\bf{u}}_{,tt}},
\end{equation}
where ${{\dot{\mathcal{{\bm{{\cal D}}}}}}_{0}}$, ${{\dot{\mathcal{\bm{{\cal E}}}}}_{0}}$ and ${{\mathbf{\dot{T}}}_{0}}$ are the `push forward' versions of corresponding Lagrangian increments that update the reference configuration from the original unstressed configuration ${{\mathcal{B}}_{r}}$ to the initial static deformed configuration ${{\mathcal{B}}}$, the subscript 0 identifies the resulting `push forward' variables, and $\mathbf{u}(\mathbf{x},t)=\dot{\mathbf{x}}\left( \mathbf{X},t \right)$ is the incremental displacement vector. For a \emph{compressible} soft electroelastic material, the linearized incremental constitutive laws in terms of the increments ${{\mathbf{\dot{T}}}_{0}}$ and ${{\dot{\mathcal{{\bm{{\cal E}}}}}}_{0}}$ can be obtained from Eq.~\eqref{5} and the push-forward operation as
\begin{equation} \label{9}
{{\bf{\dot T}}_0} = {{\bm{{\cal A}}}_0}{\bf{H}} + {{\bm{{\cal M}}}_0}{\dot {\bm{{\cal D}}}_0},\quad{\rm{   }}{\dot {\bm{{\cal E}}}_0} = {\bm{{\cal M}}}_0^{\rm{T}}{\bf{H}} + {{\bm{{\cal R}}}_0}{\dot {\bm{{\cal D}}}_0},
\end{equation}
where $\mathbf{H}=\text{grad}\mathbf{u}$ represents the incremental displacement gradient tensor with `grad' being the gradient operator in ${{\mathcal{B}}}$. It should be emphasized that the superscript $^{\text{T}}$ in Eq.~\eqref{9} denotes the transpose of a third-order tensor between the first two subscripts (that always go together) and the third subscript (i.e. ${\bm{{\cal M}}}_0^{\rm{T}}{\bf{H}}={{\cal M}_{0ijk}}H_{ij}$). The three \emph{instantaneous} electroelastic moduli tensors ${{\bm{{\cal A}}}_{0}}$, ${{\bm{{\cal M}}}_{0}}$ and ${{\bm{{\cal R}}}_{0}}$ are given in component form by
\begin{equation} \label{10}
\begin{split}
{{\cal A}_{0piqj}} &= {J^{ - 1}}{F_{p\alpha }}{F_{q\beta }}{{\cal A}_{\alpha i\beta j}}={{\cal A}_{0qjpi}},\quad {{\cal R}_{0ij}} = JF_{\alpha i}^{ - 1}F_{\beta j}^{ - 1}{{\cal R}_{\alpha \beta }}={{\cal R}_{0ji}}, \\
{{\cal M}_{0piq}} &= {F_{p\alpha }}F_{\beta q}^{ - 1}{{\cal M}_{\alpha i\beta }}={{\cal M}_{0ipq}},
\end{split}
\end{equation}
where $\bm{\mathcal{A}}$, $\bm{\mathcal{M}}$ and $\bm{\mathcal{R}}$ denote the \emph{referential} electroelastic moduli associated with $\Omega (\mathbf{F},\bm{\mathcal{D}})$, with their components defined by ${{\mathcal{A}}_{\alpha i\beta j}}={{\partial }^{2}}\Omega /(\partial {{F}_{i\alpha }}\partial {{F}_{j\beta }})$, ${{\mathcal{M}}_{\alpha i\beta }}={{\partial }^{2}}\Omega /(\partial {{F}_{i\alpha }}\partial {{\mathcal{D}}_{\beta }})$ and ${{\mathcal{R}}_{\alpha \beta }}={{\partial }^{2}}\Omega /(\partial {{\mathcal{D}}_{\alpha }}\partial {{\mathcal{D}}_{\beta }})$. 
Moreover, the additional incremental incompressibility condition is given by
\begin{equation}\label{11}
\mbox{div\hskip 1pt}\mathbf{u} =\mbox{tr\hskip 1pt}\mathbf{H}=0.
\end{equation}

Finally, the updated Lagrangian incremental forms of electric and mechanical boundary conditions, which hold on $\partial {{\mathcal{B}}}$, will become
\begin{equation} \label{13}
{\dot {\bm{{\cal E}}}_0} \times {\bf{n}} = 0,\quad {\dot {\bm{{\cal D}}}_0} \cdot {\bf{n}} =  - {\dot \sigma _{{\rm{F0}}}},\quad {\bf{\dot T}}_0^{\rm{T}}{\bf{n}} = {\bf{\dot t}}_0^A,
\end{equation}
where the electrical variables and their increments in the surrounding vacuum have been discarded, and ${{\dot{\sigma }}_{\text{F0}}}$ and $\mathbf{\dot{t}}_{0}^{A}$ are the incremental surface charge density and mechanical traction vector per unit area of $\partial {{\mathcal{B}}}$, respectively.


\section{Nonlinear deformation of an infinite periodic DE laminate}\label{Sec3}

 
\begin{figure}[htbp]
 	\centering	
 	\includegraphics[width=0.75\textwidth]{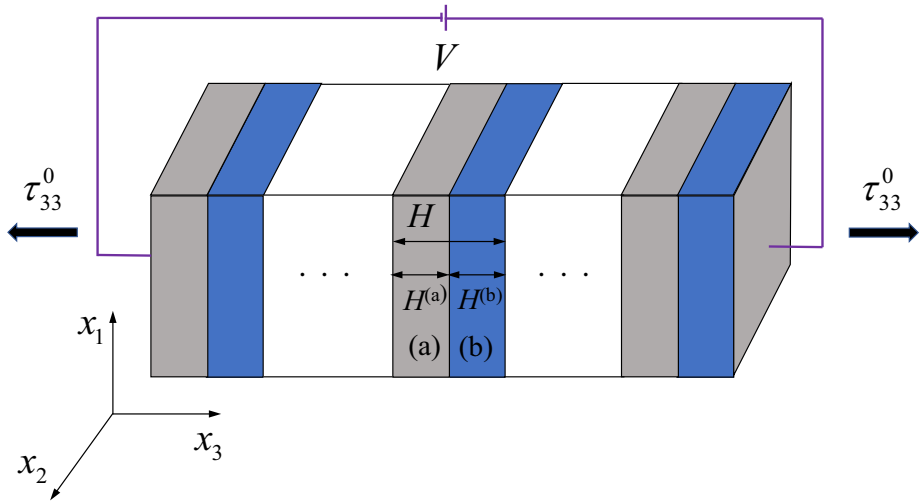}
 	\caption{Sketch of an infinite two-phase periodic DE laminate subjected to voltage $V$ and prestress $\tau _{33}^{0}$ in the thickness direction.}
 	\label{Fig1}
\end{figure}

As shown in Fig.~\ref{Fig1}, we consider an infinite periodic DE laminate composed of two compressible alternating phases denoted by $a$ and $b$, with the thickness of undeformed unit cell being $H={{H}^{\left( a \right)}}+{{H}^{\left( b \right)}}$ along the ${{x}_{3}}$ direction. In the following, the physical quantities of the $p$-th ($p=a, b$) phase are indicated by the superscript ${{\left( \cdot  \right)}^{\left( p \right)}}$. Geometrically, the initial volume fractions of phase $a$ and phase $b$ are ${{\nu }^{\left( a \right)}}={{H}^{\left( a \right)}}/H$ and ${{\nu }^{\left( b \right)}}={{H}^{\left( b \right)}}/H$, respectively. It is assumed that, when subjected to an electric voltage $V$ between the two mechanically negligible electrodes on the leftmost and rightmost surfaces of the laminate as well as a prestress $\tau _{33}^{0}$ along the thickness direction, each layer of the DE laminate is allowed to \emph{expand freely} in the ${{x}_{1}}$ and ${{x}_{2}}$ directions. Now the length of the deformed unit cell and the thicknesses of each phase become
\begin{equation} \label{14}
h = {h^{\left( a \right)}} + {h^{\left( b \right)}},\quad {h^{\left( a \right)}} = \lambda _3^{\left( a \right)}{H^{\left( a \right)}},\quad {h^{\left( b \right)}} = \lambda _3^{\left( b \right)}{H^{\left( b \right)}},
\end{equation}
where $\lambda _{3}^{\left( a \right)}$ and $\lambda _{3}^{\left( b \right)}$ are the stretch ratios of the two phases in the ${{x}_{3}}$ direction, and $h$ is the length of the unit cell in the deformed configuration. In virtue of the perfect bonding condition between the two homogeneous phases and the symmetry of the problem in the $x_1$-$x_2$ plane, we obtain the following relations
\begin{equation} \label{15}
\lambda _1^{\left( a \right)} = \lambda _2^{\left( a \right)} = \lambda _1^{\left( b \right)} = \lambda _2^{\left( b \right)} = \lambda ,
\end{equation}
where $\lambda _{1}^{\left( p \right)}$ and $\lambda _{2}^{\left( p \right)}$ denote the stretch ratios of phase $p$ in the ${{x}_{1}}$ and ${{x}_{2}}$ directions, respectively. Therefore, the deformation gradient tensor of phase $p$ can be expressed as ${{\mathbf{F}}^{\left( p \right)}}=\text{diag}[ \lambda ,\lambda ,\lambda _{3}^{\left( p \right)} ]$ with its components to be solved by boundary conditions. Since the voltage is applied along the ${{x}_{3}}$ direction, it can be derived that $D_{1}^{\left( p \right)}=D_{2}^{\left( p \right)}=0$ and the only nonzero component  $D_{3}^{\left( p \right)}$ of the electric displacement vector is a constant in each phase using Gauss's law [see Eq.~\eqref{1}$_1$]. Combining the continuity condition at the interface between phases $a$ and $b$ with the electric boundary condition [see Eq.~\eqref{3}$_2$] on the leftmost and rightmost surfaces of the laminate, we have
\begin{equation} \label{16}
{D_3} = D_3^{(a)} = D_3^{(b)} =  - {\sigma _{\rm{fR}}}.
\end{equation}
where $\sigma _{\rm{fR}}$ is the free surface charge density on the rightmost deformed surface.

{\color{red}It is well known that the rubber-like materials used in engineering are often slightly compressible and associated with dilatational deformations, such as foamed polyurethane elastomer, silicon rubber (Elite Double 32, Zhermarck) and 3D-print digital material (A85) \citep{holzapfelnonlinear, galich2017elastic}. Moreover, incompressible materials can only support shear waves and thus compressible periodic layered materials should be taken into account to analyze the longitudinal waves \citep{galich2016manipulating, wu2018tuning}.} In order to investigate both the longitudinal and shear wave propagation characteristics in the DE laminate, here we consider the following \emph{compressible enriched} DE model {\color{red} with electrostriction} \citep{galich2016manipulating}
\begin{equation} \label{17}
{\Omega}({\bf{F}},\bm{\mathcal{D}}) = \Omega ^{\text{elas}}({I_1},{I_2},{I_3}) + \frac{1}{{2{\varepsilon}J}}(\gamma _0{I_4} + \gamma _1{I_5} + \gamma _2{I_6}),
\end{equation}
where $\Omega ^{\text{elas}}({{I}_{1}},{{I}_{2}},{{I}_{3}})$ is a purely elastic energy function (such as neo-Hookean, Mooney-Rivlin and Gent {\color{red} or Gent-Gent} models), and in order to incorporate the effect of electric field, the energy density function $\Omega$ also includes those terms linear in the invariants $I_4$, $I_5$ and $I_6$. In Eq.~\eqref{17}, ${{\varepsilon }}={{\varepsilon }_{0}}\varepsilon _{r}$ is the material permittivity of phase $p$ in the undeformed state with ${{\varepsilon }_{0}}(=8.85 \text{pF/m)}$ and $\varepsilon _{r}$ being the vacuum permittivity and the relative permittivity, respectively. In addition, $\gamma _{i}$ $\left( i=0,1,2 \right)$ are dimensionless parameters which satisfy $\gamma _{0}+\gamma _{1}+\gamma _{2}=1$. Note that the compressible enriched DE model \eqref{17} can well capture the behaviour of ideal dielectrics and electrostrictive materials \citep{gei2014role, galich2016manipulating}, and the \emph{ideal} dielectric model can be recovered by imposing $\gamma _{0}=\gamma _{2}=0$ and $\gamma _{1}=1$ in Eq.~\eqref{17}. {\color{red}Note that some representative electrostrictive models have been proposed in the literature, which are discussed and concluded in \ref{AppeA}.}

Substituting Eq.~\eqref{17} into Eq.~\eqref{7} will yield the nonlinear constitutive relations as
\begin{align} \label{consti}
\bm{\tau }&=2{{J}^{-1}}\left[ \Omega _{1}^{\text{elas}}\mathbf{b}+\Omega _{2}^{\text{elas}}({{I}_{1}}\mathbf{b}-{{\mathbf{b}}^{2}})+{{I}_{3}}\Omega _{3}^{\text{elas}}\mathbf{I} \right] \notag \\ 
& -{{\varepsilon }^{-1}}\left[ \left( {{\gamma }_{0}}{{a}_{4}}+{{\gamma }_{1}}{{a}_{5}}+{{\gamma }_{2}}{{a}_{6}} \right)\mathbf{I}/2+{{\gamma }_{1}}\mathbf{D}\otimes \mathbf{D}+{{\gamma }_{2}}(\mathbf{D}\otimes \mathbf{bD}+\mathbf{bD}\otimes \mathbf{D}) \right], \\ \notag
\mathbf{E}&=\frac{1}{\varepsilon}({{\gamma }_{0}}{{\mathbf{b}}^{-1}}+{{\gamma }_{1}}\mathbf{I}+{{\gamma }_{2}}\mathbf{b})\mathbf{D},
\end{align}
where $a_i=I_i/J^2 \text{ } (i=4,5,6)$. Furthermore, for the energy density function $\Omega$ given in Eq.~\eqref{17}, the explicit expressions of instantaneous electroelastic moduli tensors ${{\bm{{\cal A}}}_{0}}$, ${{\bm{{\cal M}}}_{0}}$ and ${{\bm{{\cal R}}}_{0}}$ are derived by \citet{galich2016manipulating} and not shown here for brevity. {\color{red} Note that \citet{vertechy2013continuum} developed a nonlinear fully-coupled thermo-electro-elastic continuum model with electrostriction and their nonlinear constitutive relations are provided in \ref{AppeA}.}

To account for the particular strain-stiffening effect {\color{red} and the second strain invariant}, we adopt the compressible {\color{red}Gent-Gent model \citep{ ogden2004fitting, chen2019dielectric}} for $\Omega ^{\text{elas}} ({{I}_{1}},{{I}_{2}},{{I}_{3}})$ in the following form
\begin{equation} \label{18}
{\Omega ^\text{elas}} =  - \frac{{\mu {J_m}}}{2}\ln(1 - \frac{{{I_1} - 3}}{{{J_m}}}) {\color{red}+ {c_2}\ln \left( {\frac{{{I_2}}}{3}} \right)} - \mu \ln J + \left( {\frac{{{\Lambda _0}}}{2} - \frac{\mu }{{{J_m}}}} \right){(J - 1)^2},
\end{equation}
{\color{red}where ${{\mu }}$, $c_2$ (related to the contribution of $I_2$) and $\Lambda _{0}$ are the elastic moduli and the first Lam$\acute{\text{e}}$'s parameter in the undeformed state, $I_{1}=2{{\lambda }^{2}}+{\lambda _{3}^2 }$ and $I_{2}={{\lambda }^{4}}+2{{\lambda }^{2}{\lambda _{3}^2 }}$ represent the first and second strain invariants, and $\mu'=\mu+2{{c_2 }}/3$ is the infinitesimal initial shear modulus, which results in the bulk modulus calculated as ${{K}}=\Lambda _{0}+2{{\mu' }}/3$.} The parameter $J_{m}$ in Eq.~\eqref{18} is the dimensionless Gent constant reflecting the strain-stiffening effect in the DE laminate and defining the lock-up stretch ratio ${\lambda _{lim}}$ related to the limiting chain extensibility of rubber networks via the relation $J_{m}=2 {\lambda _{lim}^2}+{\lambda _{lim}^{-4}}-3$. Recall that {\color{red} the original Gent model \citep{gent1996new, wang2013effects} can be recovered from Eq.~\eqref{18} if $c_2=0$ and} the Gent model further reduces to the neo-Hookean model when ${{J}_{m}}\to \infty $.
 
Inserting Eqs.~\eqref{17} and \eqref{18} into Eq.~\eqref{consti} and utilizing the relations $\mathbf{b}=\text{diag}[ \lambda^2 ,\lambda^2 ,\lambda _{3}^2]$ as well as $\mathbf{D}=[0,0,D_3]^{\text{T}}$, we derive the principal Cauchy stress components and the nonzero component of Eulerian electric field vector for the $p$ phase as
\begin{equation} \label{19}
\begin{split}
\tau _{11}^{(p)} &=\tau _{22}^{(p)}= {\frac{{{\mu ^{(p)}}}}{{{J^{(p)}}}}\left( {\frac{{J_m^{\left( p \right)}}}{{J_m^{\left( p \right)} - I_1^{\left( p \right)} + 3}}{\lambda ^2} - 1} \right) + 2\left( {\frac{{\Lambda _0^{\left( p \right)}}}{2} - \frac{{{\mu ^{(p)}}}}{{J_m^{\left( p \right)}}}} \right)\left( {{J^{(p)}} - 1} \right)} \\
&{\color{red}+\frac{{2c_2^{\left( p \right)}}}{{{J^{\left( p \right)}}I_2^{\left( p \right)}}}(I_1^{\left( p \right)}{\lambda ^2} - {\lambda ^4})}- \frac{{D_3^2}}{{2{\varepsilon ^{(p)}}}}\left[ {\gamma _0^{(p)}{{(\lambda _3^{(p)})}^{ - 2}} + \gamma _1^{(p)} + \gamma _2^{(p)}{{(\lambda _3^{(p)})}^2}} \right],\\
\tau _{33}^{(p)} &=  {\frac{{{\mu ^{(p)}}}}{{{J^{(p)}}}}\left( {\frac{{J_m^{\left( p \right)}}}{{J_m^{\left( p \right)} - I_1^{\left( p \right)} + 3}}{{(\lambda _3^{(p)})}^2} - 1} \right) + 2\left( {\frac{{\Lambda _0^{\left( p \right)}}}{2} - \frac{{{\mu ^{(p)}}}}{{J_m^{\left( p \right)}}}} \right)\left( {{J^{(p)}} - 1} \right)}\\
&{\color{red}+\frac{{4\lambda _3^{\left( p \right)}c_2^{\left( p \right)}}}{{I_2^{\left( p \right)}}}}+ \frac{{D_3^2}}{{2{\varepsilon ^{(p)}}}}\left[ { - \gamma _0^{(p)}{{(\lambda _3^{(p)})}^{ - 2}} + \gamma _1^{(p)} + 3\gamma _2^{(p)}{{(\lambda _3^{(p)})}^2}} \right],
\end{split}
\end{equation}
and
\begin{equation} \label{20}
E_3^{(p)} = \frac{1}{{{\varepsilon ^{(p)}}}}\left[ {\gamma _0^{(p)}{{(\lambda _3^{(p)})}^{ - 2}} + \gamma _1^{(p)} + \gamma _2^{(p)}{{(\lambda _3^{(p)})}^2}} \right]{D_3},
\end{equation}
where ${{J}^{\left( p \right)}}={{\lambda }^{2}}\lambda _{3}^{\left( p \right)}$ and $I_{1}^{\left( p \right)}=2{{\lambda }^{2}}+{(\lambda _3^{(p)})}^2$. Considering the prestress $\tau _{33}^0$ applied along the ${{x}_{3}}$ direction as well as the traction-free condition in the ${{x}_{1}}$-${{x}_{2}}$ plane, we have 
\begin{equation} \label{21}
\begin{split}
&\tau _{33}^{\left( a \right)} = \tau _{33}^{\left( b \right)} = \tau _{33}^0,\\
&\lambda _{3}^{\left( a \right)}{\nu ^{\left( a \right)}}\tau _{11}^{\left( a \right)} + \lambda _{3}^{\left( b \right)}{\nu ^{\left( b \right)}}\tau _{11}^{\left( b \right)} = 0.
\end{split}
\end{equation}
{\color{red}Here, to derive Eq.~\eqref{21}$_2$, we have followed \citet{shmuel2012band} and \citet{galich2017shear} to assume that the consultant lateral force applied to a unit cell vanishes (instead of vanishing of the lateral normal stresses everywhere) in the sense of Saint-Venant's principle.} In addition, Eq.~\eqref{1}$_2$ indicates a curl-free electric field that can be expressed as $\mathbf{E}=-\text{grad}\phi $ where $\phi $ is an electrostatic potential. In this case, the electric potential drop $\Delta V$ over \emph{each unit cell} of the periodic laminate is calculated as
\begin{equation} \label{22}
\Delta V = \int_0^h {\text{d}{\phi ^{\left( p \right)}} = {V^{\left( a \right)}} + {V^{\left( b \right)}} =-\int_0^h {E_3^{\left( p \right)}\text{d}x_3}=-(\lambda _3^{(a)}E_3^{(a)}{H^{(a)}} + \lambda _3^{(b)}E_3^{(b)}{H^{(b)}}}),
\end{equation}
where ${V^{\left( p \right)}}$ is the electric potential drop in the $p$ phase. By defining the \emph{effective} referential electric field of each unit cell as $E_3^{\text{eff}}=-\Delta V/H$, we obtain from Eq.~\eqref{22}
\begin{equation} \label{equi-E}
E_3^{\text{eff}} = \lambda _3^{(a)}E_3^{(a)}{{\nu }^{\left( a \right)}} + \lambda _3^{(b)}E_3^{(b)}{{\nu }^{\left( b \right)}}.
\end{equation}
Thus, the \emph{overall} electric potential drop (i.e. the electric voltage $V$) between the two electrodes of the laminate is equal to $E_3^{\text{eff}}$ times the total thickness of the periodic laminate.

It is obvious from Eqs.~\eqref{20} and \eqref{equi-E} that $E_3^{\text{eff}}$ depends linearly on ${{D}_{3}}$. Thus, once the applied prestress $\tau _{33}^{0}$ and $E_3^{\text{eff}}$ are given, the stretch ratios $\lambda _{3}^{\left( a \right)}$, $\lambda _{3}^{\left( b \right)}$, $\lambda $ and the electric displacement ${{D}_{3}}$ can be completely determined from Eqs.~\eqref{19}-\eqref{equi-E}.
 

\section{Incremental wave propagation and dispersion relations}\label{Sec4}

 
In this section, the incremental longitudinal and shear waves propagating perpendicular to the DE laminate are superposed upon the finitely deformed configuration induced by the combined action of an electric voltage and a prestress in the thickness direction. In this paper, it is assumed that all physical fields depend on ${{x}_{3}}$ and $t$ only. In order to make the incremental Faraday's law in Eq.~\eqref{8}$_2$ satisfied automatically, an incremental electric potential $\dot{\phi }$ may be introduced such that ${{\dot{\bm{\mathcal{E}}}}_{0}}=-\text{grad}\dot{\phi }$, with the components being ${{\dot{\mathcal{E}}}_{0i}}=-{{\dot{\phi }}_{,i}}$. As a result, the time-harmonic incremental displacement and electric potential can be supposed as
\begin{equation} \label{23}
{u_1} = {U_1}({x_3}){e^{ - \text{i}\omega t}},\quad {u_2} = {U_2}({x_3}){e^{ - \text{i}\omega t}},\quad {u_3} = {U_3}({x_3}){e^{ - \text{i}\omega t}},\quad \dot \phi  = \Phi ({x_3}){e^{ - \text{i}\omega t}},
\end{equation}
where ${U_1}({x_3})$, ${U_2}({x_3})$, ${U_3}({x_3})$ and $\Phi({x_3})$ are the unknown functions, and $\omega $ is the circular frequency.
 
Making use of Eq.~\eqref{23}, the nonzero components of ${\dot {\bm{{\cal E}}}_0}$ and incremental displacement gradient tensor $\mathbf{H}=\text{grad}\mathbf{u}$ are obtained as
\begin{equation} \label{24}
{\dot {\cal E}_{03}} =  - {\dot \phi _{,3}},\quad {H_{13}} = {u_{1,3}},\quad {H_{23}} = {u_{2,3}},\quad {H_{33}} = {u_{3,3}}.
\end{equation}
Consequently, the incremental motion equation~\eqref{8}$_3$ and incremental Gauss's law~\eqref{8}$_1$ are simplified as
\begin{equation} \label{25}
{\dot T_{0pi,p}} = {\dot T_{03i,3}} = \rho {u_{i,tt}},\quad {\dot {\cal D}_{0i,i}} = {\dot {\cal D}_{03,3}} = 0.
\end{equation}
For the compressible Gent DE model in Eqs.~\eqref{17} and \eqref{18}, the expressions of nonzero components of instantaneous electroelastic moduli tensors ${{\bm{\mathcal{A}}}_{0}}$, ${{\bm{\mathcal{M}}}_{0}}$ and ${{\bm{\mathcal{R}}}_{0}}$ are explicitly provided in \ref{AppeB} for reference. Thus, the incremental constitutive relations [see Eq.~\eqref{9}] become
\begin{equation} \label{26}
\begin{split}
{{\dot T}_{031}} &= {{\cal A}_{03131}}{H_{13}} + {{\cal M}_{0311}}{{\dot {\cal D}}_{01}},\\
{{\dot T}_{032}} &= {{\cal A}_{03232}}{H_{23}} + {{\cal M}_{0322}}{{\dot {\cal D}}_{02}},\\
{{\dot T}_{033}} &= {{\cal A}_{03333}}{H_{33}} + {{\cal M}_{0333}}{{\dot {\cal D}}_{03}},\\
{{\dot {\cal E}}_{01}} &= {{\cal M}_{0311}}{H_{13}} + {{\cal R}_{011}}{{\dot {\cal D}}_{01}}{\rm{ = 0}},\\
{{\dot {\cal E}}_{02}} &= {{\cal M}_{0322}}{H_{23}} + {{\cal R}_{022}}{{\dot {\cal D}}_{02}}{\rm{ = 0}},\\
{{\dot {\cal E}}_{03}} & = {{\cal M}_{0333}}{H_{33}} + {{\cal R}_{033}}{{\dot {\cal D}}_{03}}.
\end{split}
\end{equation}

Solving for ${{\dot{\mathcal{D}}}_{0i}}$ in Eqs.~\eqref{26}$_{4-6}$ in terms of ${{\dot{\mathcal{E}}}_{0i}}$, then substituting the resulting expressions into Eqs.~\eqref{26}$_{1-3}$ and using Eq.~\eqref{24}, the incremental constitutive relations are rewritten in terms of $\mathbf{u}$ and $\dot{\phi }$ as
\begin{equation} \label{27}
\begin{split}
{{\dot T}_{031}} &= {c_{55}^*}{u_{1,3}},\quad
{{\dot T}_{032}} = {c_{55}^*}{u_{2,3}},\\
{{\dot T}_{033}} &= {c_{33}^*}{u_{3,3}} + {e_{33}}{{\dot \phi }_{,3}},\quad
{{\dot {\cal D}}_{03}} =   {e_{33}}{u_{3,3}} - {\varepsilon _{33}}{{\dot \phi }_{,3}},
\end{split}
\end{equation}
where ${{c}_{55}^*}={{\mathcal{A}}_{03131}}-\mathcal{M}_{0311}^{2}\text{/}{{\mathcal{R}}_{011}}$, ${{c}_{33}^*}={{\mathcal{A}}_{03333}}-\mathcal{M}_{0333}^{2}\text{/}{{\mathcal{R}}_{033}}$, ${{e}_{33}}=-{{\mathcal{M}}_{0333}}\text{/}{{\mathcal{R}}_{033}}$ and ${{\varepsilon }_{33}}=\text{1/}{{\mathcal{R}}_{033}}$. Substitution of Eq.~\eqref{27} into Eq.~\eqref{25} yields the incremental governing equations as
\begin{equation} \label{28}
\begin{split}
 {{\dot T}_{031,3}} &= {c_{55}^*}{u_{1,33}} = \rho {u_{1,tt}},\quad
 {{\dot T}_{032,3}} = {c_{55}^*}{u_{2,33}} = \rho {u_{2,tt}},\\
 {{\dot T}_{033,3}} &= {c_{33}^*}{u_{3,33}} + {e_{33}}{{\dot \phi }_{,33}} = \rho {u_{3,tt}},\quad
 {{\dot {\cal D}}_{03,3}} =   {e_{33}}{u_{3,33}} - {\varepsilon _{33}}{{\dot \phi }_{,33}} = 0.
\end{split}
\end{equation}
Obviously, the incremental transverse displacements ${{u}_{1}}$ and ${{u}_{2}}$ (defined by Eqs.~\eqref{28}$_{1,2}$) are decoupled from the incremental longitudinal displacement ${{u}_{3}}$ and electric potential $\dot{\phi }$ (governed by Eqs.~\eqref{28}$_{3,4}$), which means that shear waves are decoupled from longitudinal waves that are \emph{directly} coupled with the incremental electric field.

For further simplification, we introduce the following variable
\begin{equation} \label{29}
\dot \psi  = \frac{{{e_{33}}}}{{{\varepsilon _{33}}}}{u_3}-\dot \phi,
\end{equation}
which is substituted into Eq.~\eqref{28}$_{3,4}$ to obtain
\begin{equation} \label{30}
\begin{split}
{{c}_{33}}{u_{3,33}} = \rho {u_{3,tt}},\quad {{\dot \psi }_{,33}} = 0,
\end{split}
\end{equation}
where ${{{c}}_{33}}={{c}_{33}^*}+e_{33}^{2}/{{\varepsilon }_{33}}={{\mathcal{A}}_{03333}}$. Accordingly, inserting Eq.~\eqref{23} into Eqs.~\eqref{28}$_{1,2}$ and \eqref{30}, we obtain the solutions of $\mathbf{u}$ and $\dot{\psi }$ as
\begin{equation} \label{31}
\begin{split}
 {u_1}({x_3},t) &= \left( {{A_1}{e^{{\rm{i}}{k_T}{x_3}}} + {A_2}{e^{ - {\rm{i}}{k_T}{x_3}}}} \right){e^{ - {\rm{i}}\omega t}},\quad
 {u_2}({x_3},t) = \left( {{B_1}{e^{{\rm{i}}{k_T}{x_3}}} + {B_2}{e^{ - {\rm{i}}{k_T}{x_3}}}} \right){e^{ - {\rm{i}}\omega t}},\\
 {u_3}({x_3},t) &= \left( {{P_1}{e^{{\rm{i}}{k_L}{x_3}}} + {P_2}{e^{ - {\rm{i}}{k_L}{x_3}}}} \right){e^{ - {\rm{i}}\omega t}},\quad
 \dot \psi  = \left( {{Q_1}{x_3} + {Q_2}} \right){e^{ - {\rm{i}}\omega t}},
\end{split}
\end{equation}
where ${{k}_{T}}=\sqrt{\rho {{\omega }^{2}}/{{c}_{55}^*}}$ and ${{k}_{L}}=\sqrt{\rho {{\omega }^{2}}/{{{{c}}}_{33}}}$ denote the incremental shear and longitudinal wave numbers, respectively, and $A_1$, $B_1$, etc. are arbitrary constants. Thus, the expression of $\dot{\phi }$ is derived as
\begin{equation} \label{32}
 \dot \phi  = \frac{{{e_{33}}}}{{{\varepsilon _{33}}}}{u_3} -\dot \psi = \left[ {\frac{{{e_{33}}}}{{{\varepsilon _{33}}}}\left( {{P_1}{e^{{\rm{i}}{k_L}{x_3}}} + {P_2}{e^{ - {\rm{i}}{k_L}{x_3}}}} \right)-({Q_1}{x_3} + {Q_2})} \right]{e^{ - {\rm{i}}\omega t}}.
\end{equation}
Furthermore, inserting Eqs.~\eqref{31} and \eqref{32} into Eq.~\eqref{27}, we rewrite the corresponding incremental stresses and electric displacement as
\begin{equation} \label{33}
\begin{split}
 {{\dot T}_{031}} &= {\rm{i}}{k_T}{c_{55}^*}\left( {{A_1}{e^{{\rm{i}}{k_T}{x_3}}} - {A_2}{e^{ - {\rm{i}}{k_T}{x_3}}}} \right){e^{ - {\rm{i}}\omega t}},\\
 {{\dot T}_{032}} &= {\rm{i}}{k_T}{c_{55}^*}\left( {{B_1}{e^{{\rm{i}}{k_T}{x_3}}} - {B_2}{e^{ - {\rm{i}}{k_T}{x_3}}}} \right){e^{ - {\rm{i}}\omega t}},\\
 {{\dot T}_{033}} &= \left[{{c}_{33}}{\rm{i}}{k_L}\left( {{P_1}{e^{{\rm{i}}{k_L}{x_3}}} - {P_2}{e^{ - {\rm{i}}{k_L}{x_3}}}} \right) - {e_{33}}{Q_1}\right]{e^{ - {\rm{i}}\omega t}},\\
 {{\dot {\cal D}}_{03}} & ={\varepsilon _{33}}{Q_1}{e^{ - {\rm{i}}\omega t}},
\end{split}
\end{equation}

Next we shall employ the \emph{transfer-matrix method} \citep{adler1990matrix, shmuel2012band} in conjunction with the Bloch-Floquet theorem to derive the dispersion relations governing the incremental wave motions in the infinite periodic DE laminate. For incremental shear waves, we can choose $\mathbf{S}_n^{\left( p \right)} = {[{u_{1n}^{(p)},\dot{T}_{031n}^{(p)}}]^{\text{T}}}$ or $\mathbf{S}_n^{\left( p \right)} = {[{u_{2n}^{(p)},\dot{T}_{032n}^{(p)}}]^{\text{T}}}$ as the incremental state vector of phase $p$ in the $n$th unit cell of the laminate that should be continuous across the interfaces between two adjacent layers. Based on the Bloch-Floquet theorem, the relation of incremental state vectors of the same phase in adjacent unit cells takes the following form:
\begin{equation} \label{34}
\mathbf{S}_{n+1}^{\left( a \right)} = {e^{ {\rm{i}}k_B h}} \mathbf{S}_n^{\left( a \right)},
\end{equation}
where ${{k}_{B}}$ denotes the Bloch wave number. Thus, through some simple derivations, Eq.~\eqref{34} combined with Eqs.~\eqref{31}$_{1,2}$ and \eqref{33}$_{1,2}$ leads to the dispersion relation of incremental shear waves as
\begin{equation} \label{35}
\cos ({k_B}h) = \cos (k_T^{(a)}{h^{(a)}})\cos (k_T^{(b)}{h^{(b)}}) - \frac{1}{2}\left( {{F_T} + \frac{1}{{{F_T}}}} \right)\sin (k_T^{(a)}{h^{(a)}})\sin (k_T^{(b)}{h^{(b)}}),
\end{equation}
where ${{F}_{T}}=c_{55}^{*(a)}k_{T}^{(a)}/(c_{55}^{*(b)}k_{T}^{(b)})$.
 
For incremental longitudinal waves, the corresponding incremental state vector is taken as $\mathbf{S}_n^{\left( p \right)}={[{u_{3n}^{(p)},\dot{\phi }_{n}^{(p)},\dot{T}_{033n}^{(p)},\dot{\mathcal{D}}_{03n}^{(p)}}]^{\text{T}}}$ and in this case, Eqs.~\eqref{31}$_{3}$, \eqref{32} and \eqref{33}$_{3,4}$ are rewritten in matrix form as
\renewcommand{\arraystretch}{1.4}
\begin{equation} \label{36}
\left[ {\begin{array}{*{20}{c}}
	{{u_3}}\\
	{\dot \phi }\\
	{{{\dot T}_{033}}}\\
	{{{\dot {\cal D}}_{03}}}
	\end{array}} \right] = \left[ {\begin{array}{*{20}{c}}
	{{e^{{\rm{i}}{k_L}{x_3}}}}&{{e^{ - {\rm{i}}{k_L}{x_3}}}}&0&0\\
	{ \frac{{{e_{33}}}}{{{\varepsilon _{33}}}}{e^{{\rm{i}}{k_L}{x_3}}}}&{ \frac{{{e_{33}}}}{{{\varepsilon _{33}}}}{e^{ - {\rm{i}}{k_L}{x_3}}}}&-{{x_3}}&-1\\
	{{\rm{i}}{k_L}{{ c}_{33}}{e^{{\rm{i}}{k_L}{x_3}}}}&{ - {\rm{i}}{k_L}{{ c}_{33}}{e^{ - {\rm{i}}{k_L}{x_3}}}}&{ - {e_{33}}}&0\\
	0&0&{  {\varepsilon _{33}}}&0
	\end{array}} \right]\left[ {\begin{array}{*{20}{c}}
	{{P_1}}\\
	{{P_2}}\\
	{{Q_1}}\\
	{{Q_2}}
\end{array}} \right]{e^{ - {\rm{i}}\omega t}}.
\end{equation}
Similarly, using the continuity condition across the interfaces and the Bloch-Floquet theorem [see Eq.~\eqref{34}], we derive the dispersion relation of incremental longitudinal waves as follows: 
\begin{equation} \label{37}
\cos ({k_B}h) = \cos (k_L^{(a)}{h^{(a)}})\cos (k_L^{(b)}{h^{(b)}}) - \frac{1}{2}\left( {{F_L} + \frac{1}{{{F_L}}}} \right)\sin (k_L^{(a)}{h^{(a)}})\sin (k_L^{(b)}{h^{(b)}}),
\end{equation}
where ${{F}_{L}}={{c}_{33}^{(a)}k_{L}^{(a)}}/({{c}_{33}^{(b)}k_{L}^{(b)}})$.
 
We observe from Eqs.~\eqref{35} and \eqref{37} that except for the difference between $(k_T,{{F}_{T}})$ and $(k_L,{{F}_{L}})$, the dispersion relations of incremental shear and longitudinal waves in the compressible periodic DE laminate are in the same form, which was also shown by \citet{galich2017shear} for the compressible hyperelastic laminate. By numerically solving these two equations, the dispersion relations between ${{k}_{B}}$ and $\omega $ for shear and longitudinal waves can be achieved and the effect of applied voltage and prestress on their band structures can be investigated, which is to be shown in Sec.~\ref{section5}.
 
\section{Numerical examples and discussions}\label{section5}

We now conduct numerical calculations to elucidate how the applied electric field and prestress affect the incremental shear and longitudinal wave propagation in the compressible periodic DE laminate. In particular, attention is mainly focused on the influence of strain-stiffening, {\color{red}the second strain invariant} and electrostriction of different DEs on the tunable elastic waves. {\color{red} We will use the normalized quantities to conduct numerical calculations and} the commercial product Silicone CF19-2186 \citep{shmuel2016manipulating} is just chosen as the referential DE with its material properties given as ${{\rho }_{\text{0}}}=1100$~kg/m$^3$, ${{\mu }_{0}}=333$~kPa and ${{\varepsilon }_{r}}=2.8$, which is used to normalize the related physical quantities. Specifically, the normalized physical quantities indicated by a bar on them are defined in the following form:
\begin{equation} \label{38}
	\begin{split}
		{{\overline \mu }^{\left( p \right)}}& = {\mu ^{\left( p \right)}}/{\mu _0},\quad {{\overline \rho }^{\left( p \right)}} = {\rho ^{\left( p \right)}}/{\rho _0},\quad \overline \Lambda _0^{\left( p \right)} = \Lambda _0^{\left( p \right)}/{\mu _0},
		\\
		 \quad {{\overline \varepsilon }^{\left( p \right)}}& = {\varepsilon ^{\left( p \right)}}/\varepsilon,\quad \bar c_2^{\left( p \right)} = c_2^{\left( p \right)}/{\mu _0},\quad \overline D_{\rm{3}} = D_3/\sqrt {{\mu _0}\varepsilon},
		\\
		 \quad \overline E_3^{\left( p \right)} &= E_3^{\left( p \right)}\sqrt {\varepsilon/{\mu _0}}, \quad \overline E_3^{\text{eff}}=E_3^{\text{eff}}\sqrt {\varepsilon/{\mu _0}}, \quad \overline \tau _{{\rm{33}}}^{\rm{0}}=\tau _{{\rm{33}}}^{\rm{0}}{\rm{/}}{\mu _{\rm{0}}},
	\end{split}
\end{equation}
where $\varepsilon=\varepsilon_0 \varepsilon_r$.

In the following numerical examples, the normalized initial densities, shear moduli, and material permittivities of phases $a$ and $b$ are taken to be ${{\overline{\rho }}^{\left( a \right)}}={{\overline{\rho }}^{\left( b \right)}}=1$, ${{\overline{\mu }}^{\left( a \right)}}=1$, ${{\overline{\mu }}^{\left( b \right)}}=5$, ${{\overline{\varepsilon }}^{\left( a \right)}}=1$ and ${{\overline{\varepsilon }}^{\left( b \right)}}=2$, respectively. The first Lam$\acute{\text{e}}$'s parameter $\Lambda _{0}^{\left( p \right)}$ of phase $p$ is set to be $\Lambda _{0}^{\left( p \right)}=1000{{\mu }^{\left( p \right)}}$ in order to account for material compressibility. {\color{red}The normalized second moduli of phases $a$ and $b$ are taken to be
$\bar c_2^{\left( a \right)} = \bar c_2^{\left( b \right)}={\bar c_2}$.} In addition, the initial volume fractions of phases $a$ and $b$ are fixed as ${{\nu }^{\left( a \right)}} = {{\nu }^{\left( b \right)}} = 0.5$ and the Gent constants of both phases are equal, i.e., $J_{m}^{\left( a \right)}=J_{m}^{\left( b \right)}={{J}_{m}}$.

Furthermore, as in \citet{shmuel2012band} and \citet{galich2017shear}, the normalized Bloch wave number and circular frequency are defined as
\begin{equation} \label{39}
\begin{split}
{{\overline k}_B} &= {k_B}h,\quad \overline \omega  = \omega H/\sqrt {{\mu_0}/{\rho_0}},
\end{split}
\end{equation}
where ${{k}_{B}}$ indicates the Bloch wave number in the so-called first irreducible Brillouin zone, varying from 0 to $\pi /h$ and giving the smallest region where wave propagation is unique \citep{kittel1996introduction}.
 
\subsection{Influence of the strain-stiffening effect}\label{Sec5-1}
  
\begin{figure}[htbp]
  	\centering
 	\setlength{\abovecaptionskip}{0pt}
    \setlength{\belowcaptionskip}{0pt}
  	\includegraphics[width=0.5\textwidth]{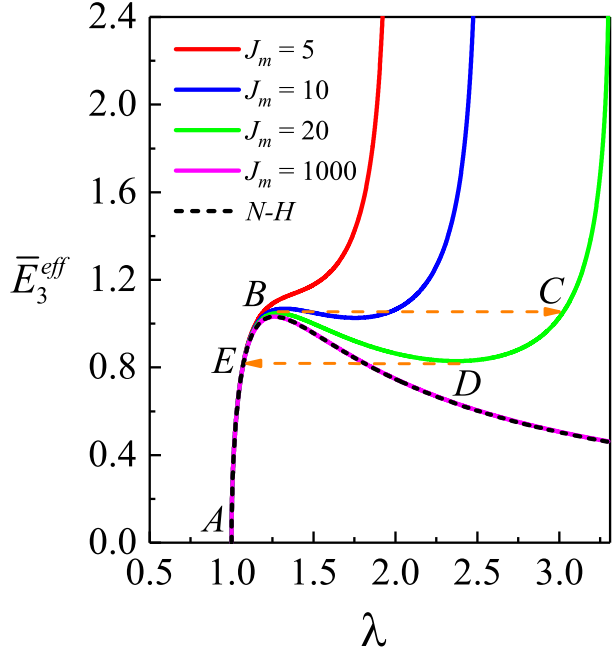}
  	\caption{Nonlinear response of the lateral stretch $\lambda $ to the normalized effective referential electric field $\overline E_3^{\text{eff}}$ in the DE laminate for the Gent model with ${{J}_{m}}=$ 5, 10, 20, 1000 and the neo-Hookean (N-H) model. The snap-through instaibility occurs when the effective electric field reaches a critical value $\overline E_3^{\text{cr}}$.}
  	\label{Fig2}
\end{figure} 
  
Before studying the tunable band structures of incremental waves propagating in the periodic DE laminate, we first examine the influence of strain-stiffening effect on the nonlinear static response of ideal dielectric materials subjected to the electric field only (the prestress, the electrostriction effect {\color{red}and the contribution of $I_2$} are not considered at this moment, i.e., $\tau_{33}^0=\gamma _{0}=\gamma _{2}=0$, $\gamma _{1}=1$ {\color{red}and  ${\bar c_2}=0$}). According to Eqs.~\eqref{21} and \eqref{equi-E}, the nonlinear responses of lateral stretch $\lambda $ versus the normalized effective {\color{red} referential} electric field $\overline E_3^{\text{eff}}$ in the periodic DE laminate are illustrated in Fig.~\ref{Fig2} for both the neo-Hookean (N-H) and Gent models (${{J}_{m}}=$ 5, 10, 20, 1000). As mentioned in Sec.~\ref{Sec3}, the N-H model can be recovered from the Gent model when ${{J}_{m}}$ approaches infinity, which can be verified in Fig.~\ref{Fig2} that the nonlinear curve of the N-H model coincides with that of the Gent model with a very large value ${{J}_{m}}=$ 1000. It can be seen in Fig.~\ref{Fig2} that the onset of snap-through instability exists when the dimensionless Gent parameter ${{J}_{m}}=$ 10 and 20, and the dashed orange arrows in Fig.~\ref{Fig2} indicate the paths of snap-through transition for ${{J}_{m}}=10$ (path $B$-$C$ and path $D$-$E$). Specifically, as the applied electric field $\overline E_3^{\text{eff}}$ increases from zero to a critical value $\overline E_3^{\text{cr}}$ (from state $A$ to state $B$) which is dependent on the material properties, the DE laminate shrinks due to the electro-mechanical coupling effect. Then, a further ascending change in the electric field will trigger the so-called snap-through transition from state $B$ to a stable state $C$, which leads to a drastic change in geometrical configuration. Conversely, a contracting snap-through process from state $D$ to state $E$ may also be generated by a consecutive fall in the effective electric field to the state $D$.

However, the results of the N-H model and the Gent model with ${{J}_{m}}=5$ in Fig.~\ref{Fig2} indicate that the snap-through transition is not accessible. The underlying mechanism is explained as follows. The occurrence of snap-through transition results from the competition between the change in mechanical stress owing to the strain-stiffening effect and that in electrostatic stress counterpart induced by the electric stimuli. Thus, the DE laminate with a larger ${{J}_{m}}$ shows a weaker strain-stiffening effect, and the N-H model presents the electromechanical instability that results in ultimate electric breakdown \citep{zhao2008electrostriction, wu2018tuning}. In turn, with a smaller ${{J}_{m}}$ (for example,${{J}_{m}}=5$) that means a more pronounced strain-stiffening effect, the snap-through transition may not take place under the application of electric field as well, which is due to the fact that DE material has reached the strain-stiffening stage in advance before the electric field achieves the aforementioned critical value $\overline E_3^{\text{cr}}$. It is also noted from Fig.~\ref{Fig2} that, for the Gent model, the critical electric field where the snap-through transition occurs from $B$ to $C$ is almost independent of ${{J}_{m}}$, and the DE laminate will arrive at a stable state $C$ in the vicinity of the lock-up stretch state $\lambda _{lim}$.

\begin{figure}[htbp]
	\centering
 	\setlength{\abovecaptionskip}{0pt}
    \setlength{\belowcaptionskip}{0pt}
	\includegraphics[width=0.45\textwidth]{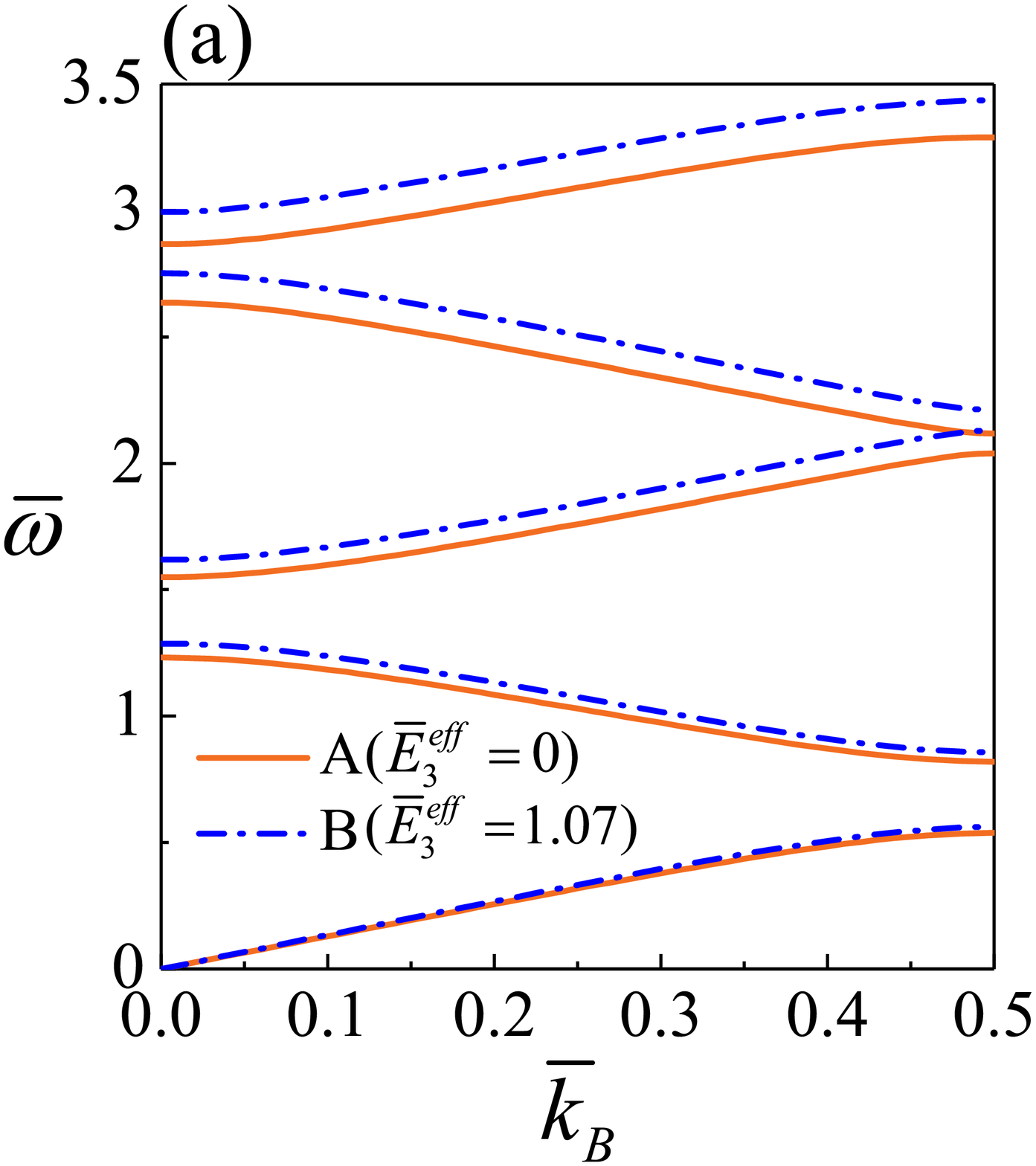}
	\includegraphics[width=0.45\textwidth]{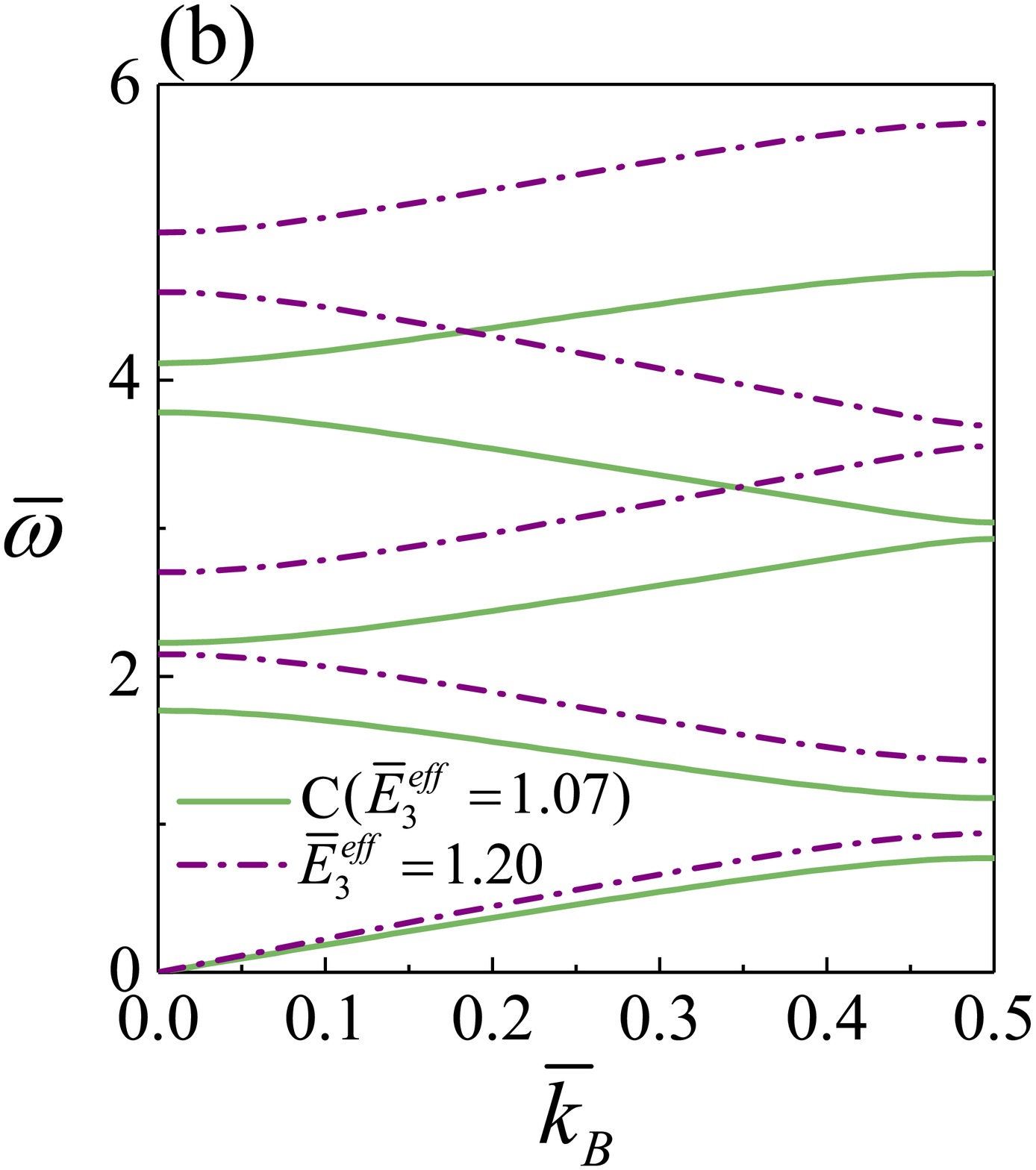}
	\caption{Evolution of the band structures of \emph{shear} waves before (a) and after (b) the snap-through transition from states $B$ to $C$ shown in Fig.~\ref{Fig2} (${{J}_{m}}=10$).}
	\label{Fig3}
\end{figure}
  
\begin{figure}[htb]
  	\centering	
  	\includegraphics[width=0.45\textwidth]{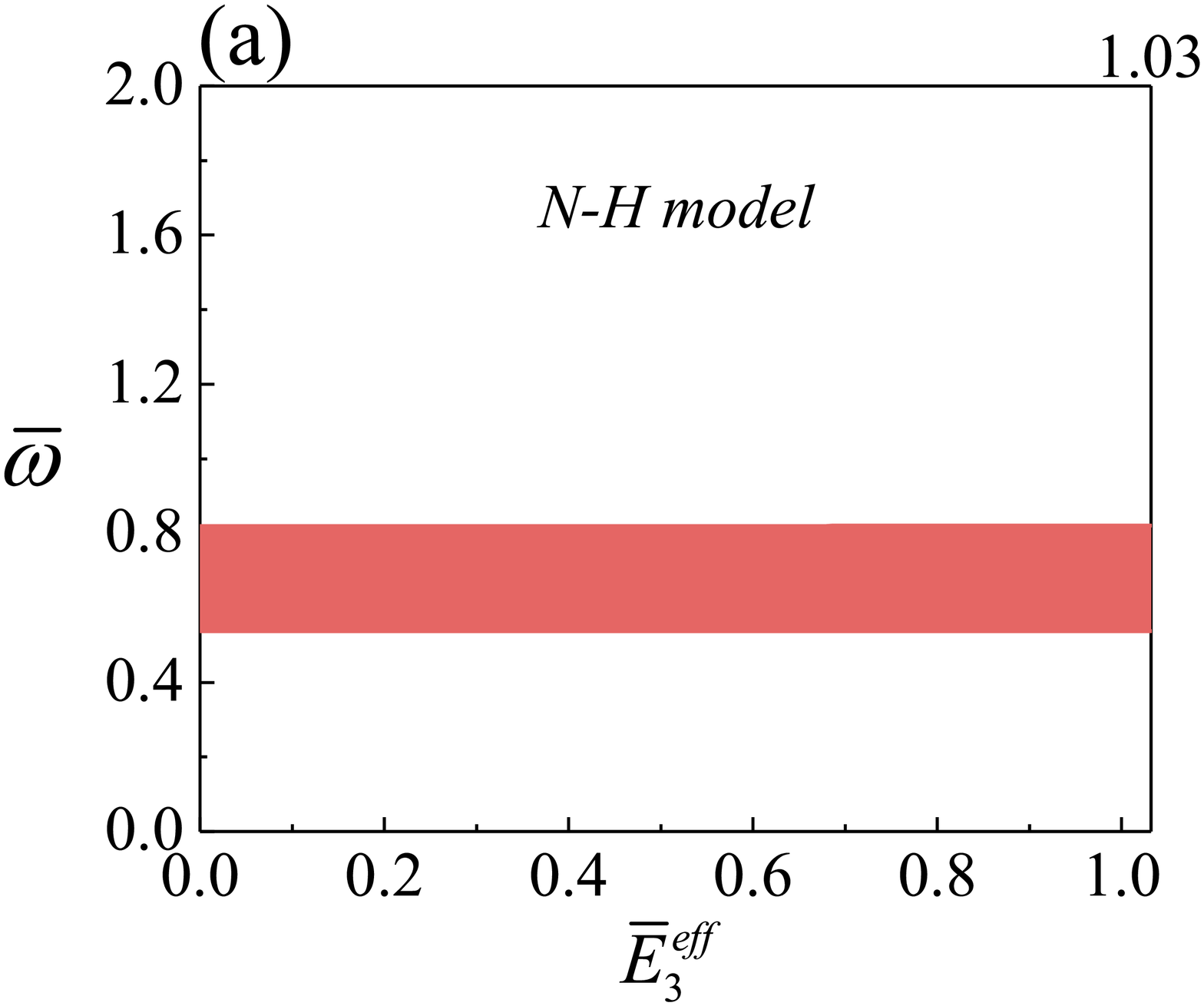} \hspace{0.03\textwidth}
  	\includegraphics[width=0.45\textwidth]{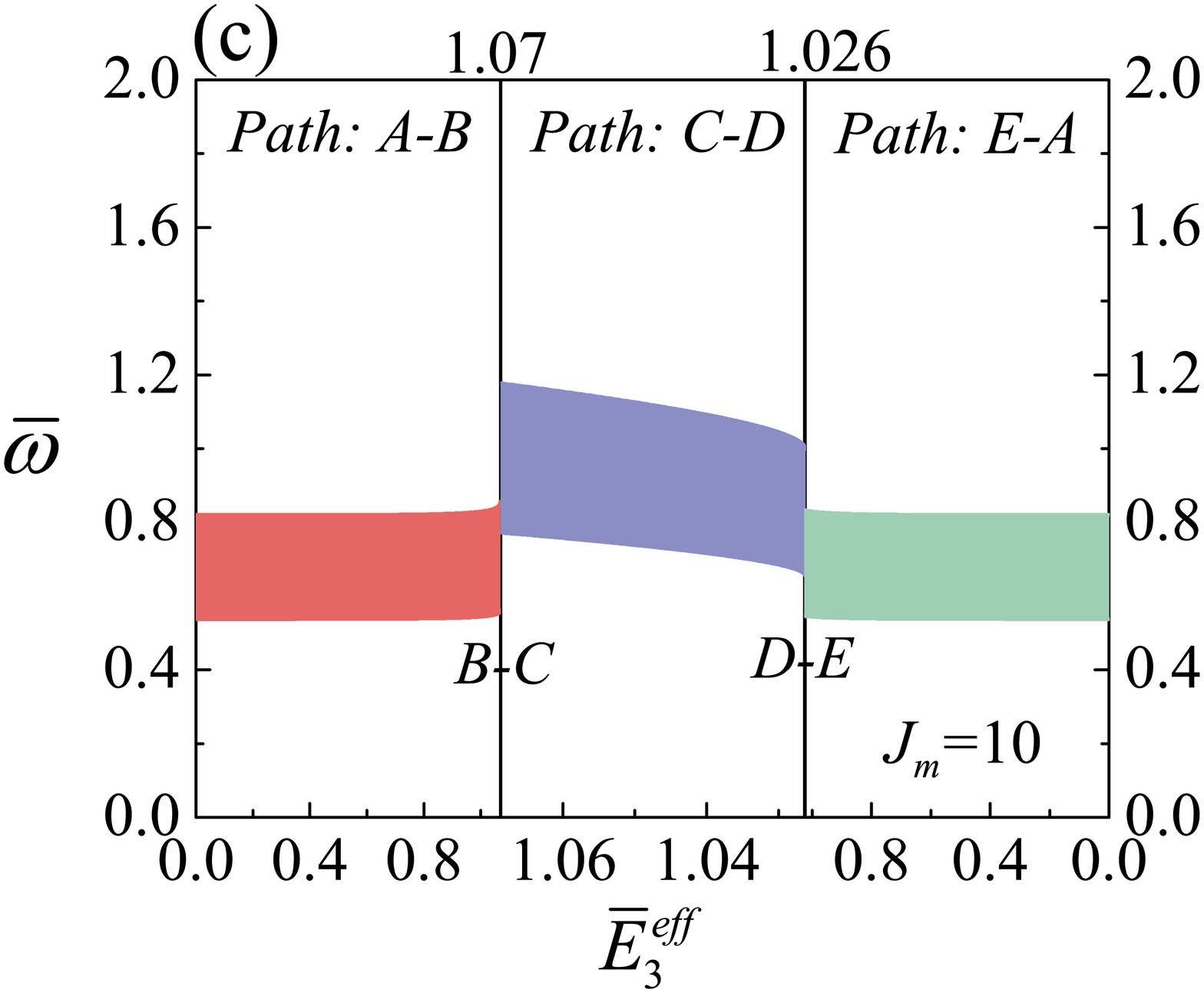}
  	\includegraphics[width=0.43\textwidth]{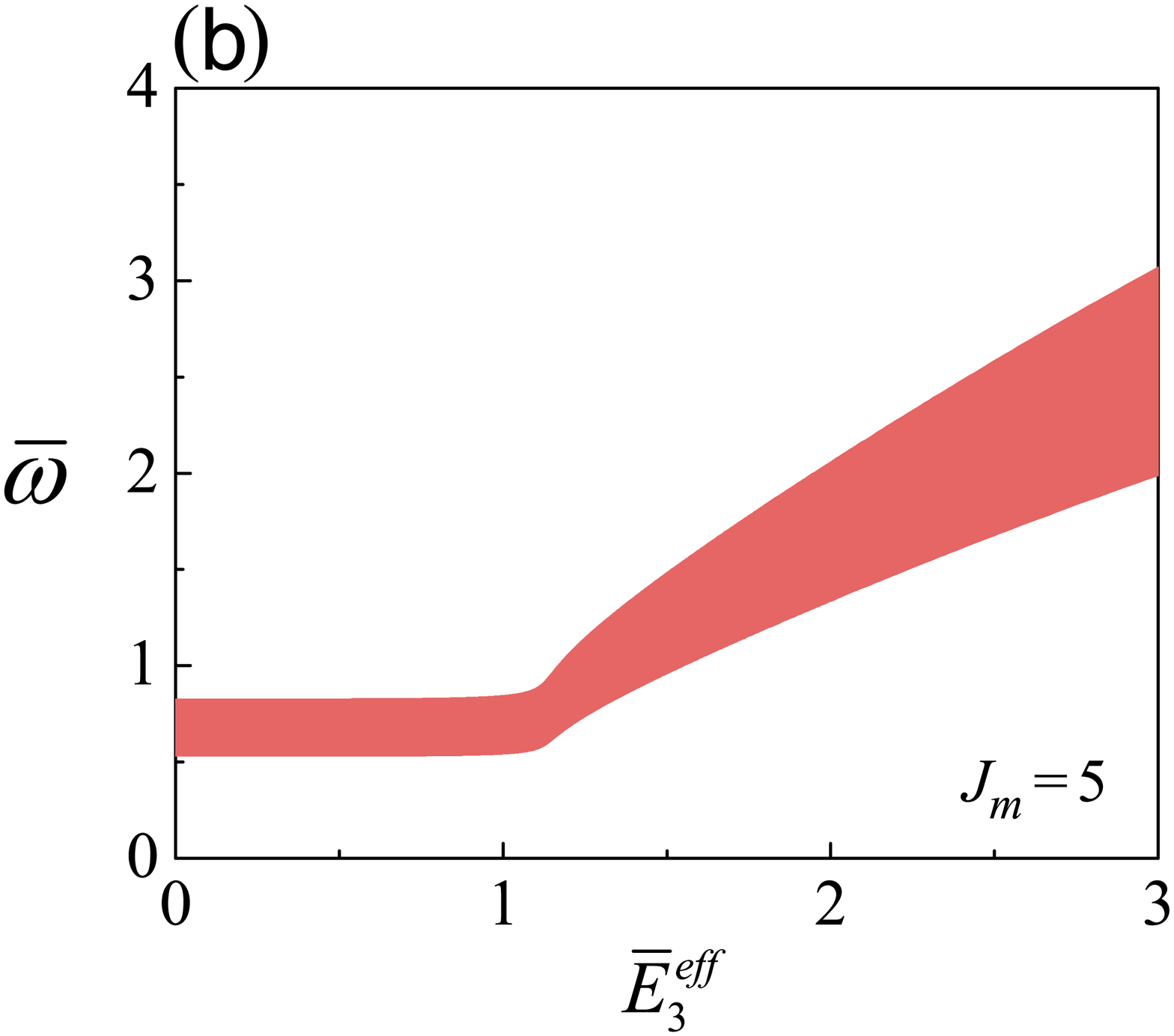} \hspace{0.045\textwidth}
  	\includegraphics[width=0.45\textwidth]{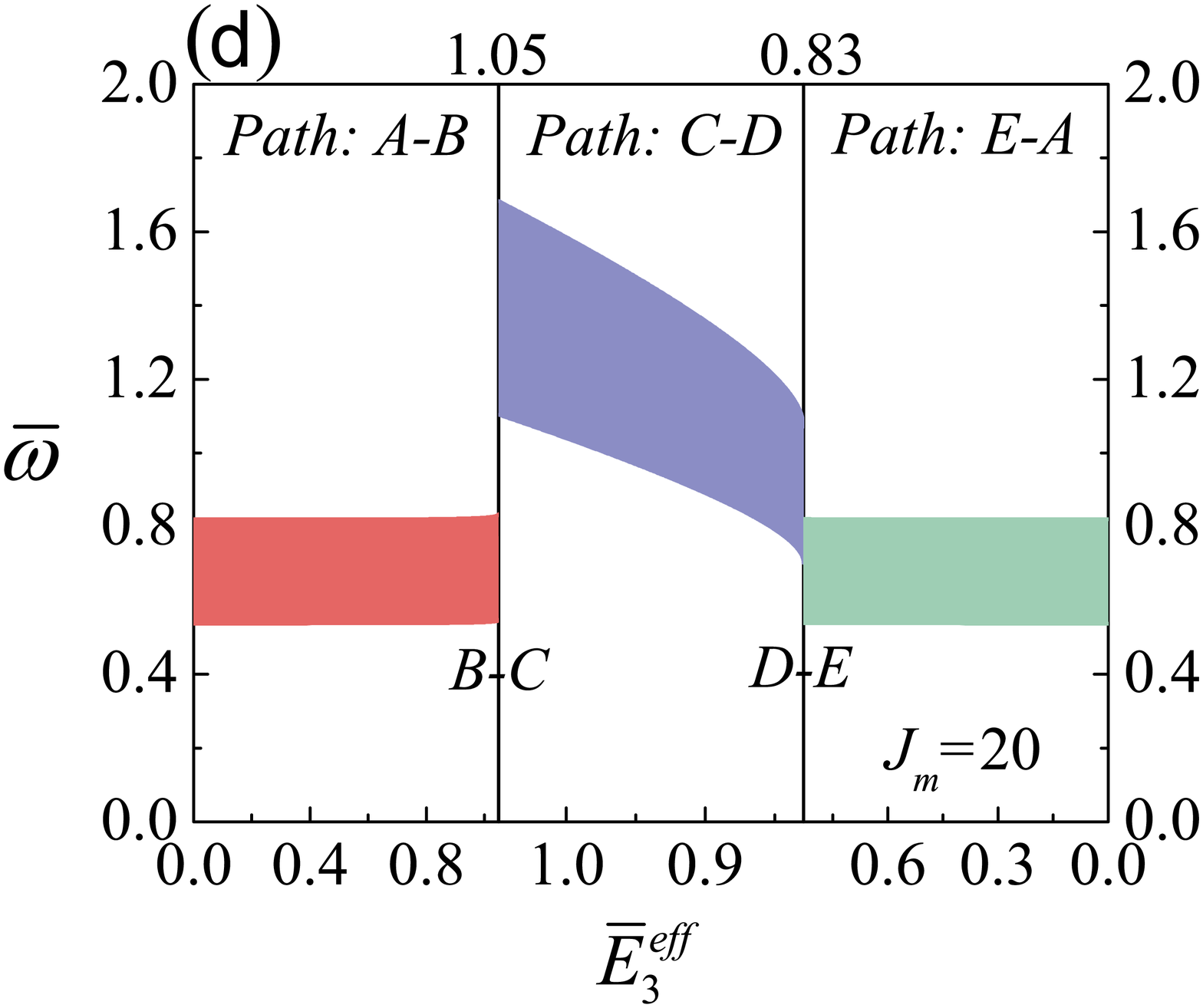}
  	\caption{The frequency limits of the first band gap of incremental \emph{shear} waves versus the dimensionless electric field $\overline E_3^{\text{eff}}$ in the periodic DE laminate for the N-H model (a) and the Gent model with (b) ${{J}_{m}}=5$, (c) ${{J}_{m}}=10$ and (d) ${{J}_{m}}=20$.}
  	\label{Fig4}
\end{figure}

  
Based on the nonlinear response of different strain-stiffening models with or without the snap-through transition, it is extremely feasible to electrostatically tune the band structures of incremental waves in DE laminates in various manners. For the Gent model with $J_m=10$ and different electric fields, the dispersion curves of incremental shear waves are depicted in Figs.~\ref{Fig3}(a) and \ref{Fig3}(b) before and after the snap-through transition, respectively. It is observed that the band gaps appear at the center and border of the first Brillouin zone and their positions move up with increasing electric field. Comparing the result in Fig.~\ref{Fig3}(a) with that in Fig.~\ref{Fig3}(b), we can see that a huge change in dispersion curves can be achieved by means of the snap-through transition.
  
To clearly demonstrate the effect of snap-through instabilities on the band gaps, the first band gap of incremental shear waves is taken as an example and its variations with the normalized effective electric field are displayed in Fig.~\ref{Fig4}(a)-(d) for the ideal N-H and Gent models (${{J}_{m}}=5$, 10 and 20). Hereafter, all the critical electric field value $\overline E_3^{\text{cr}}$ will be marked out on the top of corresponding figures. We observe from Fig.~\ref{Fig4}(a) that the first band gap of shear waves in ideal N-H DE laminates is not affected by the variation of electric stimuli, which agrees well with the result of \citet{galich2017elastic} for the compressible periodic \emph{hyperelastic} laminate. This phenomenon can be ascribed to the mutual cancellation between the change in geometrical and material properties via the finite deformation induced by electric stimuli. For the ideal Gent models with $J_m=20$ and 10 shown in Figs.~\ref{Fig4}(c) and~\ref{Fig4}(d), we can obtain that the snap-through transition triggered at the corresponding critical electric field leads to a remarkable change in the position and width of shear wave band gap, and the subsequent decrease in the electric field from state $C$ to state $D$ lowers the position and narrows the width of band gap continuously. Besides, the results along the loading path $E$-$A$ coincide with those along path $A$-$B$ and their variation trend of the first band gaps of shear waves for ${{J}_{m}}=10$ and 20 appears to be the same as that of the N-H model, which can be explained based on the fact that the Gent phase has not reached the strain-stiffening stage under a small electric field and resembles the N-H model. Different from Figs.~\ref{Fig4}(c) and \ref{Fig4}(d), Fig.~\ref{Fig4}(b) presents the continuous band gap variation with increasing electric field $\overline E_3^{\text{eff}}$ for the ideal Gent model ${{J}_{m}}=5$ without the snap-through instability. To be specific, the band gap widens and lifts up beyond a threshold value of the effective electric field ($\overline E_3^{\text{eff}}\simeq1.15$), where the influence of strain-stiffening effect (belonging to the material property change) exceeds that of geometrical change on the shear wave band gap.
  
\begin{figure}[htbp]
  	\centering
  	\setlength{\abovecaptionskip}{0pt}
  	\setlength{\belowcaptionskip}{0pt}
  	\includegraphics[width=0.45\textwidth]{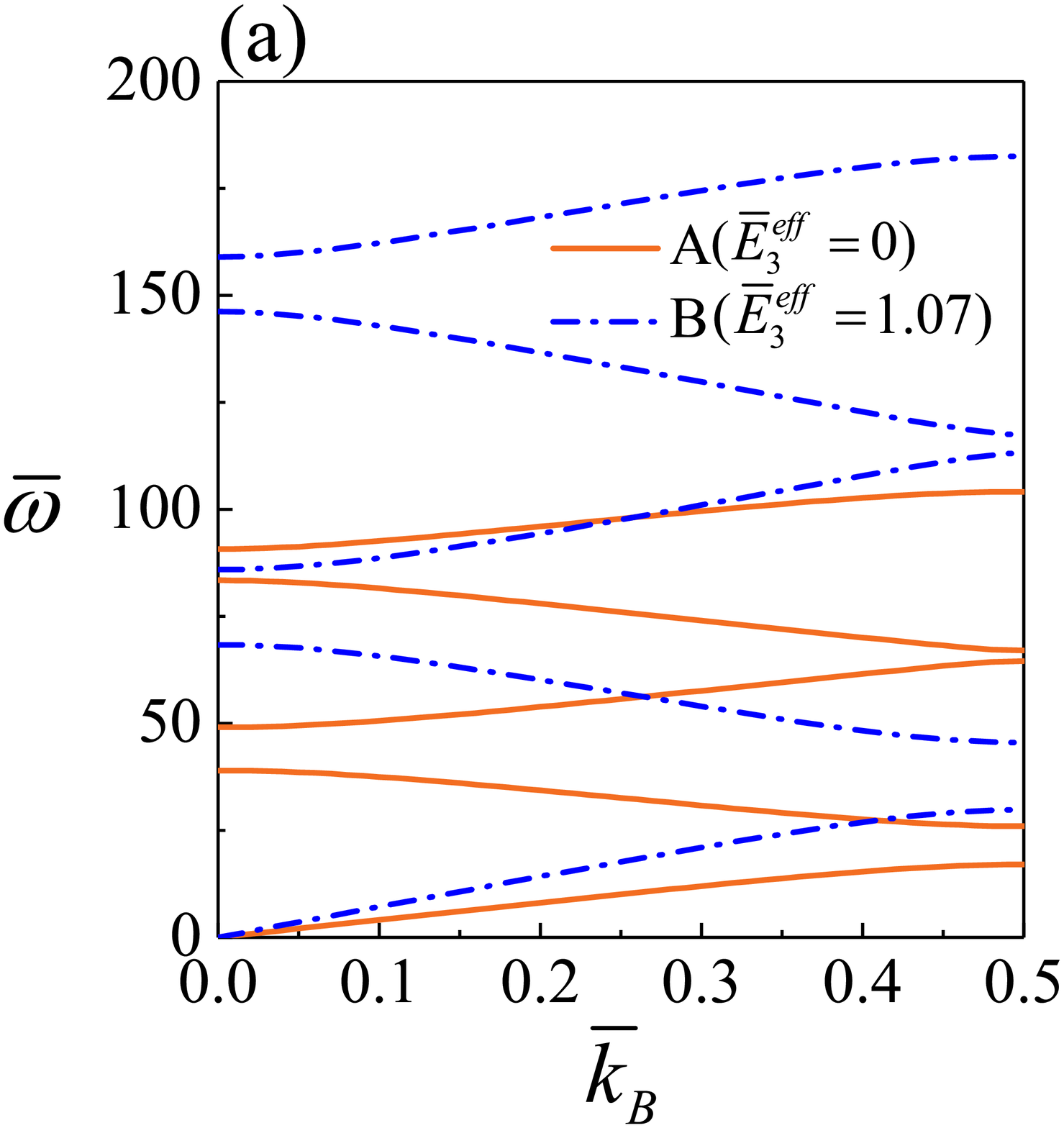}
  	\includegraphics[width=0.45\textwidth]{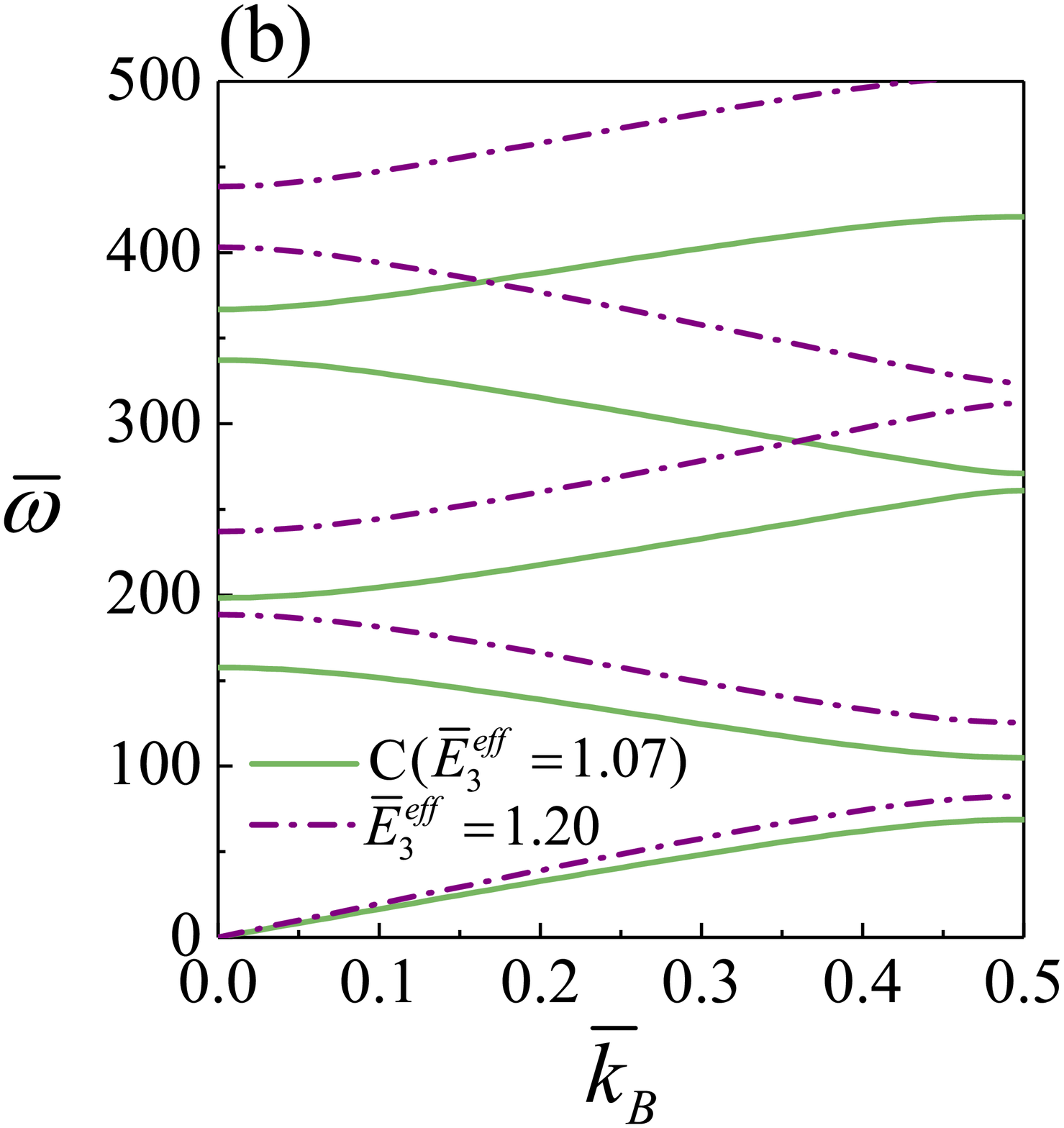}
  	\caption{Evolution of the band structures of \emph{longitudinal} waves before (a) and after (b) the snap-through transition from states $B$ to $C$ shown in Fig.~\ref{Fig2} (${{J}_{m}}=10$).}
  	\label{Fig5}
\end{figure}

\begin{figure}[htbp]
	\centering
  	\includegraphics[width=0.43\textwidth]{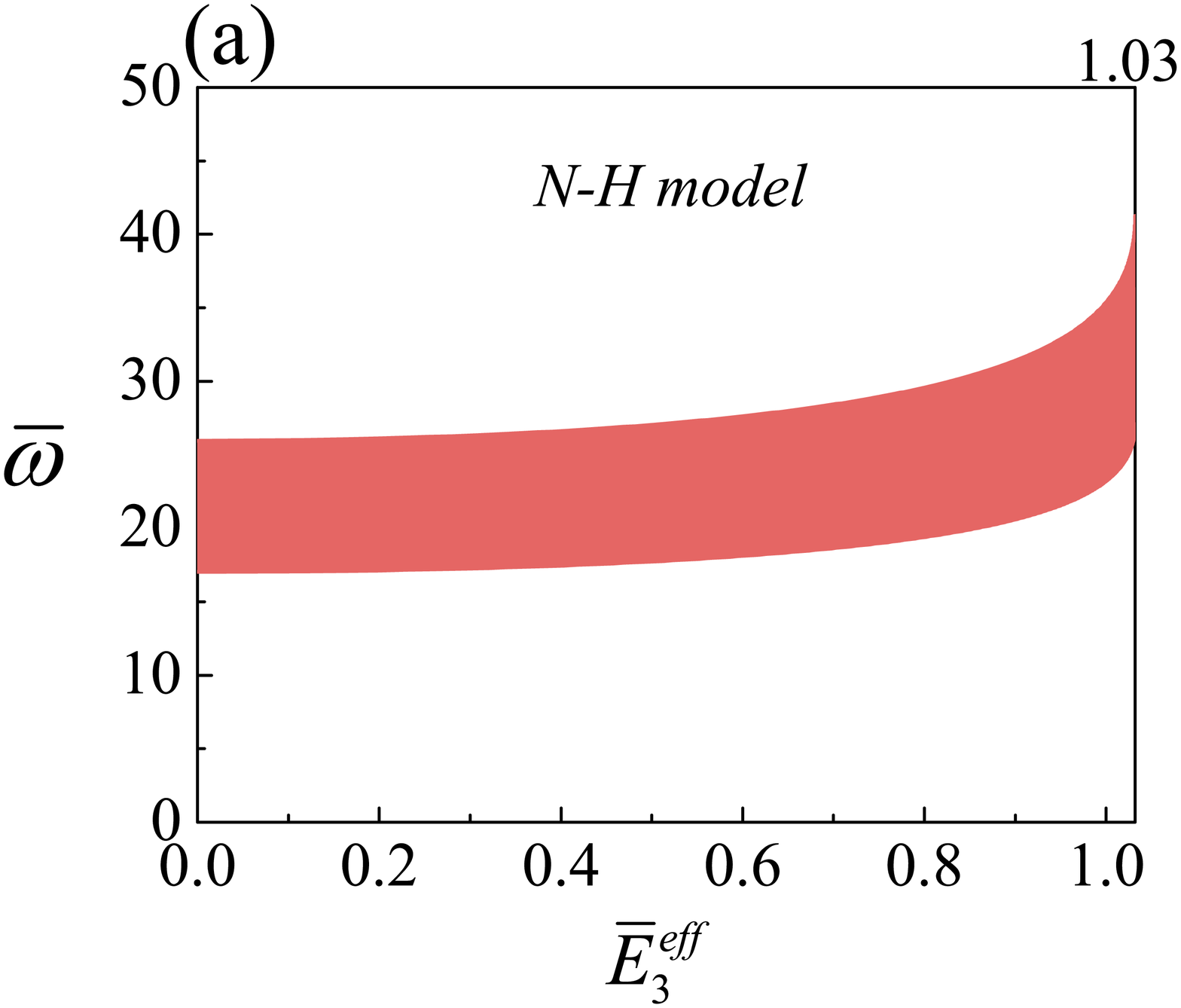} \hspace{0.03\textwidth}
    \includegraphics[width=0.45\textwidth]{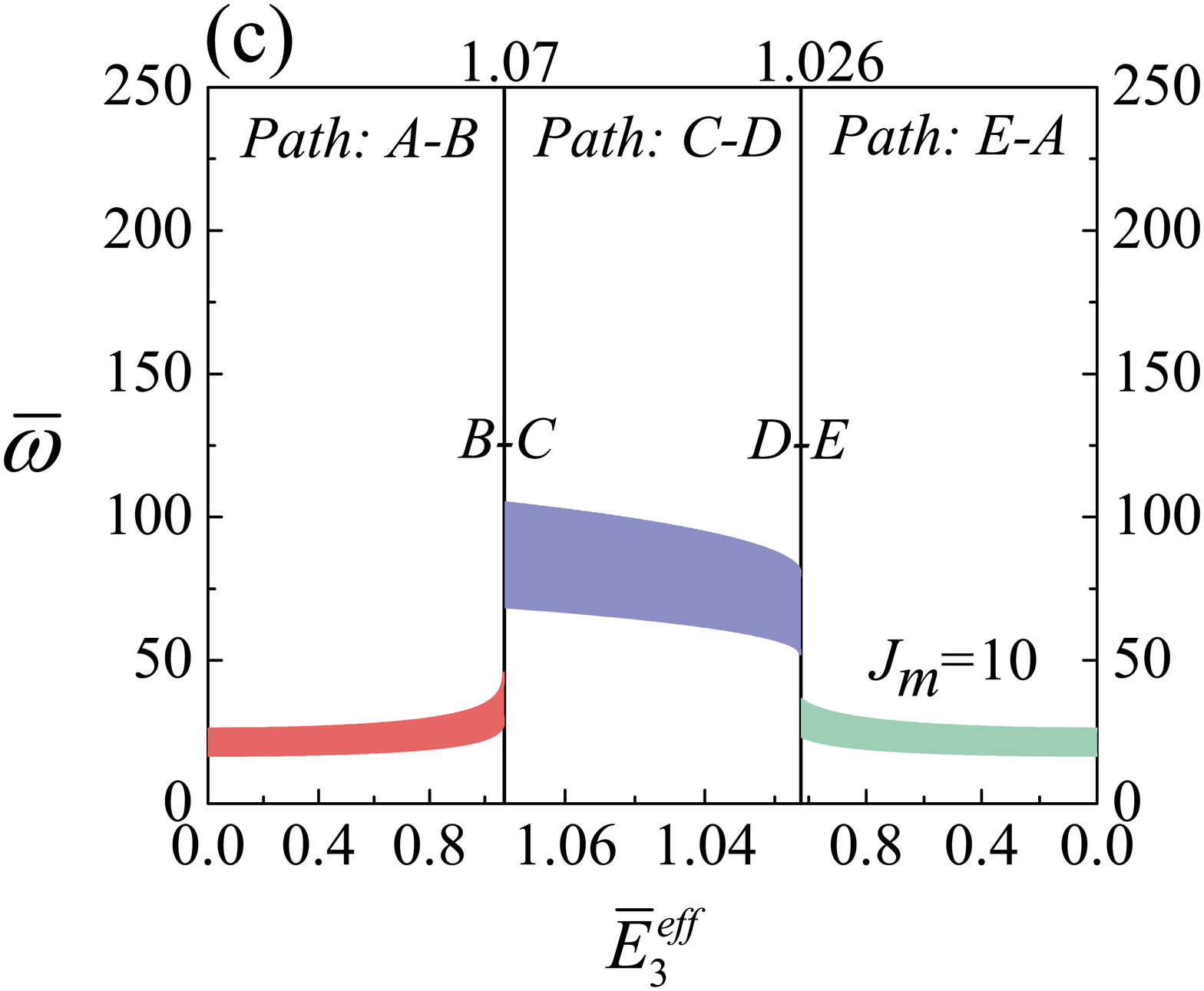}
    \includegraphics[width=0.425\textwidth]{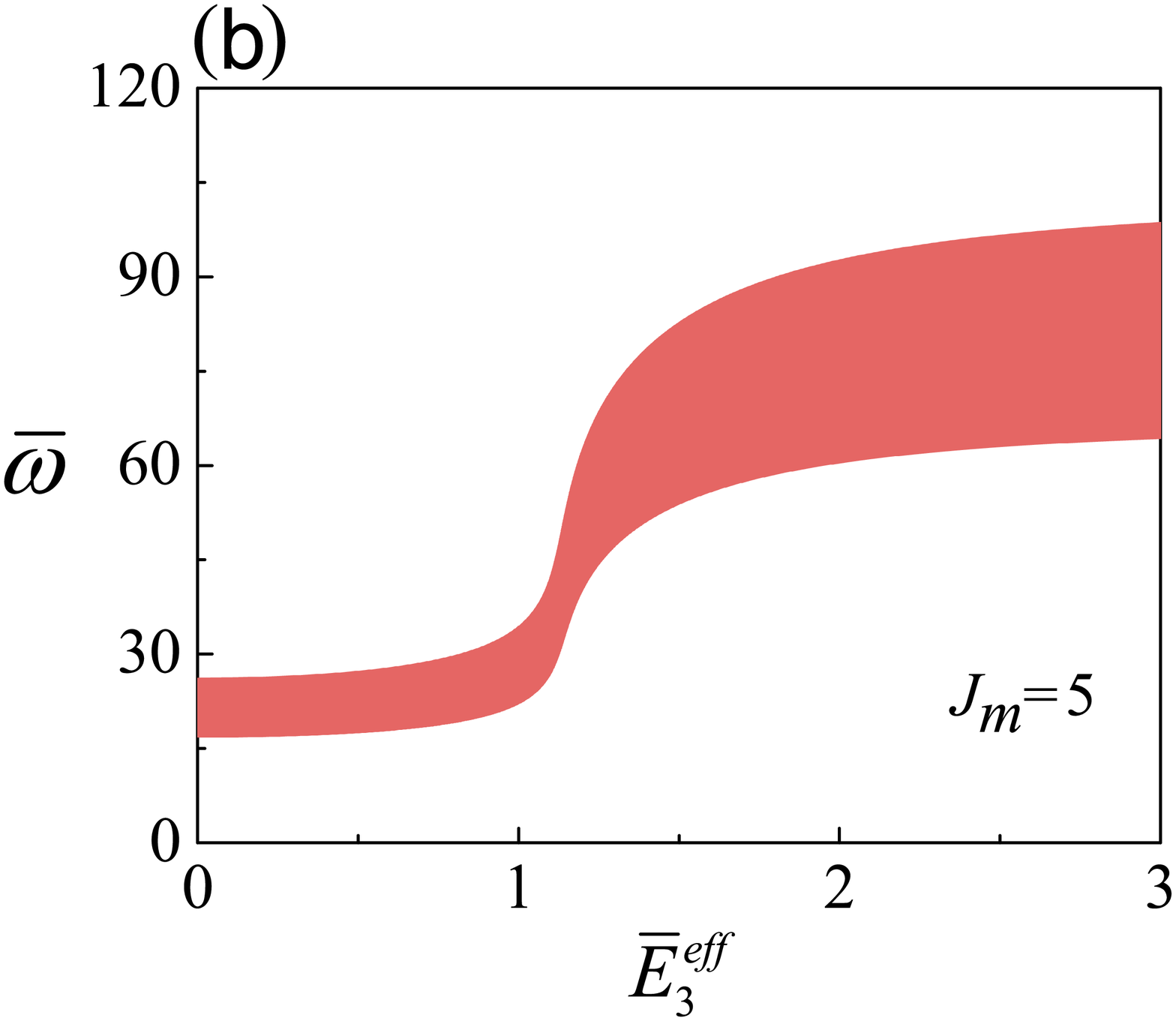} \hspace{0.03\textwidth}
    \includegraphics[width=0.45\textwidth]{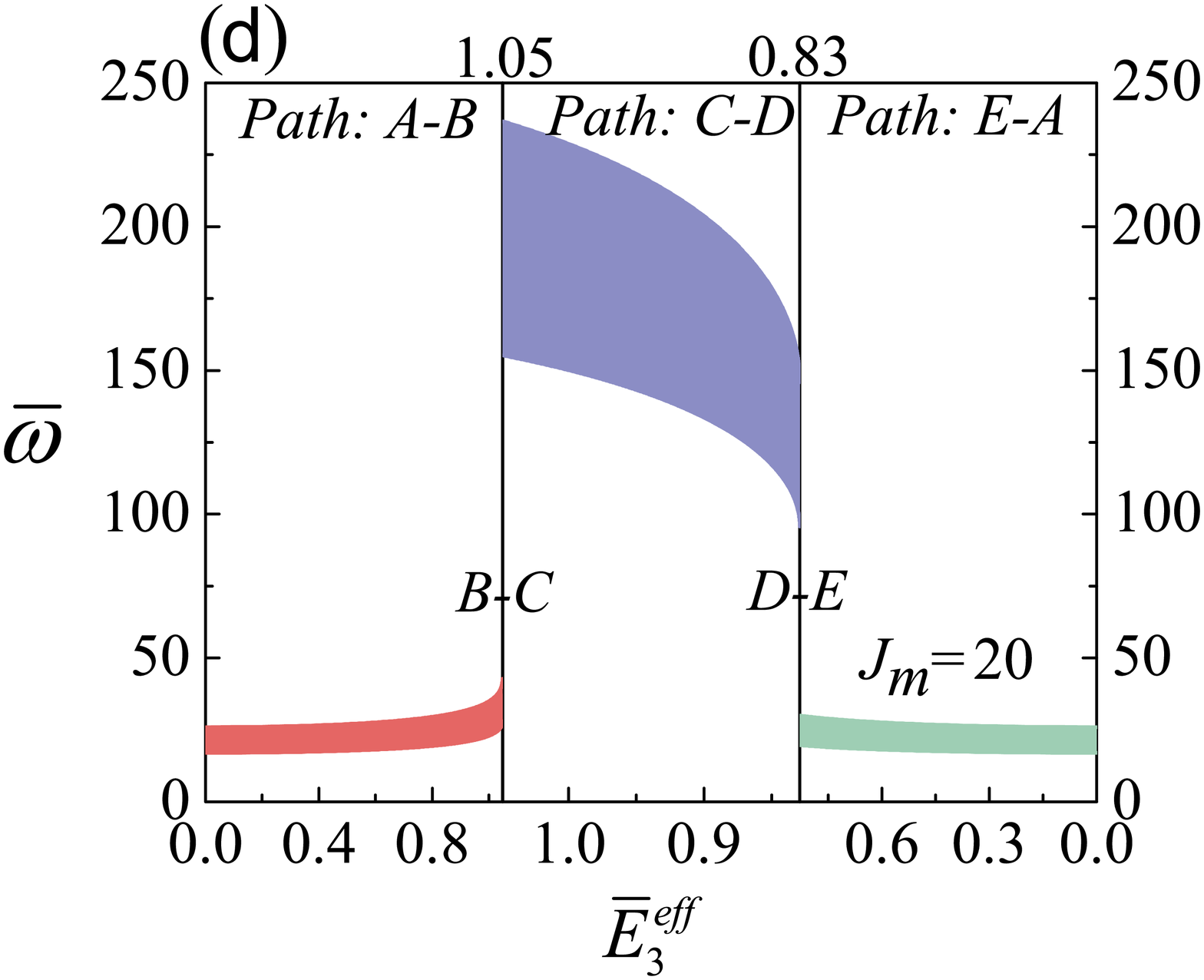}
	\caption{The frequency limits of the first band gap of incremental \emph{longitudinal} waves versus the dimensionless electric field $\overline E_3^{\text{eff}}$ in the periodic DE laminate for the N-H model (a) and the Gent model with (b) ${{J}_{m}}=5$, (c) ${{J}_{m}}=10$ and (d) ${{J}_{m}}=20$.}
	\label{Fig6}
\end{figure}
  
Similarly, for the ideal Gent model with $J_m=10$ and different electric stimuli, Figs.~\ref{Fig5}(a) and \ref{Fig5}(b) show the evolution of dispersion diagrams of incremental longitudinal (or pressure) waves propagating in the periodic DE laminate before and after snap-through transition, respectively. Analogous observations to those in Figs.~\ref{Fig3}(a) and \ref{Fig3}(b) can be obtained and hence the discussions are omitted for brevity. Nevertheless, it should be emphasized that the position or the frequency of band gaps for pressure waves is much higher than that for shear waves when comparing Figs.~\ref{Fig5}(a) and \ref{Fig5}(b) with Figs.~\ref{Fig3}(a) and \ref{Fig3}(b).

Furthermore, for incremental pressure waves, Figs.~\ref{Fig6}(a)-(d) display the frequency limits of the first band gap as functions of the normalized electric field for the ideal N-H model and the Gent models with ${{J}_{m}}=5$, 10 and 20. We find from Figs.~\ref{Fig6}(a)-(c) that the pressure wave band gap for the N-H model \emph{before} the snap-through transition can be altered by the electric stimuli, which is distinguished from the case of shear waves, since the dominant mechanism affecting the pressure wave band gaps is the geometrical change of layers via the finite deformation generated by the electric field \citep{galich2017elastic}. Specifically, the pressure wave band gap moves upward owing to the geometrical change as the electric field increases although the material may not reach the strain-stiffening stage before the snap-through transition. Comparing the variation trends at the snap-through transitions and at the loading path $C$-$D$ for pressure waves in Figs.~\ref{Fig6}(c) and \ref{Fig6}(d) with those for shear waves in Figs.~\ref{Fig4}(c) and \ref{Fig4}(d), we can obtain qualitatively similar characteristics and will not repeat here. However, we note that compared with the result of tuning incremental shear waves, a better tunable effect on pressure waves can be obtained in view of a relatively greater jump in the band gap induced by the snap-through transition. In addition, for the ideal Gent model with $J_m=5$ shown in Fig.~\ref{Fig6}(b), we see that before the same threshold value $\overline E_3^{\text{eff}}\simeq1.15$ as that in Fig.~\ref{Fig4}(b), the first band gap of longitudinal waves ascends and keeps the same width with the increase of electric field. A further increase in electric field makes the band gap move upward and become wider due to the combination effects of strain stiffening and great geometrical change in the compressible ideal DE laminate. Finally, the band gap may reach a plateau at the lock-up stretch $\lambda_{lim}$, where the layers of DE laminate have been compressed sufficiently such that the strain stiffening and the geometrical configuration almost keep unchanged. 

In a word, by {\color{red} choosing DE materials with different Gent constants} ${{J}_{m}}$ reflecting the strength of strain-stiffening effect, sharp transition and continuous control of band gaps are practicable by tuning the electric field for both shear and longitudinal waves, which should be beneficial for the design of tunable acoustic/elastic wave devices.

\subsection{{\color{red}Influence of the second strain invariant}} \label{Sec5-Supp}
 {\color{red} Now the effect of the second strain invariant $I_2$ characterized by the second modulus $c_2$ on the nonlinear response and wave propagation behavior in periodic DE laminates is considered in this subsection without electrostrictive effect and prestress ($\tau_{33}^0=\gamma _{0}=\gamma _{2}=0$ and $\gamma _{1}=1$). By virtue of fitting experimental data with the Gent-Gent constitutive model for the mechanical response of natural rubbers, \citet{ogden2004fitting} obtained two sets of parameters with ${c_2}/\mu=$ 0.42 and 0.14 for simple tension and equibiaxial tension, respectively. Recently, \citet{chen2019dielectric} performed the pure-shear tests on three different DE materials: the THERABAND YELLOW 11726 made of styrenic rubber, the OPPO BAND GREEN 8003 made of natural rubber and VHB 4905 by 3M, with ${c_2}/\mu$ measured as 0.91, 0.37 and 0.00024, respectively. In our work, the normalized material parameter $\bar c_2$ used for numerical calculations is chosen as 0, 0.37 and 0.91 in accordance with the experimental results obtained by \citet{chen2019dielectric}.}

\begin{figure}[htbp]
	\centering
	\setlength{\abovecaptionskip}{0pt}
	\setlength{\belowcaptionskip}{0pt}
	\includegraphics[width=0.5\textwidth]{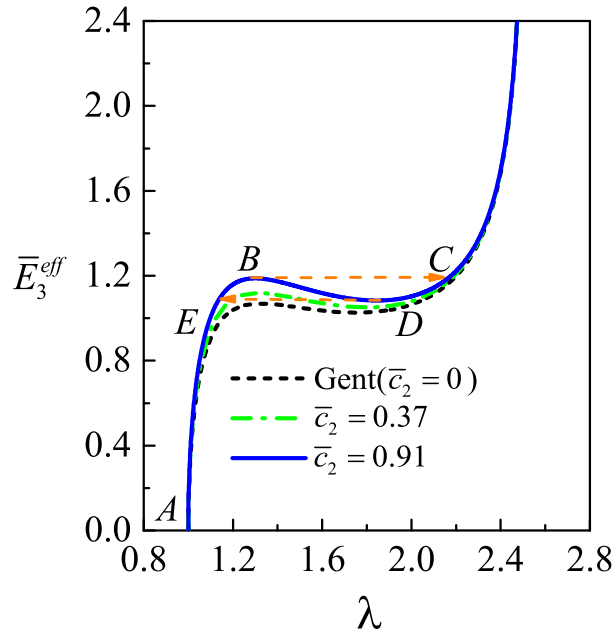}
	\caption{{\color{red}Nonlinear response of the lateral stretch $\lambda $ to the normalized effective referential electric field $\overline E_3^{\text{eff}}$ in the DE laminate for Gent-Gent model with ${\bar c_2}=$ 0 (original Gent model), 0.37 and 0.91 ($J_m=$10).}}
	\label{nonres}
\end{figure}

{\color{red} Fig.~\ref{nonres} displays the nonlinear response of the periodic ideal DE laminate under electric stimuli for the compressible Gent-Gent model \eqref{18} with different values of ${\bar c_2}=$ 0, 0.37 and 0.91 ($J_m=$ 10, $\tau_{33}^0=\gamma _{0}=\gamma _{2}=0$ and $\gamma _{1}=1$). We observe that the increase in ${\bar c_2}$ may result in larger critical electric field at which the snap-through instability from state $B$ to state $C$ takes place and the resulting lateral stretch at the stable state $C$ also becomes larger. However, the nonlinear responses at small and large effective electric fields ($\overline E_3^{\text{eff}}\lesssim1.0$ and $\overline E_3^{\text{eff}}\gtrsim1.2$) appear to be independent of the second strain invariant.}

\begin{figure}[htbp]
	\centering
	\setlength{\abovecaptionskip}{0pt}
	\setlength{\belowcaptionskip}{0pt}
	\includegraphics[width=0.44\textwidth]{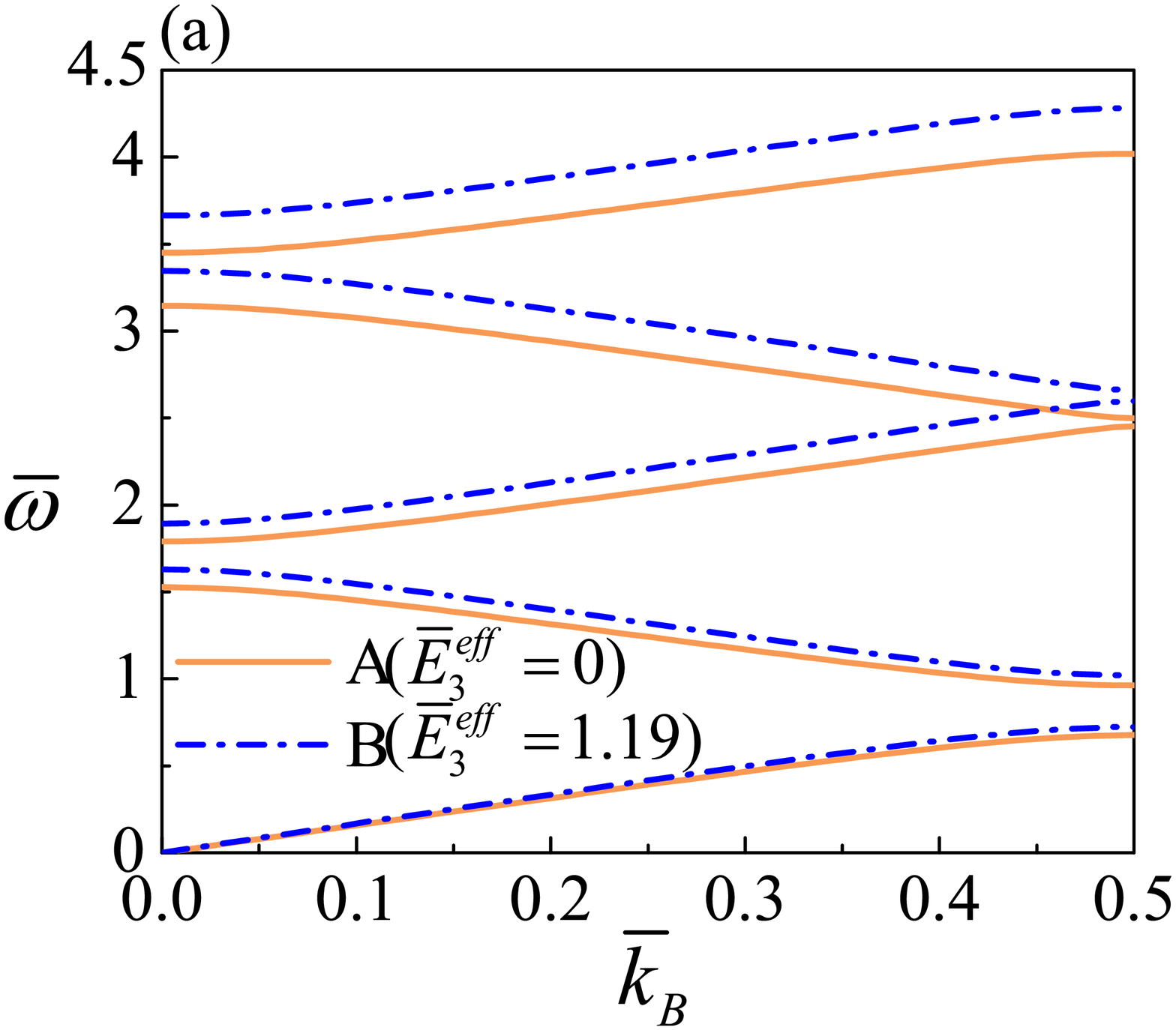}
	\includegraphics[width=0.46\textwidth]{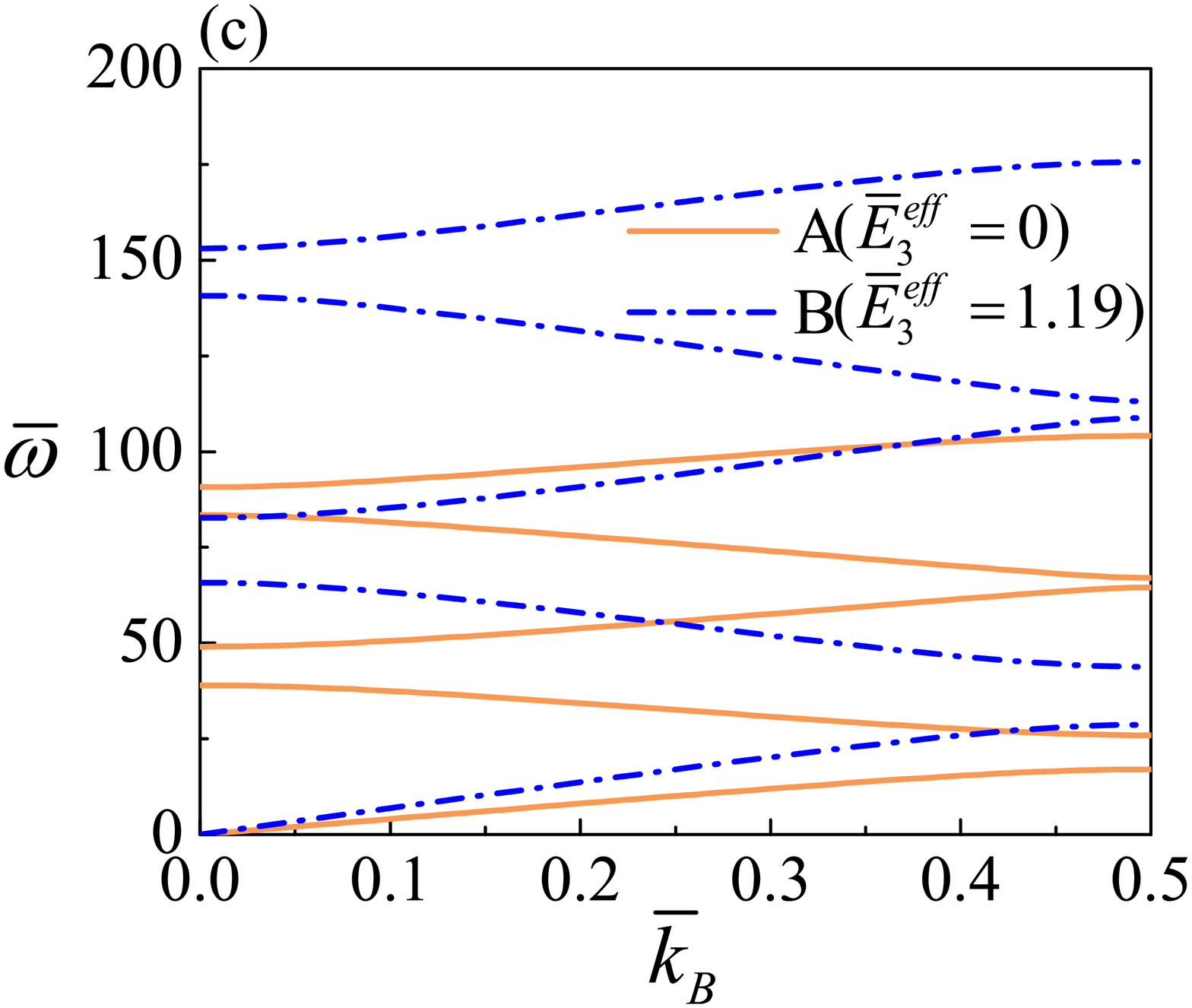}
	\includegraphics[width=0.44\textwidth]{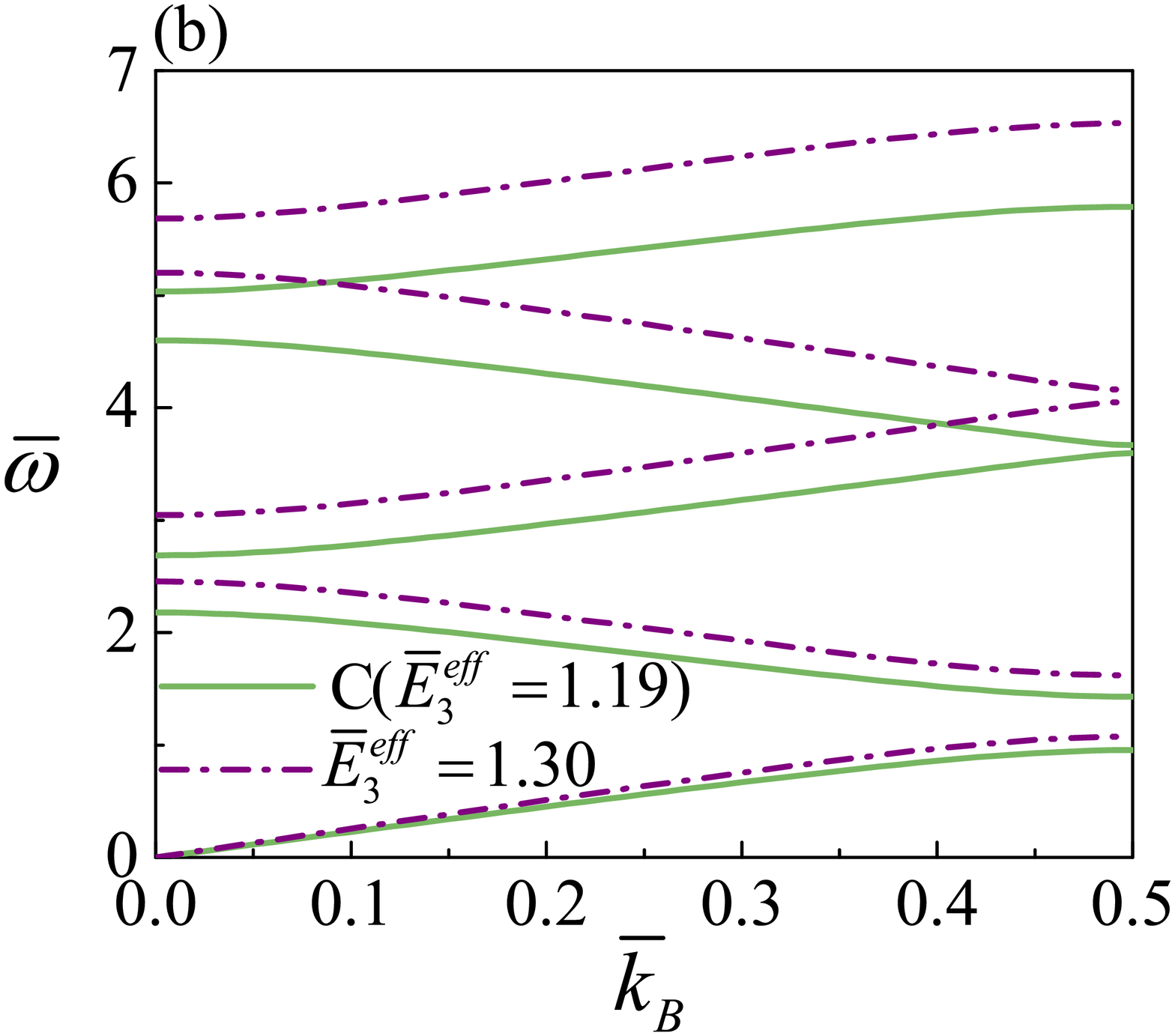}
	\includegraphics[width=0.47\textwidth]{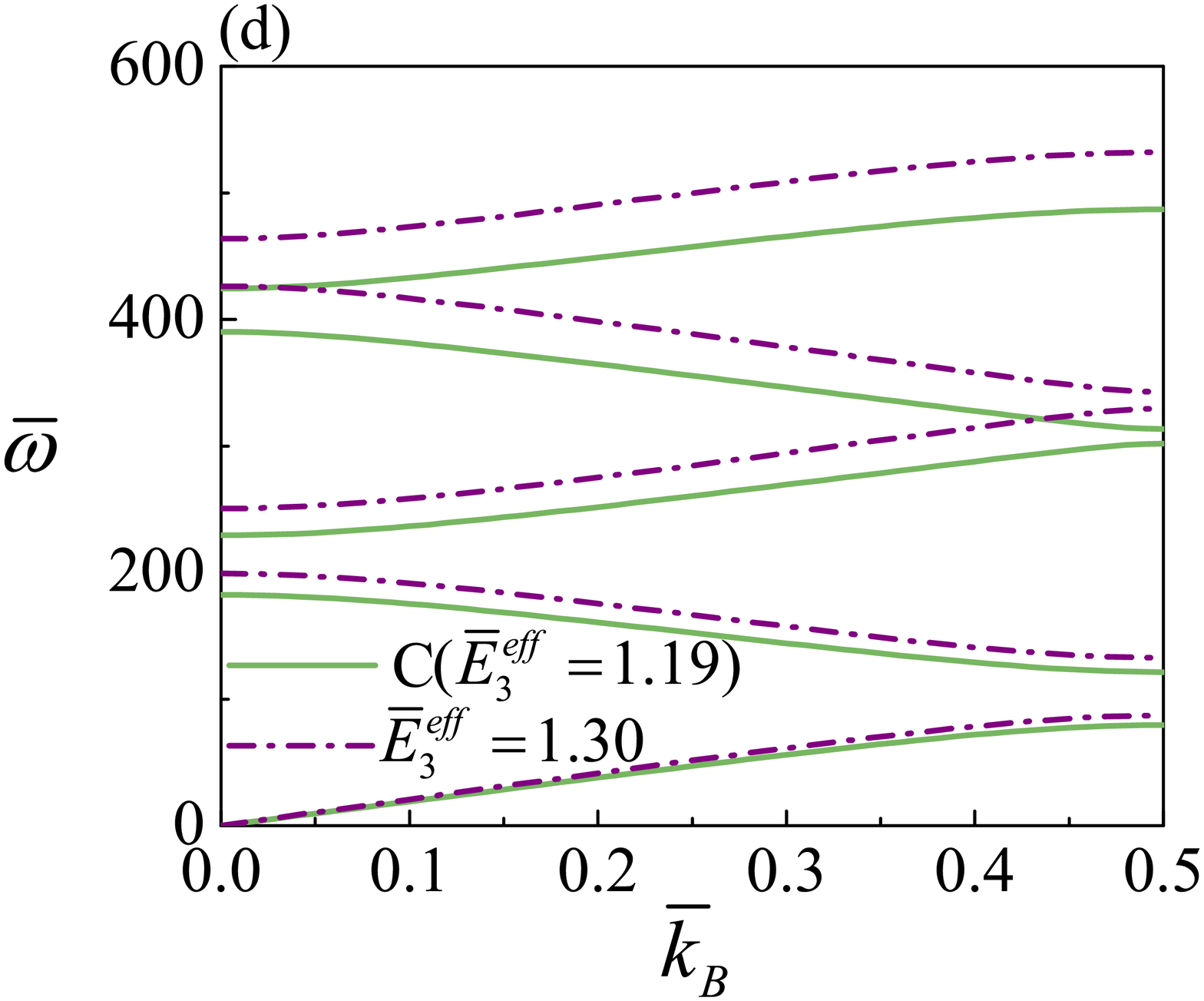}
	\caption{{\color{red}Evolution of the band structures of \emph{shear} (a, b) and \emph{longitudinal} (c, d) waves before (a, c) and after (b, d) the snap-through transition from states $B$ to $C$ shown in Fig.~\ref{nonres} for ${\bar c_2}=$ 0.91 (${{J}_{m}}=10$).}}
	\label{SBG}
\end{figure}

{\color{red}The evolution of band structures for incremental shear and longitudinal waves before and after the snap-through transition is displayed in Fig.~\ref{SBG} for ${\bar c_2}= 0.91$ and different electric fields. We find from Fig.~\ref{SBG} that the band gaps can be moved up by increasing the electric field, and the triggered snap-through transition helps to realize a drastic change in band gaps. In addition, by comparing the results in Figs.~\ref{SBG}(c) and (d) with those in Figs.~\ref{SBG}(a) and (b), we see that the band gap frequency of longitudinal waves is much larger than that of shear waves. These are qualitatively analogous to the phenomena observed in Figs.~\ref{Fig3} and~\ref{Fig5}.}

\begin{figure}[htbp]
	\centering
	\setlength{\abovecaptionskip}{0pt}
	\setlength{\belowcaptionskip}{0pt}
	\includegraphics[width=0.455\textwidth]{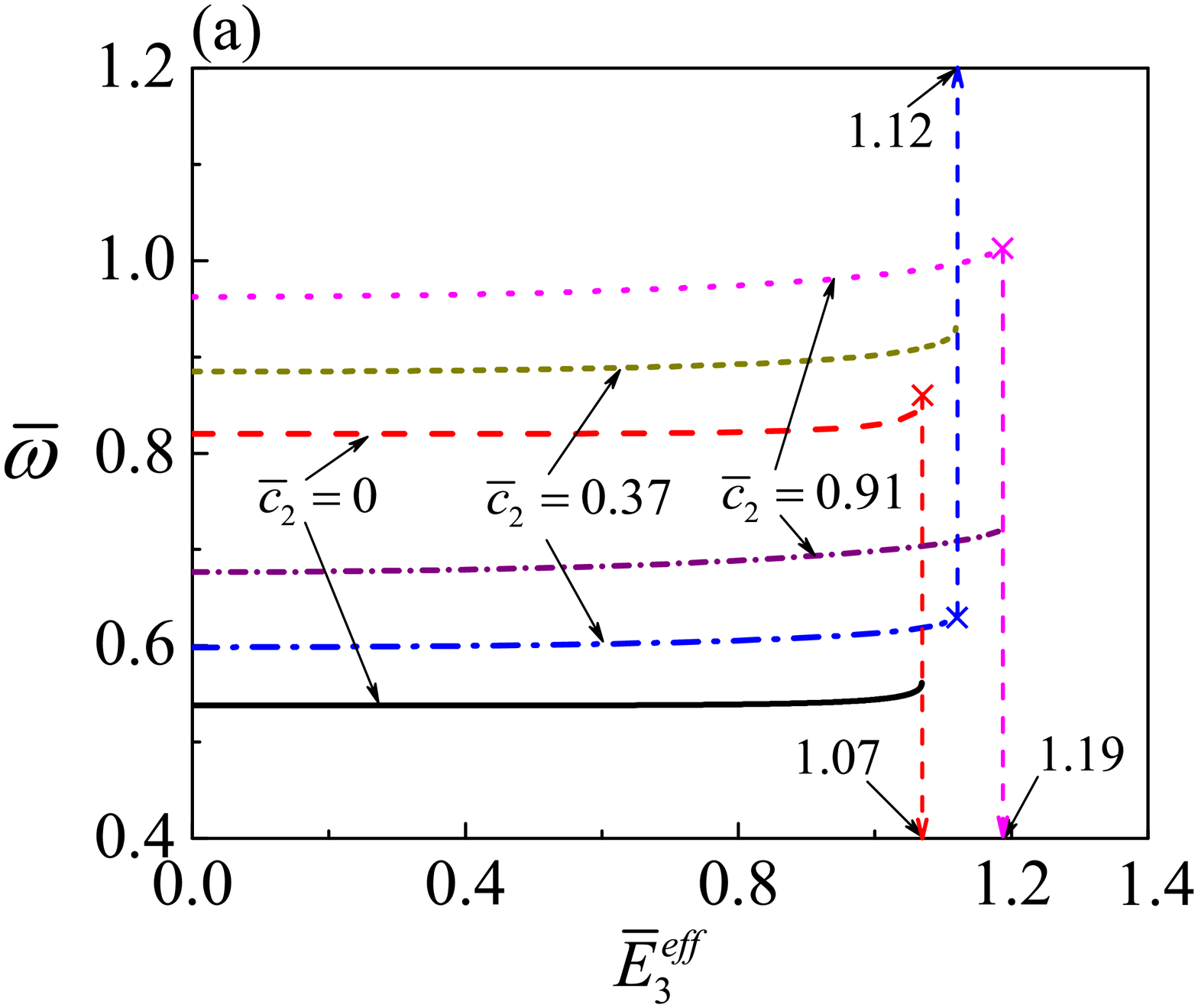}
	\includegraphics[width=0.455\textwidth]{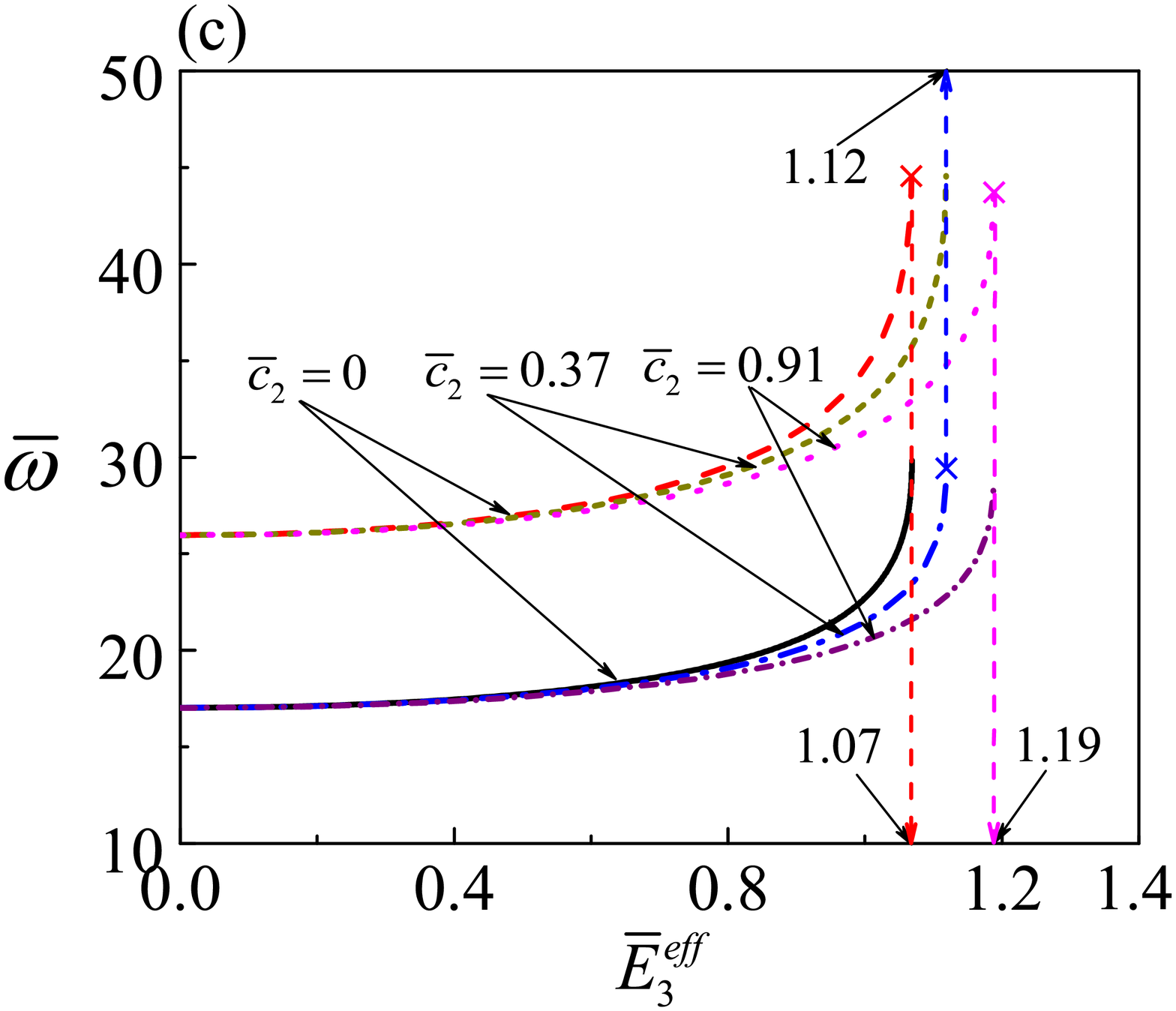}
	\includegraphics[width=0.465\textwidth]{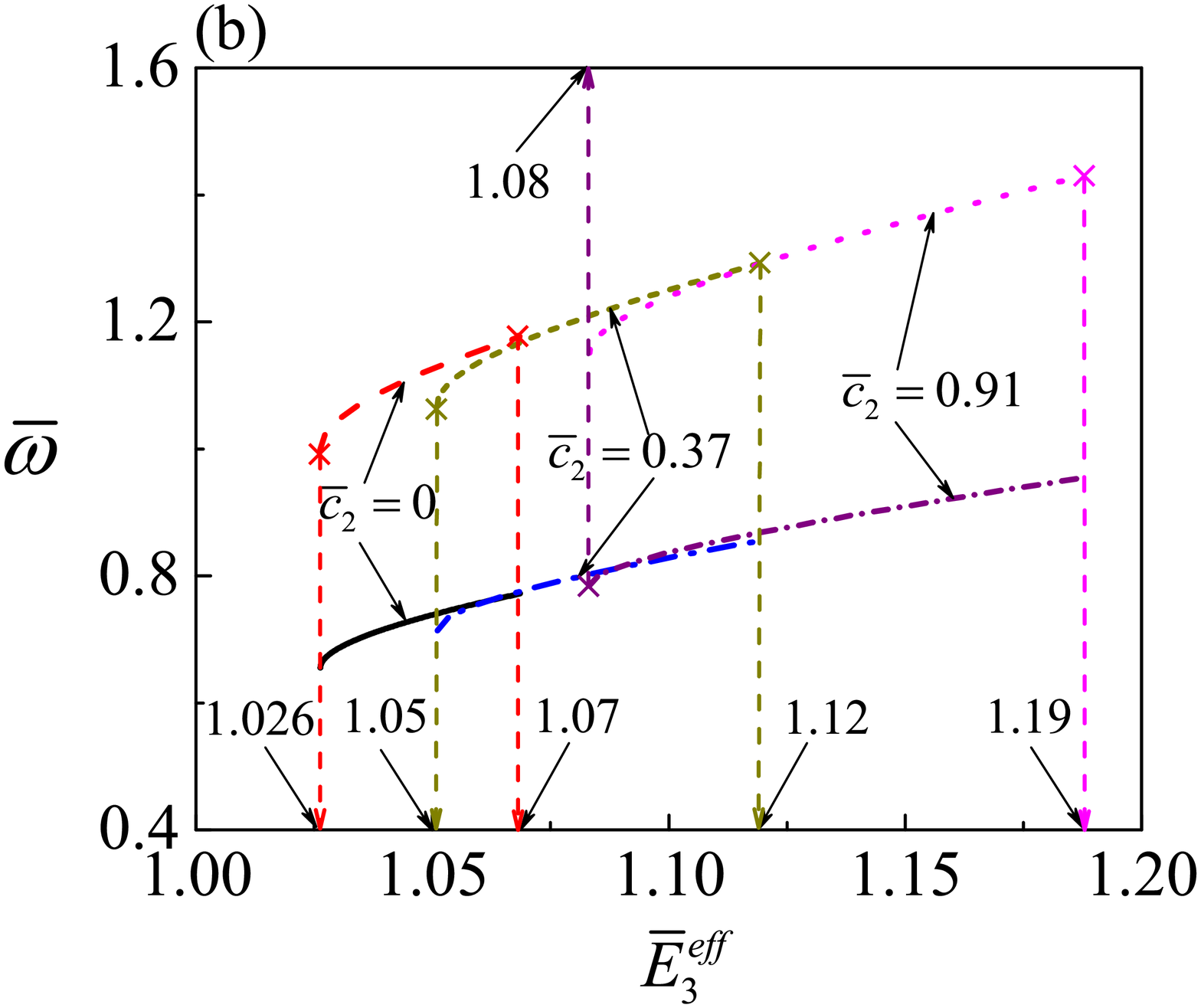}
	\includegraphics[width=0.465\textwidth]{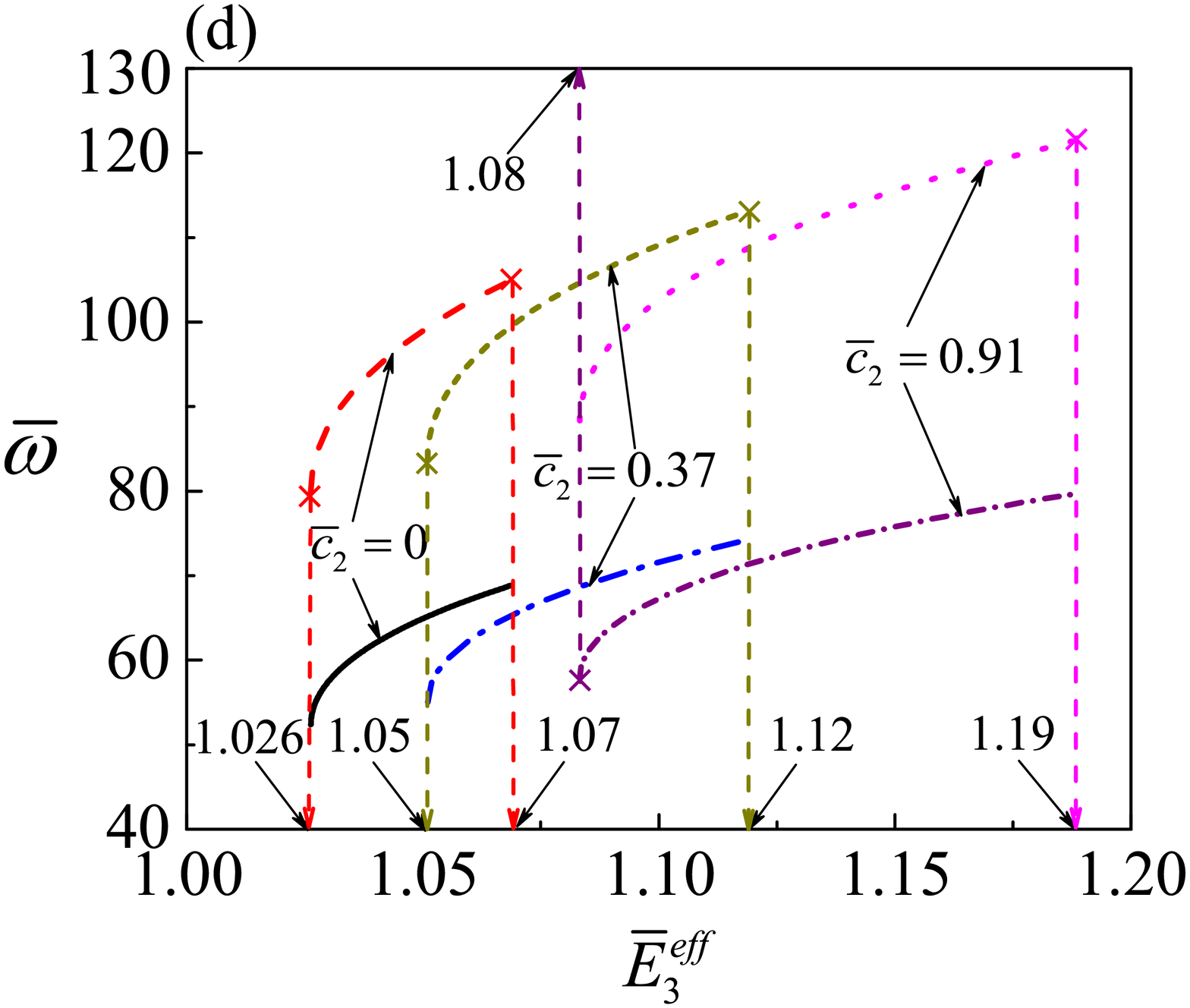}
	\caption{{\color{red}The frequency limits of the first \emph{shear} (a, b) and \emph{longitudinal} (c, d) wave band gap versus the dimensionless electric field $\overline E_3^{\text{eff}}$ in the ideal DE laminate before (a, c) and after (b, d) the snap-through transition for ${{J}_{m}}=10$ and different ${\bar c_2}$ (The corresponding critical $\overline E_3^{\text{cr}}$ is also marked by cross in this figure).}}
	\label{Svary}
\end{figure}

{\color{red} In order to clearly show the effect of the second strain invariant on wave propagation behaviors, Fig.~\ref{Svary} depicts the frequency limits of the first shear and pressure wave band gaps versus the effective electric field before (path $A$ to $B$) and after (path $C$ to $D$) the snap-through transtion at three different values of ${\bar c_2}=$ 0, 0.37 and 0.91. The result for ${\bar c_2}=$ 0 corresponds to the case of the original Gent model. For different ${\bar c_2}$, the variation trends of frequency limits with electric field $\overline E_3^{\text{eff}}$ are qualitatively consistent but quantitatively different for both loading paths. The differences are as follows: (1) the critical electric field becomes larger and the frequency limits (i.e. the position of band gaps) are lifted up with the increase of ${\bar c_2}$; (2) the band gaps of both shear and pressure waves can be tuned in a wider range of electric field for a larger ${\bar c_2}$.}

\subsection{Influence of the electrostrictive effect} \label{Sec5-2}

\begin{table}[htb]
	\centering
	\begin{tabular}{lccc}
		\hline
		\hline \specialrule{0em}{0pt}{4pt}
		Reference & ${{\gamma }_{\text{0}}}$ & ${{\gamma }_{\text{1}}}$ & ${{\gamma }_{\text{2}}}$   \\
		\specialrule{0em}{0pt}{2pt} \hline \specialrule{0em}{0pt}{8pt}
		ES-1 \citep{wissler2007electromechanical} & 0.00104  & 1.14904 & -0.15008  \\
		\specialrule{0em}{0pt}{2pt} \hline \specialrule{0em}{0pt}{8pt}
		ES-2 \citep{li2011effect} & 0.00458  & 1.3298 & -0.33438   \\
		\specialrule{0em}{0pt}{4pt}
		\hline
		\hline
	\end{tabular}
	\caption{Electrostrictive parameters of the enriched Gent DE model utilized in numerical calculations (`ES-1' denotes the first set of electrostrictive parameters from \citet{wissler2007electromechanical}, while `ES-2' represents the second one from \citet{li2011effect}).}
	\label{Table1}
\end{table}
  
Due to the large deformations of DEs induced by the application of electric fields, the dielectric permittivity is no longer a constant and dependent on the deformation process, which is referred to as the electrostrictive effect \citep{zhao2008electrostriction}. This phenomenon has been observed by the experimental works \citep{wissler2007electromechanical,li2011effect} for typical DE materials (such as 3M VHB4910). By means of the numerical fitting of the enriched material model \eqref{17} to experimental data, \citet{gei2014role} obtained the values of electrostrictive parameters for experimentally tested DE materials conducted by \citet{wissler2007electromechanical} and \citet{li2011effect}. The two sets of electrostrictive parameters marked by `ES-1' and `ES-2' are provided in Table \ref{Table1}. In this subsection, we will illustrate the effect of electrostriction on nonlinear response and incremental wave characteristics of the periodic DE laminate under electric stimuli. {\color{red}Note that \ref{AppeA} provides some other electrostrictive models available in the literature \citep{zhao2008electrostriction, dorfmann2010electroelastic, vertechy2013continuum}, numerical comparison of which is beyond the scope of this paper and worth further research.}

\begin{figure}[htbp]
	\centering
	\setlength{\abovecaptionskip}{0pt}
	\setlength{\belowcaptionskip}{0pt}
	\includegraphics[width=0.5\textwidth]{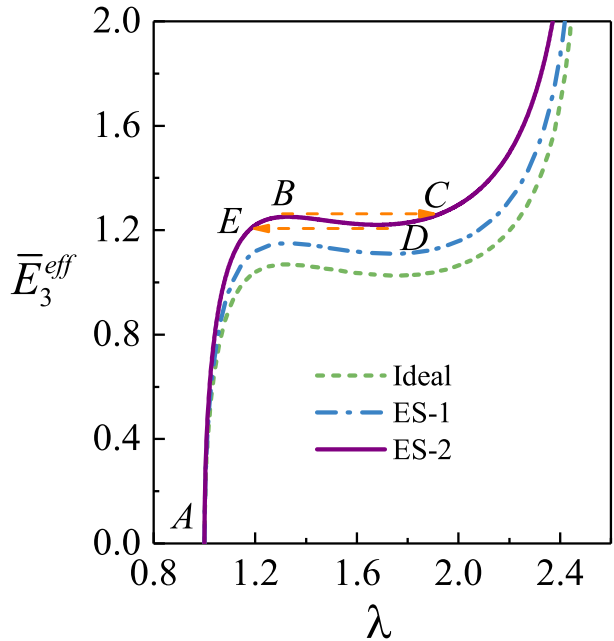}
	\caption{Nonlinear response of the periodic DE laminate to the effective electric field for the ideal, ES-1 and ES-2 Gent models (${{J}_{m}}=10$).}
	\label{Fig7}
\end{figure}

  
Based on Eqs.~\eqref{21} and \eqref{equi-E}, the nonlinear response of the periodic DE laminate to the electric stimuli are displayed in Fig.~\ref{Fig7} for the ideal ($\gamma _{0}=\gamma _{2}=0$ and $\gamma _{1}=1$) and enriched Gent models with the electrostrictive effect. Note that the prestress here is still set to be $\tau_{33}^0=0$ and the Gent constant is taken as ${{J}_{m}}=10$ for all three cases. For small effective electric fields ($\overline E_3^{\text{eff}}\lesssim0.8$) where the effect of electrostriction can be neglected, the results predicted by the electrostrictive model are the same as that based on the ideal Gent model. When considering the electrostriction effect, the critical electric field $\overline E_3^{\text{cr}}$ of the enriched Gent model where the snap-through transition from state $B$ to state $C$ occurs is higher than that given by the ideal Gent model. However, the critical stretch $\lambda_{\text{cr}}$ at state $B$ is almost independent of the electrostrictive parameters. Moreover, the enriched Gent DE laminate may achieve a smaller lateral stretch $\lambda $ at the stable state $C$ after snap-through transition.

\begin{figure}[htbp]
	\centering	
	\includegraphics[width=0.45\textwidth]{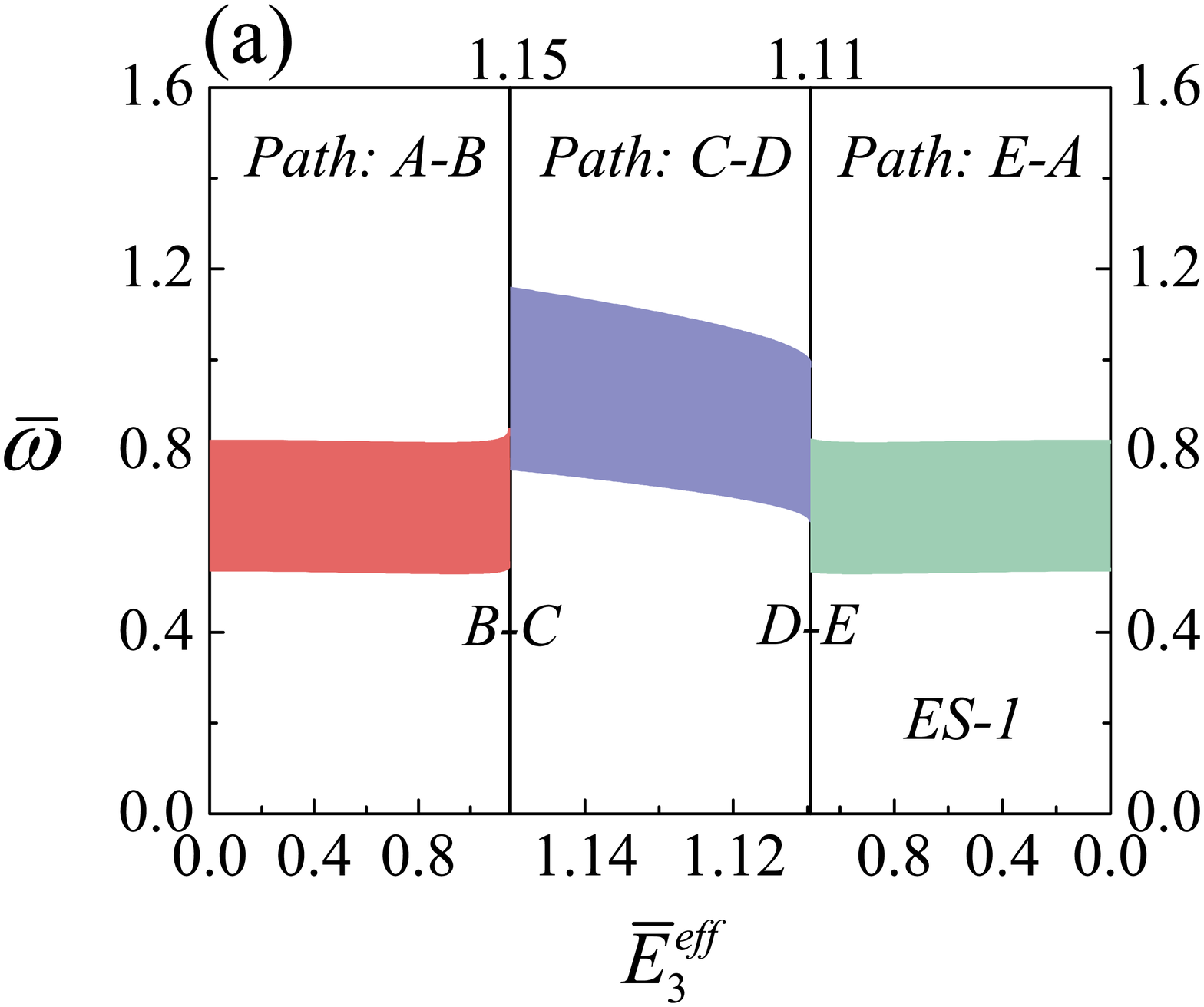}   \hspace{0.03\textwidth}
	\includegraphics[width=0.45\textwidth]{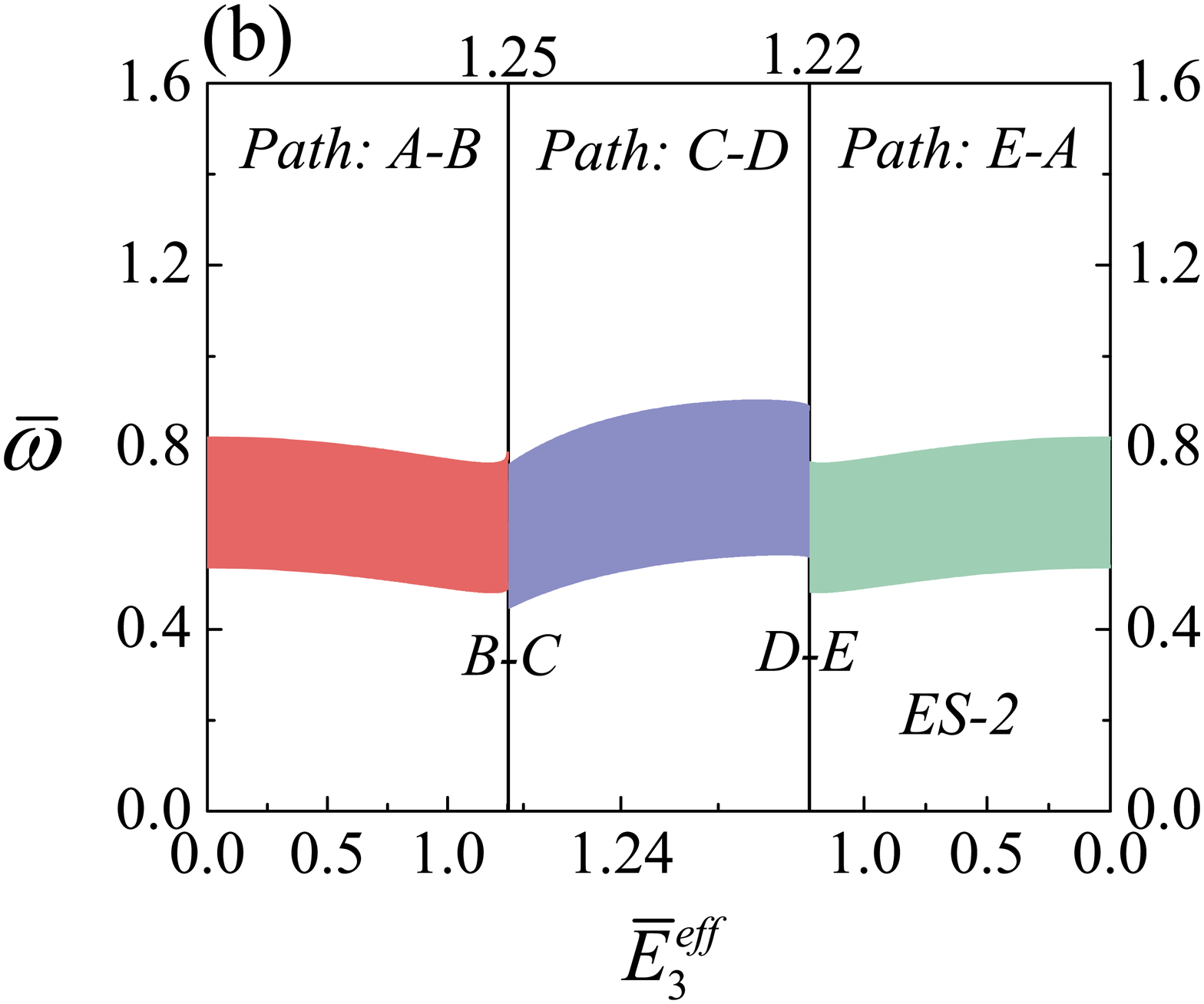}
	\caption{The frequency limits of the first band gap of incremental \emph{shear} waves versus the dimensionless electric field $\overline E_3^{\text{eff}}$ in the periodic DE laminate for the enriched Gent model ($J_m=10$) with different electrostriction parameters: (a) ES-1; (b) ES-2.}
	\label{Fig8}
\end{figure}

For the first band gap of incremental shear waves, we highlight in Figs.~\ref{Fig8}(a) and~\ref{Fig8}(b) the dependence of prohibited frequency limits on the dimensionless effective electric field for ES-1 and ES-2 enriched Gent model ($J_m=10$), respectively. We see from Fig.~\ref{Fig8}(a) that the variation trend for ES-1 parameters is qualitatively analogous to that in Fig.~\ref{Fig4}(c) for the ideal model except for the various critical electric field values $\overline E_3^{\text{cr}}$. Nonetheless, the shear wave band gap of the enriched Gent DE laminate with ES-2 parameters shown in Fig.~\ref{Fig8}(b) exhibits different variation features when subjected to the electric field. Specifically, when the applied electric field increases from zero to the critical value $\overline E_3^{\text{cr}}$ (from $A$ to $B$), {\color{red}the frequency limits first decrease and then increase slightly}. Additionally, when adopting ES-2 parameters \citep{li2011effect}, the jump of band gap during the snap-through transition from $B$ to $C$ is not so significant as those in Figs.~\ref{Fig4}(c) and \ref{Fig8}(a).

\begin{figure}[htbp]
	\centering	
	\includegraphics[width=0.45\textwidth]{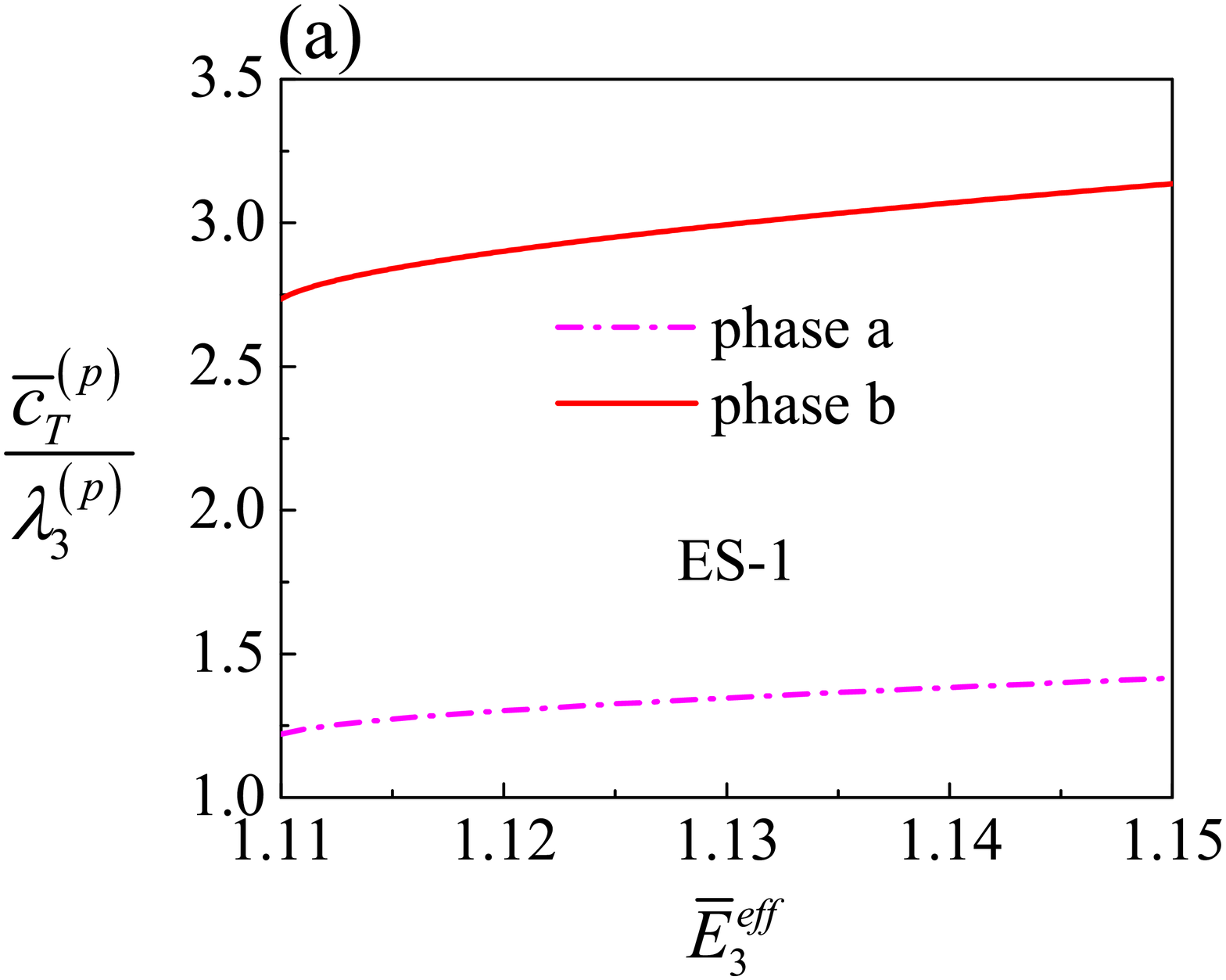}   \hspace{0.03\textwidth}
	\includegraphics[width=0.45\textwidth]{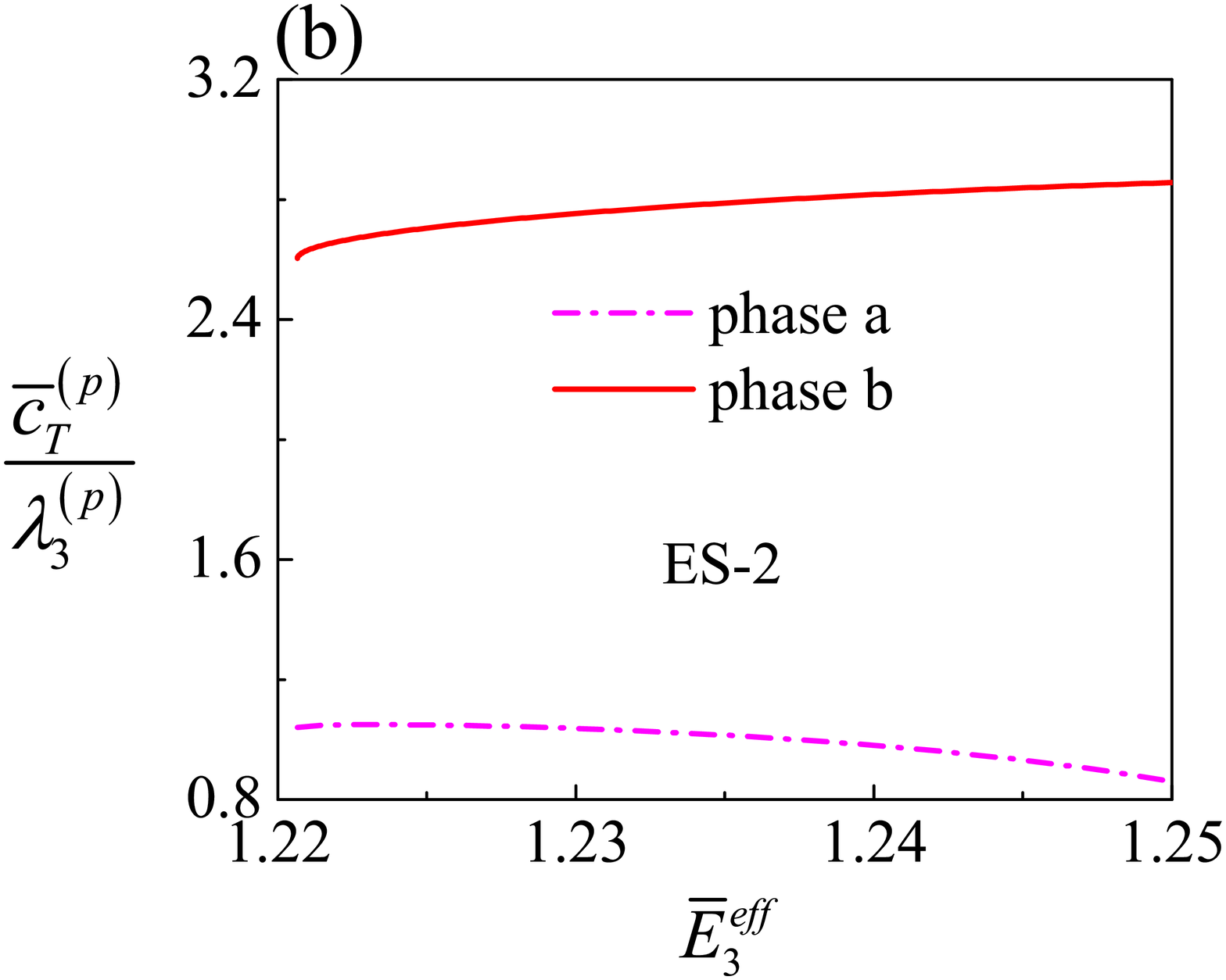}
	\caption{The variation of parameter $\bar c_T^{\left( p \right)}/\lambda _3^{\left( p \right)}$ for both DE phases with the dimensionless electric field $\overline E_3^{\text{eff}}$ for loading path $C$-$D$ and different electrostriction parameters: (a) ES-1; (b) ES-2.}
	\label{CT}
\end{figure}

Furthermore, Fig.~\ref{Fig8}(b) for ES-2 parameters shows that the forbidden frequency of incremental shear waves even increases monotonically with the decrease of $\overline E_3^{\text{eff}}$ along the path $C$-$D$, in contrast to the monotonous declining in Figs.~\ref{Fig4}(c) and \ref{Fig8}(a). {\color{red}The reason for this phenomenon can be explained as follows. First, we define the effective shear wave velocity as ${c_T} = \sqrt {c_{55}^*/\rho }$ and hence the wave number is ${k_T} = \omega /{c_T}$. Then, the terms  ${k_T}h$ in Eq.~\eqref{35} for each phase can be rewritten as $\bar \omega {\lambda _3}/{\bar c_T}$, where ${\bar c_T} = \sqrt {Jc_{55}^*/{\mu _0}} $. It is well-known that the frequency of band gaps depends on the dimensionless wave velocity ${\bar c_T}$ and stretch ratio ${\lambda _3}$ \citep{galich2017elastic}. Specifically, an increase in ${\bar c_T}$ stands for the increase of material stiffness, which is accompanied by an ascending change in frequency; conversely, with the increase of ${\lambda _3}$, the deformed size of structure becomes larger, which may result in the decrease in frequency. Since ${\bar c_T}$  and ${\lambda _3}$ affect the frequency of band gaps in the opposite way, the variation of parameter ${\bar c_T}/{\lambda _3}$ as a whole with the effective electric field for loading path $C$-$D$ is displayed in Fig.~\ref{CT} to take the effects of both phase material properties and size into consideration. As we can see from Fig.~\ref{CT}(a) for ES-1, $\bar c_T^{\left( p \right)}/\lambda _3^{\left( p \right)}$ becomes smaller for both phases with the decrease of $\overline E_3^{\text{eff}}$ from state $C$ to state $D$, which means that the shear wave band gap may be moved downwards by decreasing the electric field, as shown in Fig.~\ref{11}(a). But for ES-2, Fig.~\ref{CT}(b) demonstrates that the curves of phases $a$ and $b$ vary in opposite trends (i.e., with the decrease of $\overline E_3^{\text{eff}}$, $\bar c_T^{\left( p \right)}/\lambda _3^{\left( p \right)}$ increases for phase $a$, but decreases for phase $b$). Because the increase in $\bar c_T^{\left( a \right)}/\lambda _3^{\left( a \right)}$  exceeds the decrease in $\bar c_T^{\left( b \right)}/\lambda _3^{\left( b \right)}$, the shear wave band gap in Fig.~\ref{11}(b) moves up with the decrease of electric field. Similar explanations can be applied to other loading paths and longitudinal waves.}
  
\begin{figure}[htbp]
	\centering	
	\includegraphics[width=0.45\textwidth]{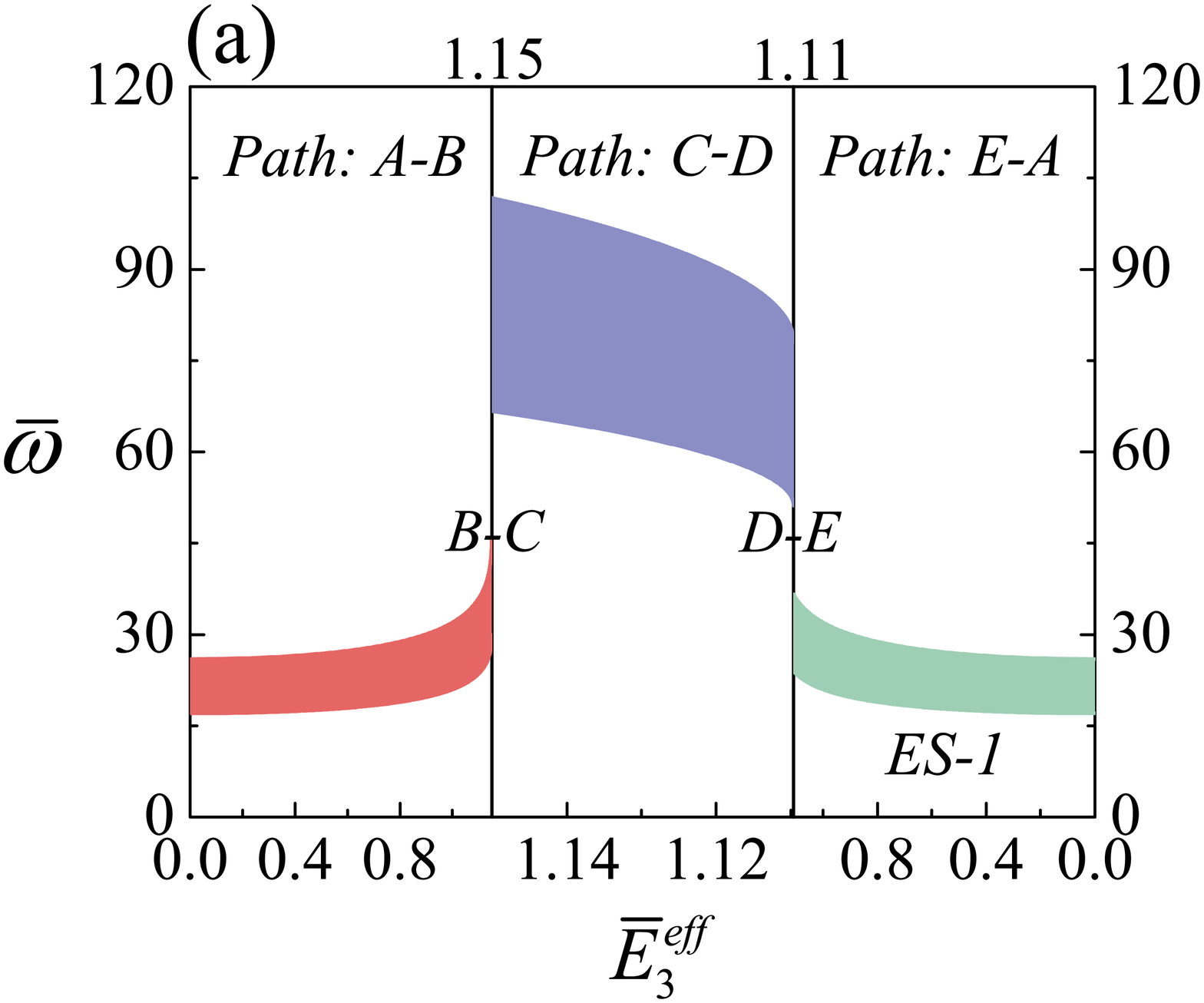} \hspace{0.03\textwidth}
	\includegraphics[width=0.45\textwidth]{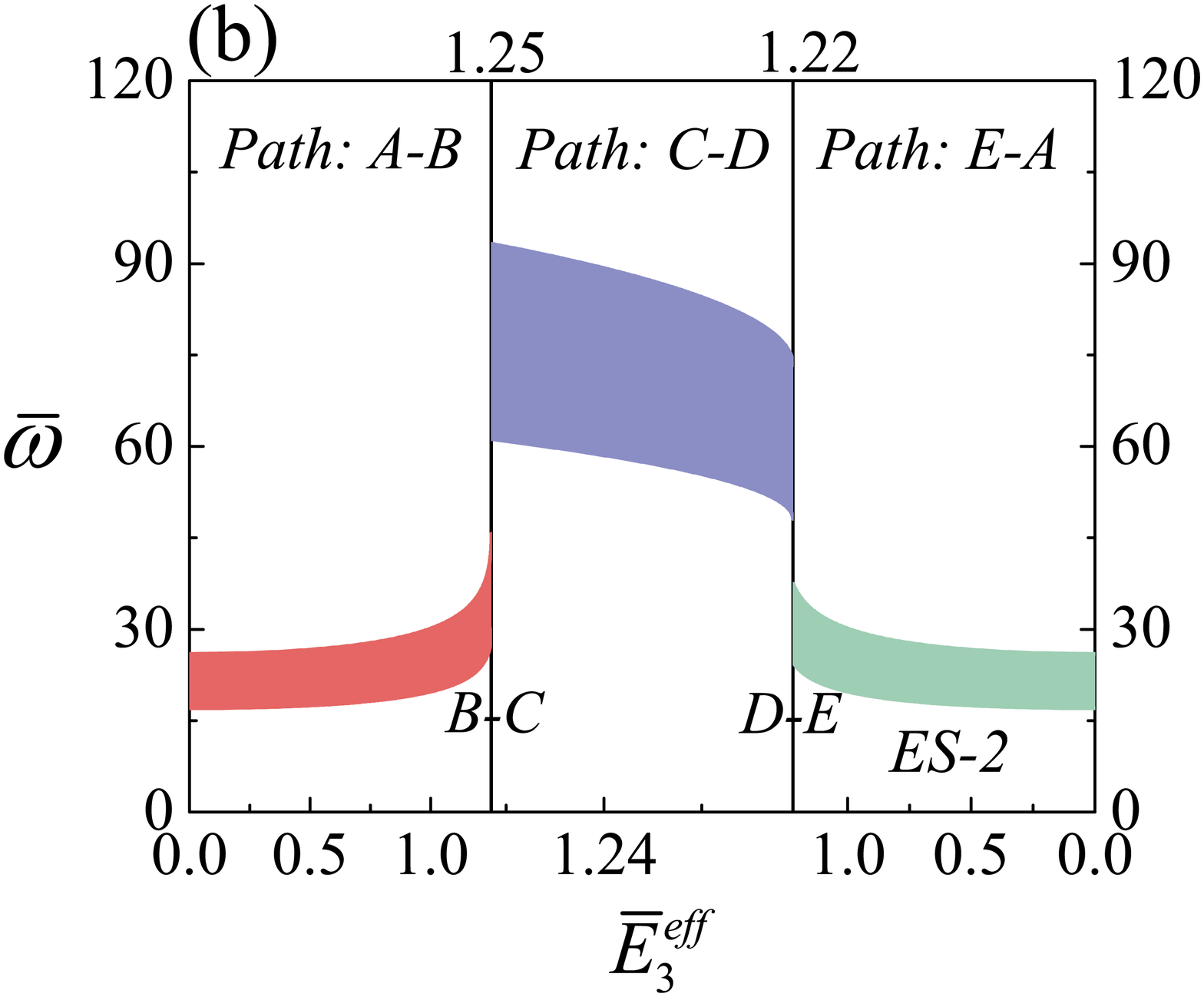}
	\caption{The frequency limits of the first band gap of incremental \emph{longitudinal} waves versus the dimensionless electric field $\overline E_3^{\text{eff}}$ in the periodic DE laminate for the enriched Gent model ($J_m=10$) with different electrostriction parameters: (a) ES-1; (b) ES-2.}
	\label{Fig9}
\end{figure}

Fig.~\ref{Fig9} illustrates the variation trend of the first band gap of longitudinal waves with the electric field $\overline E_3^{\text{eff}}$ for the enriched DE model with the ES-1 and ES-2 parameters. Along the loading path $A$-$B$, qualitatively analogous characteristics to those of the ideal Gent DE model are observed, which are different from the results of shear waves. The reason is that the geometrical change {\color{red}dominates} the longitudinal wave band gaps and the critical stretch ratio $\lambda_{\text{cr}}$ at state $B$ is almost independent of the effect of electrostriction (as shown in Fig.~\ref{Fig7}). Besides, both of the two sets of electrostrictive parameters lower the {\color{red}frequency} jump for the snap-through transition, compared with the ideal DE model.

In brief, the electrostrictive effect increases the critical electric field where the snap-through instability happens, and it {\color{red}weakens} the sharp transition of band gaps for both shear and longitudinal waves. {\color{red}Since different electrostrictive materials influence the superimposed wave propagation behaviors in different ways (see Fig.~\ref{Fig8}), a guided wave-based testing method may be developed to characterize the electrostrictive parameters of different DEs.}
  
\subsection{Influence of the prestress} \label{Sec5-3}
  
\begin{figure}[htbp]
	\centering	
	\includegraphics[width=0.5\textwidth]{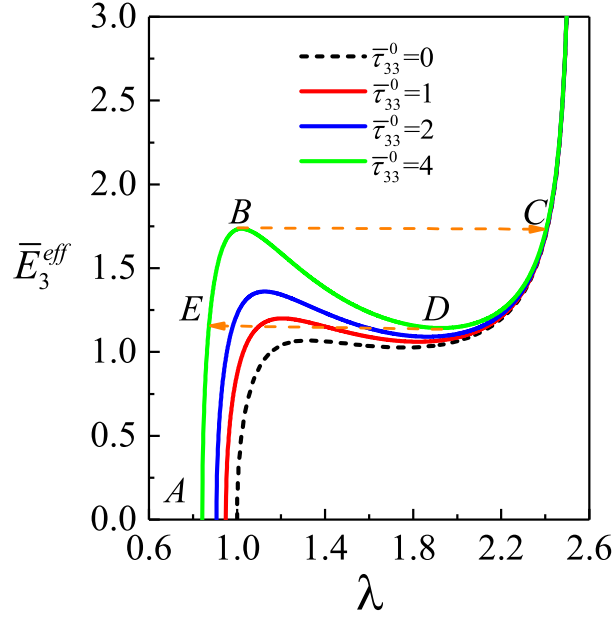}
	\caption{Nonlinear response of the lateral stretch $\lambda $ to the effective electric field $\overline E_3^{\text{eff}}$ in the periodic DE laminate subjected to different {\color{red}prestress values of} $\overline{\tau}_{\text{33}}^{\text{0}}=0, 1, 2, 4$ (the ideal Gent model with ${{J}_{m}}=10$ is adopted).}
	\label{Fig10}
\end{figure}

Except for the electrostrictive effect discussed in Sec.~\ref{Sec5-2}, the mechanical loading (here is the prestress ${\tau }_{\text{33}}^{\text{0}}$ in the thickness direction displayed in Fig.~\ref{Fig1}) can also be exploited to tune the snap-through instabilities and manipulate the wave propagation in the periodic DE laminate. For the ideal Gent model with $J_m=10$, the nonlinear response of the periodic DE laminate under electric stimuli are plotted in Fig.~\ref{Fig10} for different {\color{red}prestress values of} $\overline{\tau }_{\text{33}}^{\text{0}}=0,1,2$ and 4. We see from Fig.~\ref{Fig10} that a larger prestress is accompanied by a smaller lateral stretch $\lambda$ at the initial state $A$ ($\overline E_3^{\text{eff}}=0$). The critical electric fields $\overline E_3^{\text{cr}}$ at states $B$ and $D$ rises with the increase of prestress, which implies that the application of prestress in the thickness direction is beneficial to stabilizing the periodic DE laminate. Moreover, as the prestress increases, the resulting lateral deformation at stable state $C$ after the snap-through transition triggered by electric field becomes remarkably larger. Finally, independent of the applied prestress, the nonlinear responses approach to a vertical asymptote, where the Gent DE laminate reaches the lock-up stretch state that can be calculated as $\lambda_{lim}\simeq2.55$ according to $J_{m}=2 {\lambda _{lim}^2}+{\lambda _{lim}^{-4}}-3=10$. 

\begin{figure}[htbp]
	\centering	
	\includegraphics[width=1.02\textwidth]{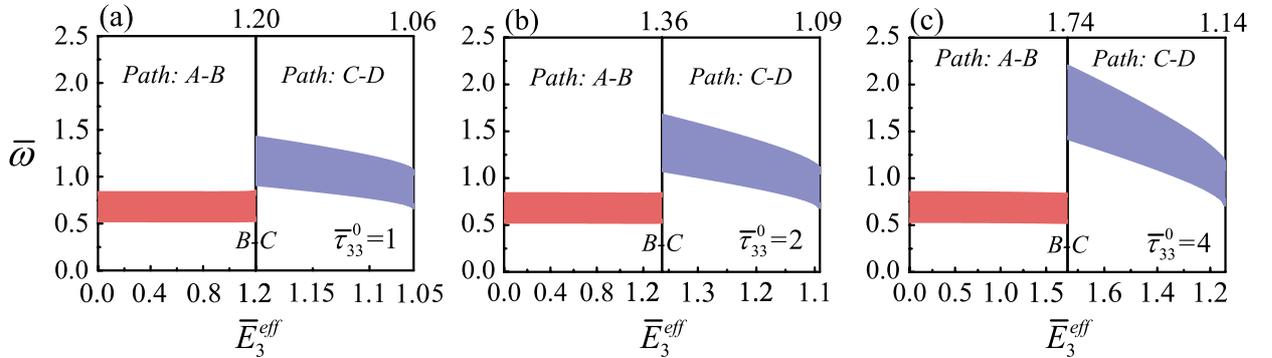}
	\caption{The frequency limits of the first band gap of incremental \emph{shear} waves versus the dimensionless electric field $\overline E_3^{\text{eff}}$ in the periodic DE laminate for the ideal Gent model ($J_m=10$) and various {\color{red}prestress values}: (a) $\overline{\tau }_{33}^{0}=1$; (b) $\overline{\tau }_{33}^{0}=2$; (c) $\overline{\tau }_{33}^{0}=4$.}
	\label{Fig11}
\end{figure}

\begin{figure}[htbp]
  	\centering	
  	\includegraphics[width=1.02\textwidth]{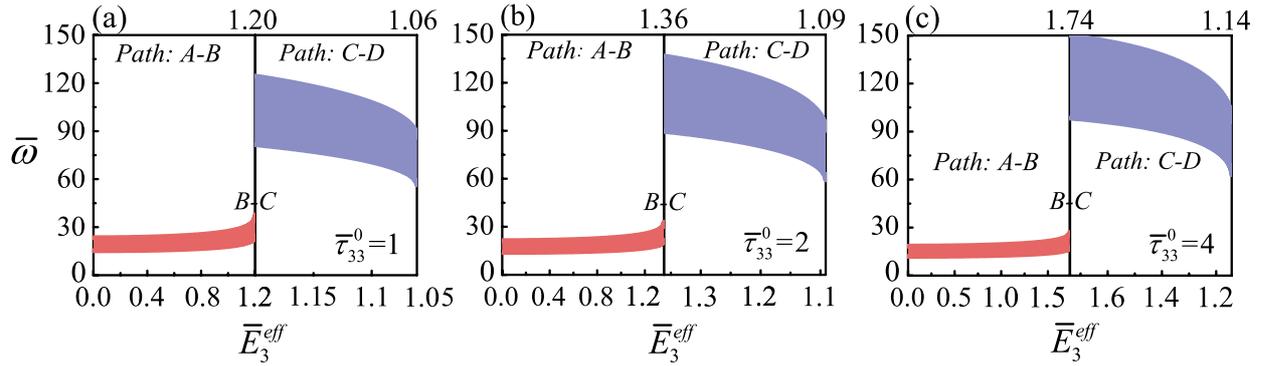}
  	\caption{The frequency limits of the first band gap of incremental \emph{longitudinal} waves versus the dimensionless electric field $\overline E_3^{\text{eff}}$ in the periodic DE laminate for the ideal Gent model ($J_m=10$) and various {\color{red}prestress values}: (a) $\overline{\tau }_{33}^{0}=1$; (b) $\overline{\tau }_{33}^{0}=2$; (c) $\overline{\tau }_{33}^{0}=4$.}
  	\label{Fig12}
\end{figure}

For different {\color{red}prestress values}, the first band gaps of shear and longitudinal waves as functions of the dimensionless electric field are demonstrated in Figs.~\ref{Fig11} and \ref{Fig12}, respectively, for the ideal Gent DE laminate with $J_m=10$. We can obtain qualitatively the same variation trends as those in Figs.~\ref{Fig4}(c) and \ref{Fig6}(c) for zero prestress state: as the electric field ascends to the critical value, the shear wave band gap in path $A$-$B$ may hardly be changed, while the longitudinal one increases monotonically; along path $C$-$D$, the band gaps for both shear and longitudinal waves are shifted to lower frequencies with their width becoming narrower. Nevertheless, it should be emphasized that the jump of band gaps from state $B$ to state $C$ via the snap-through transition becomes more significant with increasing prestress for both types of waves. What's more, the band gaps in path $C$-$D$ can be continuously adjusted in a wider range of the electric field. We can take the shear waves in Fig.~\ref{Fig11} as the example: for $\overline{\tau }_{33}^{0}=1$, the applied electric field can be tuned from 1.20 to 1.06 (i.e., from state $C$ to state $D$ in
Fig.~\ref{Fig10}) with its central frequency varying from 1.17 to 1.02 , while for $\overline{\tau }_{33}^{0}=4$, the electric field changes from 1.74 to 1.14 with the tuning range of central {\color{red}frequency} from 1.81 to 0.91.

\section{Conclusions}\label{section6}

  
The nonlinear response and superimposed elastic wave motions in an infinite periodic DE laminate subjected to electromechanical loadings are theoretically investigated in this work. First of all, the theory of nonlinear electroelasticity proposed by Dorfmann and Ogden is employed to describe the nonlinear static deformations in periodic DE laminates under combined action of prestress and electric field, where the enriched {\color{red}Gent-Gent} DE model is used to characterize the strain-stiffening, {\color{red}the second strain invariant} and electrostrictive effects. Utilizing the transfer matrix method as well as the Bloch-Floquet theorem, the dispersion relations of decoupled shear and longitudinal waves superimposed on finite deformations are obtained, which are in the same form as the purely elastic counterpart. At last, numerical investigations are conducted to demonstrate how the material properties and external stimuli affect the nonlinear static response and wave propagation characteristic. What follows is a list of useful numerical findings from the present investigation: 

\begin{enumerate}[(1)]
	\item For the periodic DE laminate described by ideal Gent model, the snap-through instability originating from geometrical and material nonlinearities can be harnessed to achieve the drastic change in geometrical configuration and band gaps (including the position and width).
	
	\item The snap-through transition may not appear when the strain-stiffening effect is strong enough (such as ${{J}_{m}}=5$) or when the neo-Hookean model is adopted. Thus, the shear and pressure wave band gaps can be tuned continuously by {\color{red}adjusting} the electric field.
	
    {\color{red}\item Increasing the second modulus raises the critical electric field and lifts up the position of band gaps for both types of waves;}
		
	\item Both the electrostrictive effect and applied prestress contribute to stabilizing the periodic DE laminate. However, they affect the band structure in different manners: the former weakens the band gap jump induced by the snap-through transition, while the latter strengthens the tunable frequency range for both types of waves.
	
	\item Due to the difference in physical mechanism dominating the wave motions, the electrostatically tunable effect on the band gap of longitudinal waves is more evident than that on the shear wave band gap.
\end{enumerate}

{\color{red}The results obtained here may provide a strong theoretical guidance for the design and fabrication of electrostatically tunable DE wave devices and for the wave-based characterization method of electrostrictive materials. Experiments will definitely be beneficial to understand the wave propagation behaviors in periodic DE laminates, but they are out of the scope of this paper. 
	
Note that viscoelasticity neglected in this work is indeed an important factor in the analysis of wave propagation behaviors. However, it should be emphasized that the dissipation/damping effect of soft materials on wave propagation is generally significant in the high frequency range, but is usually negligible at low frequencies (see, for example, \citet{li2017guided, li2018observation, li2019harnessing} and \citet{gao2019harnessing}). Here we mainly focus on the first-order band gap in periodic DE laminates, which is the most important one in real-life situations. Numerical calculations indicate that the frequency of the first-order shear wave band gap is relatively low, but that of the pressure wave band gap is a little higher. As a result, the dissipation/damping in DE materials will hardly affect the first-order shear wave band gap while it will have remarkable suppressive effect on the first-order pressure wave band gap. Thus, consideration of dissipation can potentially improve the predictions of the first-order pressure wave band gap, which demands further research.
}
  

\section*{Acknowledgements}

This work was supported by a Government of Ireland Postdoctoral Fellowship from the Irish Research Council (Nos. GOIPD/2019/65 and GOIPG/2016/712) and the National Natural Science Foundation of China (Nos. 11872329, 11621062 and 11532001). Partial supports from the Fundamental Research Funds for the Central Universities, PR China (No. 2016XZZX001-05) and the Shenzhen Scientific and Technological Fund for R\&D, PR China (No. JCYJ20170816172316775) are also acknowledged.
  
\appendix

{\color{red}\section{Representative electrostrictive models}\label{AppeA}

This appendix will list the specific forms of energy density function for several representative electrostrictive models suggested in the literature to characterize isotropic DE materials. The energy density function $\Omega ^{\text{elec}}({\bf{F}},\bm{\mathcal{D}})$ related to the electrostrictive effect and purely electrostatic field is usually assumed to linearly depend on the three invariants ${I_4}$, ${I_5}$ and ${I_6}$ as 
\begin{equation} \label{A.1}
\Omega ^{\text{elec}}({\bf{F}},\bm{\mathcal{D}})= \frac{1}{{2{\varepsilon}J}}(\gamma _0{I_4} + \gamma _1{I_5} + \gamma _2{I_6}),
\end{equation}
where ${{\varepsilon }}={{\varepsilon }_{0}}\varepsilon _{r}$ is the material permittivity in the undeformed state. In general, the three coefficients $\gamma _0$, $ \gamma _1$ and $ \gamma _2$ are functions of the other three invariants ${I_1}$, ${I_2}$ and ${I_3}$ associated with the deformation. For constant coefficients, Eq.~\eqref{A.1} recovers the compressible enriched DE model \eqref{17} \citep{galich2016manipulating} and the incompressible electrostrictive model with $J=1$ \citep{gei2014role}. Furthermore, if $\gamma _2=0$ and $J=1$, Eq.~\eqref{A.1} reduces to the incompressible electrostrictive model used in \citet{dorfmann2010electroelastic}.

Moreover, neglecting the dependence on ${I_4}$ and ${I_6}$, Eq.~\eqref{A.1} becomes
\begin{equation} \label{A.2}
\Omega ^{\text{elec}}({\bf{F}},\bm{\mathcal{D}})= \frac{\gamma _1({\bf{F}})}{{2{\varepsilon}J}}{I_5} \equiv \frac{{I_5}}{{2J{\varepsilon_d}({\bf{F}})}},
\end{equation}
where ${\varepsilon_d}({\bf{F}})= {\varepsilon}/{\gamma _1({\bf{F}})}$ is the deformation-dependent material permittivity. The soft electro-active material characterized by Eq.~\eqref{A.2} is called \emph{quasilinear dielectrics} \citep{zhao2008electrostriction}.

 
Based on the free energy density function $\Psi({\mathbf{F}},\mathbf{E})$ per unit mass rather than the total energy density function $\Omega (\mathbf{F},\bm{\mathcal{D}})$ and using Eulerian electric field vector $\mathbf{E}$ as independent electric variable instead of Lagrangian electric displacement vector $\bm{\mathcal{D}}$ adopted in this work, \citet{vertechy2013continuum} obtained the nonlinear constitutive equations (neglecting terms related to temperature field in their work) as
\begin{equation} \label{A.3}
{\bm{\tau }} =\rho \frac{\partial \Psi }{\partial \mathbf{F}}{{\mathbf{F}}^{\text{T}}}+{{\bm{\sigma }}_{\text{M}}}+\mathbf{E}\otimes \mathbf{P}, \quad
\mathbf{P} =-\rho \frac{\partial \Psi }{\partial \mathbf{E}},
\end{equation}
where ${{\bm{\sigma }}_{\text{M}}}={{\varepsilon }_{0}}[\mathbf{E}\otimes \mathbf{E}-(\mathbf{E}\cdot \mathbf{E})\mathbf{I}/2]$ is the well-known Maxwell's stress tensor in vacuum and $\mathbf{P} =\mathbf{D}-{{\varepsilon }_{0}}\mathbf{E}$ is the electric polarization vector. It has been verified (see Secs. 2.4 and 2.5 in the review work by \citet{wu2016theory}) that the nonlinear constitutive equation \eqref{A.3} can be transformed to Eq.~\eqref{5} by introducing the following augmented free energy density function ${{\Omega }^{*}}\left( \mathbf{F},\bm{\mathcal{E}} \right)$ and conducting the following Legendre transformation
\begin{align}
{{\Omega }^{*}}\left( \mathbf{F},\bm{\mathcal{E}} \right)={{\rho }_{0}}\Psi \left( \mathbf{F},\mathbf{E} \right)-\pi \left( \mathbf{F},\mathbf{E} \right), \quad 
\Omega \left( \mathbf{F},\bm{\mathcal{D}} \right)={{\Omega }^{*}}\left( \mathbf{F},\bm{\mathcal{E}} \right)+\bm{\mathcal{E}}\cdot \bm{\mathcal{D}},
\end{align}
where
\begin{equation}
\pi \left( \mathbf{F},\mathbf{E} \right)={{\varepsilon }_{0}}J\bm{\mathcal{E}}\cdot \left( {{\mathbf{C}}^{-1}}\bm{\mathcal{E}} \right)/2={{\varepsilon }_{0}}J\mathbf{E}\cdot \mathbf{E}/2,\quad \bm{\mathcal{E}}\cdot \bm{\mathcal{D}}=J\mathbf{E}\cdot \mathbf{D}.
\end{equation}
For isotropic DE materials, $\Psi({\mathbf{F}},\mathbf{E})$ depends on the following six invariants:
\begin{align}\label{A.6}
{{\bar{I}}_{1}}&=\text{tr}\mathbf{b}=I_1,\quad {{\bar{I}}_{2}}=\frac{1}{2}\left[ {{\left( \text{tr}\mathbf{b} \right)}^{2}}-\text{tr}\left( {{\mathbf{b}}^{2}} \right) \right]=I_2,\quad {{\bar{I}}_{3}}=\det \mathbf{b}={{J}^{2}}=I_3, \notag \\
{{\bar{I}}_{4}}&=\mathbf{E}\cdot \mathbf{E},\quad {{\bar{I}}_{5}}=\mathbf{E}\cdot \left( \mathbf{bE} \right),\quad {{\bar{I}}_{6}}=\mathbf{E}\cdot \left( {{\mathbf{b}}^{2}}\mathbf{E} \right),
\end{align}
substitution of which into Eq.~\eqref{A.3} yields
\begin{align} \label{A.7}
& {\bm{\tau }}=2\rho \left[ \left( {{\Psi }_{1}}+{{\Psi }_{2}}{{{\bar{I}}}_{1}} \right)\mathbf{b}-{{\Psi }_{2}}{{\mathbf{b}}^{2}}+{{\Psi }_{3}}{{{\bar{I}}}_{3}}\mathbf{1}-{{\Psi }_{4}}\mathbf{E}\otimes \mathbf{E}+{{\Psi }_{6}}\left( \mathbf{bE} \right)\otimes \left( \mathbf{bE} \right) \right]+{{\bm{\sigma }}_{\text{M}}}, \notag \\ 
& \mathbf{D}=-2\rho \left( {{\Psi }_{4}}\mathbf{E}+{{\Psi }_{5}}\mathbf{bE}+{{\Psi }_{6}}{{\mathbf{b}}^{2}}\mathbf{E} \right), 
\end{align}
where ${{\Psi }_{i}}=\partial \Psi /\partial {{\bar{I}}_{i}}(i=1-6)$. In particular, \citet{vertechy2013continuum} proposed a \emph{compressible Mooney-Rivlin} electrostrictive model with $\Psi({\mathbf{F}},\mathbf{E})$ decomposed in the following three parts
\begin{align} \label{A.8}
& \Psi ={{\Psi }^{\text{hyel}}}\left( {{{\bar{I}}}_{1}},{{{\bar{I}}}_{2}},J \right)+{{\Psi }^{\text{estr}}}\left( J,{{{\bar{I}}}_{4}},{{{\bar{I}}}_{5}} \right)+{{\Psi }^{\text{esta}}}\left( J,{{{\bar{I}}}_{4}} \right), \notag \\ 
& {{\Psi }^{\text{hyel}}}=\frac{1}{{{\rho }_{0}}}\left[ {{c}_{10}}\left( {{{\bar{I}}}_{1}}-3 \right)+{{c}_{20}}\left( {{{\bar{I}}}_{2}}-3 \right)+{{c}_{30}}{{\left( J-1 \right)}^{2}}-2\left( {{c}_{10}}+2{{c}_{20}} \right)\ln J \right], \notag \\ 
& {{\Psi }^{\text{estr}}}=-\frac{1}{2{{\rho }_{0}}}\left[ \frac{{{\beta }_{10}}\left( {{{\bar{I}}}_{5}}-{{{\bar{I}}}_{4}} \right)J}{2}+{{\beta }_{20}}{{{\bar{I}}}_{5}}\left( J-1 \right) \right], \quad {{\Psi }^{\text{esta}}}=-{{\varepsilon }_{0}}\frac{{{\chi }_{0}}{{{\bar{I}}}_{4}}J}{2{{\rho }_{0}}},
\end{align}
where $c_{10}$, $c_{20}$, $c_{30}$, $\beta_{10}$ and $\beta_{20}$ are constant material parameters and ${{\chi }_{0}}={{\varepsilon }_{r}}-1$ is the electric susceptibility. The detailed explanation of the three parts in Eq.~\eqref{A.8} can be found in \citet{vertechy2013continuum}. Inserting Eq.~\eqref{A.8} into Eq.~\eqref{A.7} leads to specific constitutive relations (see Eqs.~(80)-(81) and (85)-(87) in \citet{vertechy2013continuum}). Note that due to the different choices of energy density function and independent electric variable, it is intractable to establish a direct and explicit connection between the nonlinear constitutive equations \eqref{consti} and those in \citet{vertechy2013continuum}.}
  
\section{Nonzero components of instantaneous elelctroelastic moduli used in this paper}\label{AppeB}

According to the formulations derived by \citet{dorfmann2010electroelastic} and \citet{galich2016manipulating} and for the compressible {\color{red}Gent-Gent} DE model in Eqs.~\eqref{17} and \eqref{18}, the nonzero components of instantaneous elelctroelastic moduli tensors ${{\bm{\mathcal{A}}}_{0}}$, ${{\bm{\mathcal{M}}}_{0}}$ and ${{\bm{\mathcal{R}}}_{0}}$ utilized in Sec.~\ref{Sec4} read
\begin{align}
{\cal A}_{03131} &= {\cal A}_{03232} = \frac{{{\mu }}}{{{J}}}\frac{{J_m}}{{J_m - I_1 + 3}}{\lambda _3^2}+ {\color{red}\frac{{2{\lambda _3}{c_2}}}{{{I_2}}}}+ \frac{{D_3^2}}{{{\varepsilon}}} \left[ {\gamma _1 + \left( {{\lambda ^2}+2{\lambda _3^2} } \right)\gamma _2} \right], \nonumber \\
{\cal A}_{03333} &= {\lambda ^2}\lambda _3\left( {\Lambda _0 - \frac{{2{\mu}}}{{J_m}}} \right) + \frac{{{\mu}}}{{{\lambda ^2}\lambda _3}}\left[ {\frac{{J_m}}{{J_m - I_1 + 3}}{\lambda _3^2} + 1} \right] + {\color{red} \frac{{4{\lambda _3}{c_2}}}{{{I_2}}}\left( {1 - \frac{{4{\lambda ^2}\lambda _3^2}}{{{I_2}}}} \right)} \nonumber\\
&+ \frac{{2{\mu}}}{{{\lambda ^2}\lambda _3}}\frac{{J_m}}{{{{\left( {J_m - I_1 + 3} \right)}^2}}}{\lambda _3^4} + \frac{{D_3^2}}{{{\varepsilon }}}\left( {{\lambda _3^{ - 2}}\gamma _0 + 3{\lambda _3^2}\gamma _2} \right),\nonumber \\
{\cal M}_{0311} &= {\cal M}_{0322}= \frac{{{D_3}}}{{{\varepsilon }}}\left[ {\gamma _1 + \left( {{\lambda ^2} + {\lambda _3^2}} \right)\gamma _2} \right] ,\nonumber \\
{\cal R}_{011} &= \frac{1}{{{\varepsilon}}}\left( {{\lambda ^{ - 2}}\gamma _0 + \gamma _1 + {\lambda ^2}\gamma _2} \right).\nonumber
\end{align}
 


\section*{References}

\bibliographystyle{elsarticle-harv.bst}
\nocite{*}
\bibliography{DE_PC.bib}







\end{document}